\documentclass[prb,amsmath,amssymb,lengthcheck,showpacs,floatfix,groupedaddress,superscriptaddress,longbibliography]{revtex4-1}
\usepackage{graphicx}
\usepackage[english]{babel}
\usepackage{times}
\usepackage{dcolumn}
\usepackage[usenames,dvipsnames]{color}
\usepackage{units}
\usepackage{bm}
\usepackage{soul}
\usepackage[utf8]{inputenc}
\usepackage{float}

\graphicspath{ {./figures/} }

\begin{document}

\newcommand{\EC}[1]{E_C^{( #1 )}}
\newcommand{\ECL}{\EC{L}}
\newcommand{\ECR}{\EC{R}}

\newcommand{\GL}{\Gamma_L}
\newcommand{\GR}{\Gamma_R}
\newcommand{\dg}{\delta\Gamma}

\newcommand{\nn}[1]{n_0^{( #1 )}}
\newcommand{\nL}{\nn{L}}
\newcommand{\nR}{\nn{R}}    

\newcommand{\ket}[1]{\vert #1 \rangle}
\newcommand{\bra}[1]{\langle #1 \vert}

\newcommand{\ketg}{\ket{\psi_g}}
\newcommand{\ketu}{\ket{\psi_u}}

\newcommand{\NN}{\mathcal{N}}
\newcommand{\ketref}{\ket{\psi_\mathrm{ref}}}
\newcommand{\nref}{n_\mathrm{ref}}
\newcommand{\expv}[1]{\langle #1 \rangle}
\newcommand{\phiL}{\ket{\phi_L}}
\newcommand{\phiR}{\ket{\phi_R}}

\newcommand{\cisL}{ c_{i, \sigma, L } }
\newcommand{\ciupL}{ c_{i, \uparrow, L } }
\newcommand{\cidnL}{ c_{i, \downarrow, L } }
\newcommand{\cjupL}{ c_{j, \uparrow, L } }
\newcommand{\cjdnL}{ c_{j, \downarrow, L } }

\newcommand{\cisR}{ c_{i, \sigma, R } }
\newcommand{\ciupR}{ c_{i, \uparrow, R } }
\newcommand{\cidnR}{ c_{i, \downarrow, R } }
\newcommand{\cjupR}{ c_{j, \uparrow, R } }
\newcommand{\cjdnR}{ c_{j, \downarrow, R } }

\newcommand{\gis}{ g_{i, \sigma } }
\newcommand{\gjs}{ g_{j, \sigma } }
\newcommand{\giup}{ g_{i, \uparrow } }
\newcommand{\gidn}{ g_{i, \downarrow } }
\newcommand{\gjup}{ g_{j, \uparrow } }
\newcommand{\gjdn}{ g_{j, \downarrow } }

\newcommand{\uis}{ u_{i, \sigma } }
\newcommand{\ujs}{ u_{j, \sigma } }
\newcommand{\uiup}{ u_{i, \uparrow } }
\newcommand{\uidn}{ u_{i, \downarrow } }
\newcommand{\ujup}{ u_{j, \uparrow } }
\newcommand{\ujdn}{ u_{j, \downarrow } }

\title{Qubit based on spin-singlet Yu-Shiba-Rusinov states}
\author{Luka Pave\v{s}i\'{c}}
\author{Rok \v{Z}itko}
\email{rok.zitko@ijs.si}
\affiliation{Jo\v{z}ef Stefan Institute, Jamova 39, SI-1000 Ljubljana, Slovenia}
\affiliation{Faculty of Mathematics and Physics, University of Ljubljana, Jadranska 19, SI-1000 Ljubljana, Slovenia}

\begin{abstract}
The local magnetic moment of an interacting quantum dot occupied by a single electron can be screened by binding a Bogoliubov quasiparticle from a nearby superconductor. This gives rise to a long-lived discrete spin-singlet state inside the superconducting gap, known as the Yu-Shiba-Rusinov (YSR) state. We study the nature of the subgap states induced by a quantum dot embedded between two small superconducting islands. We show that this system has two spin-singlet subgap states with different spatial charge distributions. These states can be put in a linear superposition and coherently manipulated using electric-field pulses applied on the gate electrode. Such YSR qubit could be implemented using present-day technology.
\end{abstract}

\maketitle

\section{INTRODUCTION}
The coupling of an impurity carrying a local magnetic moment to a
superconductor (SC) produces discrete subgap states, known as the
Yu-Shiba-Rusinov states \cite{Yu, Shiba, Rusinov}. These are spin-singlet bound states of Bogoliubov quasiparticles screening the impurity spin through antiferromagnetic Kondo exchange interactions. The energy gain from this coupling allows them to descend deep into the superconducting gap. 
Such subgap states appear in the spectra of many modern superconducting devices, where local magnetic moments arise in multiple ways, such as in adsorbed magnetic atoms or molecules, or in semiconductor quantum dots (QD)\cite{Balatasky2006_review, MartinRodero_2011, Heinrich_2018, Lutchyn_2018, Meden2019, Prada_2020, Frolov_2020}. 

Systems of this kind are commonly modeled by extending the Anderson impurity model \cite{SIAM} to the case of a superconducting bath, typically described by the BCS mean-field theory \cite{BCS}. Such Hamiltonians can be reliably solved by various numerical approaches, with the numerical renormalization group (NRG) proving the most successful \cite{Wilson_NRG, Krishnamurthy_NRG, Bulla2008, Sakai_1993, Yoshioka_2000, Satori_1992}.

Recent developments of experimental techniques enabled the fabrication of epitaxial SC islands in hybrid semiconductor-superconductor devices that are small enough for the Coulomb repulsion between the electrons to be important \cite{Krogstrup_2015, Chang_2015}.
This is taken into account by an effective charging term $E_C \hat{n}_\mathrm{SC}^2$, where $ \hat{n}_\mathrm{SC}$ is the SC electron number operator and $E_C = e_0^2/2C$ is the charging energy with $C$ the island's total capacitance \cite{Flensberg_1993, Matveev_1995, vonDelft2001, Wiel_2002, Yuval_2003, Frithjof_2004, Frithjof_2005, Zhang_2019, Mitchell_2021}.
Incorporating this term into existing numerical techniques proves very difficult for two reasons.
First, the electron number is not a conserved quantity in the BCS theory, which makes the implementation of an electron number operator problematic \cite{Matveev1991}.
Second, the Coloumbic repulsion of the SC electrons makes the bath an
interacting system, resulting in a problem which cannot be solved by any traditional impurity solver.
Most theoretical treatments were based on ad hoc pictures relying on physical intuition rather than solving microscopic models. 

In a recent paper \cite{paper1} we presented a numerical method which does not suffer from the described problems.
We introduced a model that builds on previous work in the context of ultrasmall superconducting grains \cite{vonDelft2001, RBT1995, RBT1997, Gobert_2004}, where the Richardson-Gaudin charge conserving model of superconductivity \cite{Richardson1963, Richardson1964, vonDelft1999} was successfully applied.
This model describes the SC as a set of single particle energy levels with an attractive all-to-all pairing interaction and is well suited for the description of small SC systems.
Using such model with a few hundred energy levels is very close to the BCS description appropriate in the thermodynamic limit, but retains the advantage of a well defined number of particles and thus the ability to include in a clean and well-defined manner the Coloumbic repulsion in the superconducting island into the model. 
Using the DMRG \cite{DMRG, White_1992, White_1993} as the numerical solver, we are able to treat the charging term exactly and on the same level as other parameters and obtain accurate results in all parameter ranges.
This approach was shown to give results in remarkable agreement with experiments in a setup of a QD with a single SC island \cite{JuanCarlos2021}. 
The implementation based on matrix product states (MPS) allows us to calculate basically any observable and allows for a detailed theoretical insight into the nature of the subgap states.

In this paper, we extend our analysis to a QD embedded between two SC islands. This is motivated by the
considerable experimental interest in complex hybrid devices \cite{DeFranceschi2010, Aguado_2020},
and specifically in the two-channel problems where the impurity is coupled to two superconductors, e.g., embedded in a Josephson junction \cite{Pillet_2010, Chang_2013, Casparis_2016, Casparis_2018, Kurilovich_2021}.
In the absence of flux bias, to make the two superconductors behave as
two independent channels at least one of them needs to have a
significant charging energy. It may be noted that this is also the necessary condition for observing the two-channel Kondo effect in normal-state systems \cite{Yuval_2003, Affleck_2005, Potok2007, Iftikhar_2015, Kirchner_2020}.
Coupling an impurity spin to two independent SC islands produces two singlet subgap states, as a
Bogoliubov quasiparticle from either of the channels can screen the
impurity spin \cite{zitko_2ch}. In this work we investigate the nature of these subgap states in realistic models of hybrid devices,
focusing on the singlet symmetry sector with an even total number of electrons in the system.

This paper is organized as follows. In Sec.~\ref{sec2} we introduce the model and discuss the technical issues arising from the presence of multiple superconducting regions. 
In Sec.~\ref{sec3} we discuss the symmetric situation with two equivalent superconductors, having the same SC gap $\Delta$ and the same charging energy $E_C$. 
In Sec.~\ref{sec4} we analyze the asymmetric situation with different charging energies (e.g., one large superconductor and one small superconducting island). 
In Sec.~\ref{sec5} we model a realistic QD by reducing $U$, and thus moving away from the Kondo limit to an experimentally relevant regime. We briefly discuss the effect of decreased $U$ and then proceed in Sec.~\ref{secgate} with the gate tuning effects that ultimately reveal a regime propitious to operating such devices as a YSR qubit. 
In Sec.~\ref{sec7} we calculate the electric transition moments, quantities important for manipulation of subgap states. 
We close with a discussion and a conclusion.

\section{Model and method}
\label{sec2}

\subsection{Hamiltonian}

The Hamiltonian consists of a single impurity level coupled to two SC islands \cite{Braun_vonDelft_1, Braun_vonDelft_2, SIAM}:
\begin{equation}
H = H_\mathrm{imp} + \sum_{\beta=L,R} \left( H_\mathrm{SC}^{(\beta)} + H_\mathrm{hyb}^{(\beta)} \right),
\label{eq:hamiltonian}
\end{equation}
where 
\begin{equation*}
\begin{aligned}
H_\mathrm{imp} & = \varepsilon \hat{n}_\mathrm{imp} + U \hat{n}_{\mathrm{imp}, \uparrow} \hat{n}_{\mathrm{imp}, \downarrow} = \frac{U}{2} \large( \hat{n}_\mathrm{imp} - \nu \large)^2 + \mathrm{const.}, \\
H_\mathrm{SC}^{(\beta)} = & \sum_{i, \sigma}^N \varepsilon_i c_{i,\sigma,\beta}^\dagger c_{i,\sigma,\beta} - \alpha d \sum_{i,j}^N c_{i, \uparrow, \beta}^\dagger c_{i, \downarrow, \beta}^\dagger c_{j, \downarrow, \beta} c_{j,\uparrow,\beta} \\ 
 & + \EC{\beta} \big( \hat{n}_\mathrm{SC}^{(\beta)} - n_0^{(\beta)} \big)^2, \\
H_\mathrm{hyb}^{(\beta)} & = (v_\beta / \sqrt{N}) \sum_{i\sigma}^N \left( c_{i,\beta,\sigma}^\dagger d_\sigma + d_\sigma^\dagger c_{i,\beta,\sigma} \right).
\end{aligned}
\end{equation*}

\begin{figure}[H]
    \centering
    \includegraphics[width=0.35 \textwidth]{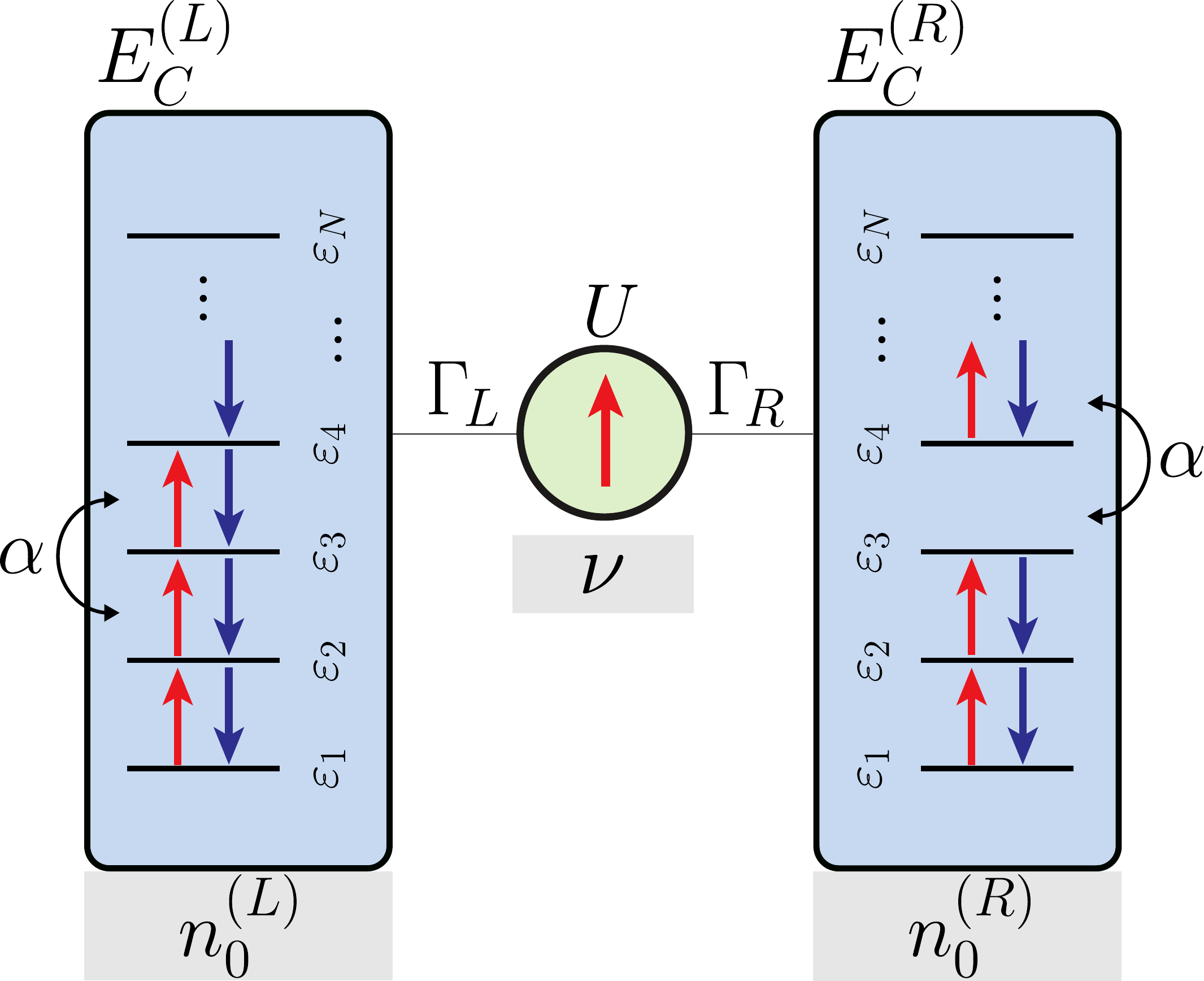}
    \caption{Schematic representation of the system: an interacting
    quantum dot embedded between two
    superconducting islands with charging energies
    $E_C^{(L)}$ and $E_C^{(R)}$. The electron occupancy in the
    different parts of the device is tunable by the gate voltages $\nL$,
    $\nR$ and $\nu$.
    }
    \label{fig:sketch}
\end{figure}

Here $\varepsilon$ is the energy level and $U$ the electron-electron repulsion on the QD. 
The impurity term can be rewritten in terms of $\nu = 1/2 - \varepsilon/U$, the impurity level in units of electron number. 
$d_\sigma$ and $c_{i,\beta,\sigma}$ are the annihilation operators corresponding to the QD and the two SC baths labeled by $\beta=L,R$ (left and right). The spin index is $\sigma=\uparrow, \downarrow$. Each SC bath is modelled by $N$ energy levels $\varepsilon_i$ spaced by a constant separation $d=2D/N$, where $2D$ is the bandwidth. The levels are coupled all-to-all by a pairing interaction with strength $\alpha$ \cite{Braun_vonDelft_1, Braun_vonDelft_2}. The SCs are coupled to the QD with the hybridisation strengths $\Gamma_\beta=\pi \rho v_\beta^2$, where $\rho=1/2D$ is the normal-state density of states in each bath. The number operators are $\hat{n}_\mathrm{imp} = \sum_\sigma d_\sigma^\dagger d_\sigma$ for the impurity, and $\hat{n}^{(\beta)}_\mathrm{SC} = \sum_{i=1,\sigma}^N c_{i,\sigma,\beta}^\dagger c_{i,\sigma,\beta}$ for each SC bath. $\EC{\beta}$ are the charging energies, with $n_0^{(\beta)}$ the optimal occupation of the SC island in units of electron charge.
A sketch of the model system is shown in Fig. \ref{fig:sketch}.
In experimental setups, $\nu$, $n_0^{(\beta)}$ and $\Gamma_\beta$ are typically continuously tunable by the voltages applied to gate electrodes, while $U$, $E^{(\beta)}_C$ and $\Delta$ are device properties only weakly affected by the gate voltages. 

If $\EC{L}=\EC{R}$, $n_0^{(L)}=n_0^{(R)}$ and $\Gamma_L=\Gamma_R$, the problem
has a mirror (left-right) symmetry. In the absence of charging terms, the problem
then simplifies to a single-channel problem. This is easy to establish using an
appropriate unitary transformation after a mean-field decoupling of the pairing terms.
The same transformation applied to the original Hamiltonian leads to terms that mix states in
both superconductors (see Appendix~\ref{appA}), but the effects of these terms become less physically significant with the increasing
number of levels $N$, and disappear in the thermodynamic limit (with the gauge symmetry breaking).

\subsection{Parameter choices}

The calculations are performed for $N=200$ energy levels in each SC. Using the half-bandwidth $D=1$ as the energy unit, this corresponds to an inter-level spacing of $d = 2D/N = 0.01$. 
We choose $\alpha = 0.4$, which gives rise to the SC gap $\Delta=0.165$ in each bath. This value is chosen so that an appropriate number of levels is engaged in the pairing interaction thus minimizing the finite-size effects, while also minimizing the finite-bandwidth effect. Disregarding the finite-size corrections, the low-energy properties of the model are universally scalable by $\Delta$. 

We have implemented the Hamiltonian in the matrix product operator (MPO) form using matrices of dimension $9\times 9$. The full expressions are given in Appendix \ref{appB}. We use the density matrix renormalization group (DMRG) \cite{DMRG} to calculate the lowest eigenstates of the system in symmetry sectors defined by the total number of particles $n$ and the total $z$ component of spin $S_z$. We denote these by $\vert n, i \rangle$, with $i=0$ the ground state (GS) in a given sector, and $i=1$ the first excited state (ES). The states with even $n$ are spin-singlets with $S_z=0$, the states with odd $n$ are spin-doublets with $S_z=\pm 1/2$. The reference state $\ketref$ is defined by filling the system with an even-integer number of electrons $\NN$ in each SC and one electron in the QD, for a total of $\nref=2\NN+1$ electrons. This spin-doublet state is the thermodynamic ground state of the system for $\nu=1$, $n_0^{(\beta)}=\NN$ in the limit $\Gamma_\beta \to 0$; it is a product state of two BCS wavefunctions in SCs (each projected to a fixed electron number $\NN$) and a local moment on the QD site.
We typically pick $\nref$ close to half filling $2N+1$ in order to minimize the finite-size effects, but the results discussed in the following do not depend on this choice.
The calculations are performed for $n$ spanning a narrow range of total-charge states
around $\nref$.

The nature and energies of the low-lying excitations are determined by the charge distribution controlled by $\big( \nL, \nu, \nR \big)$ and the impurity couplings $\GL$, $\GR$. 
The parity of the occupation number in SC islands is important for superconducting pairing, resulting in the even-odd effect \cite{Averin_1992, Lafarge_1993, vonDelft_1996, Matveev_1997, Mastellone_1998, Tuominen_1997}.
Even occupation is favoured in each SC, while in the case of odd occupation a Bogoliubov quasiparticle is formed, increasing the energy by $\Delta$. 
The charging energy $E_C$ determines the energy cost of the situation
where the filling of the superconducting island is different from
$n_0$. If $n_0$ is an even integer, all excitations with odd SC island occupancy
have a further energy cost of $E_C$. In SC islands the charge gap
therefore increases to $\Delta + E_C$. The picture is reversed if
$n_0$ is tuned to an odd integer value. The energy of the
even-occupancy superconductor GS is then increased by $E_C$, reducing the gap or even closing it completely at $E_C=\Delta$. Further increase of $E_C$ then reopens the gap which gradually develops the character of a Coulomb blockade gap. By introducing the impurity coupling $\Gamma$, the unpaired Bogoliubov quasiparticles can antiferromagnetically bind to the impurity, giving rise to the YSR subgap states. The strength of the binding, and the resulting energy reduction, is characterised by $\Gamma$. The subgap spectrum of the single-channel model with charging energy was thoroughly investigated in Ref.~\onlinecite{paper1}. 

\subsection{Notation}

The eigenstates of the full model can be qualitatively characterized
through the results of a simplified calculation with a small number of
energy levels (at a very rough level even for $N=1$), which gives
low-lying excitations that are in a one-to-one correspondence with
those of the full problem and smoothly converge to the correct result
with increasing $N$. This motivates us to introduce a simplified
notation for the basis states $\vert S_L, S_\mathrm{imp}, S_R\rangle$,
where $S_i$ denotes the spin in the left bath, at the impurity, and in
the right bath, respectively. Thus the state consisting of an electron
in the left bath bound into a spin-singlet state with the impurity
electron is written as 
\begin{equation}
    \phiL = (\vert \downarrow, \uparrow, 0\rangle
- \vert \uparrow, \downarrow, 0\rangle)/\sqrt{2},
\end{equation}
while the
spin-singlet in the right bath is 
\begin{equation}
    \phiR=(\vert 0, \uparrow,
\downarrow\rangle - \vert 0, \downarrow, \uparrow\rangle)/\sqrt{2}.
\end{equation}
It must be stressed, however, that the excitations in the large-$N$ limit are
not single-particle states, but rather collective many-body states
involving many electron and hole states around the Fermi level.

\subsection{Fixed-phase vs. fixed-charge superconducting states}

In the BCS theory, the GS in the thermodynamic limit has a broken
symmetry. In the grandcanonical picture for a fixed phase of the SC order parameter, the
GS does not have a well-defined number of particles: it is a superposition of
states with $n, n+2, n-2, \dots$ particles. Our model has, however, a finite
size and we work in the microcanonical ensemble. The GS thus has a
fixed number of particles, while the BCS phase is indeterminate. In
calculations for a single SC bath this poses no conceptual nor
technical difficulty, and the results are fully equivalent to those of the BCS theory in the limit of $N\rightarrow\infty$. 

\begin{figure}
    \centering
    \includegraphics[width=0.49 \textwidth]{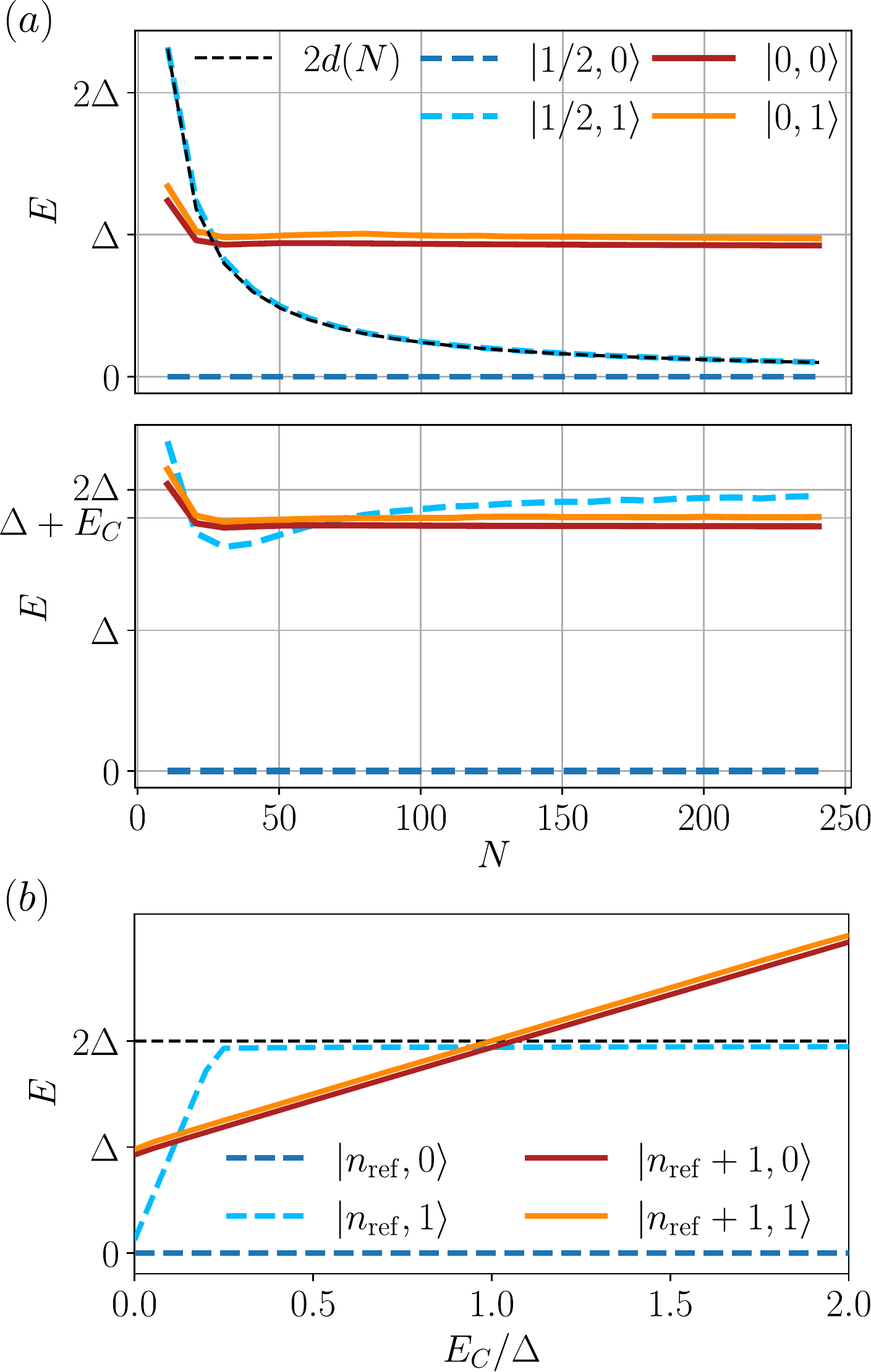}
    \caption{(a,b) System-size dependence of the subgap spectrum for small
    $\Gamma/U=0.02$, when the subgap states barely detach from the
    continuum, for (a) $E_C=0$ and (b) $E_C=0.8\Delta$. (c) Subgap
    spectrum as a function of $E_C$ at fixed system size $N=200$. 
    $U=50 \Delta$
    }
    \label{fig:N_dependence}
\end{figure}

The situation is different in the presence of two SC baths with finite
sizes. When they have small charging energies (the $E_C\to0$ limit),
one of them can act as a particle reservoir for the other. Let us
consider the situation with $\nref=2\NN+1$ electrons in the system,
for $\Gamma_i \to 0$. We find that the GS has $\NN$ particles in each SC bath, the remaining one occupying the QD level. The first ES has a configuration $(\NN+2,\NN-2)$, i.e., one Cooper pair moves from one SC to the other, at the energy cost of $2d$ and tending to zero in the thermodynamic limit, as shown in Fig.~\ref{fig:N_dependence}(a).
These (near)degenerate states are physical and not a numerical artifact. In the thermodynamic limit, the symmetry breaking would fix the phase difference between the SC baths and eliminate all states save one by singling out the suitable linear superposition of $(n_L,n_R)$ states. The lowest excitation in the equal-charge sector would then have energy $2\Delta$ (i.e., the breaking of a single Cooper pair), as expected in the BCS picture. See Ref.~\onlinecite{tinkham_book}, chapter 7, for an in-depth discussion of physics in small superconducting grains.

In a finite-size system the multitude of low-lying excitations for
very small values of $E_C$ poses a technical difficulty and, for example, hinders access
to the more important Cooper-pair-breaking excitations on the scale of
$2\Delta$ in the spin-doublet sector. The same issue also arises in the spin-singlet sector, where each YSR singlet substate would have replicas at slightly higher energies differing only in the distribution of the Cooper pairs between the superconductors.
Fortunately, here we are interested in small systems which intrinsically have  non-zero charging energies $E_C$. Moving a Cooper pair from one SC bath to another then costs $4 \ECL + 4 \ECR$. When this energy cost is larger than $2\Delta$, the $(\NN+2, \NN-2)$ states are pushed beyond the first excitation with the $(\NN, \NN)$ configuration, which will then become the first ES of the system. We therefore constrain most of our calculation to the regime with $\ECL + \ECR > \Delta/2$, where there are no technical issues and the lowest-lying ES with a broken Cooper pair appears in the doublet sector, see Fig. \ref{fig:N_dependence}b, where $\ECL = \ECR = 0.8 \Delta$. 
Fig. \ref{fig:N_dependence}c shows the evolution of the subgap spectrum with increasing $E_C$, illustrating the quick change in nature of the doublet ES when $E_C > \Delta / 4$. 
As $\Gamma\approx 0$, the singlet $\nref+1$ states are barely below the quasicontinuum and therefore have energy marginally lower than $\Delta+E_C$.  

We do not lose access to any important parts of the parameter space by the restriction to $\ECL + \ECR > \Delta/2$, as precise calculations in the thermodynamic limit are possible using the NRG for $E_C=0$. Furthermore, for left-right mirror-symmetric problems, where the most relevant levels are the lowest ones in each parity subspace, the DMRG calculation could be implemented with the parity as a conserved quantum number  after a suitable transformation of the Hamiltonian in the symmetry-adapted basis and an elaboration of its matrix-product-operator form. Alternatively, one could target excited states with appropriate parity by adding weight terms that penalize states with undesired parity in the optimization sweeps. The required parity operator could be implemented using tensor index reordering.
 
\section{Subgap states: Case of equivalent superconductors}
\label{sec3}

In this section we consider the simplest case of symmetric devices composed of two equivalent SCs, i.e., $\ECL = \ECR \equiv E_C$.
We also set $\nL=\nR=\NN$ with even integer $\NN$, and restrict the discussion to the Kondo limit of $U/\Delta=30$ and $\nu=1$, where the occupation of the QD is pinned to 1.

\subsection{Symmetric QD coupling}

\begin{figure}
    \centering
    \includegraphics[width=0.49 \textwidth]{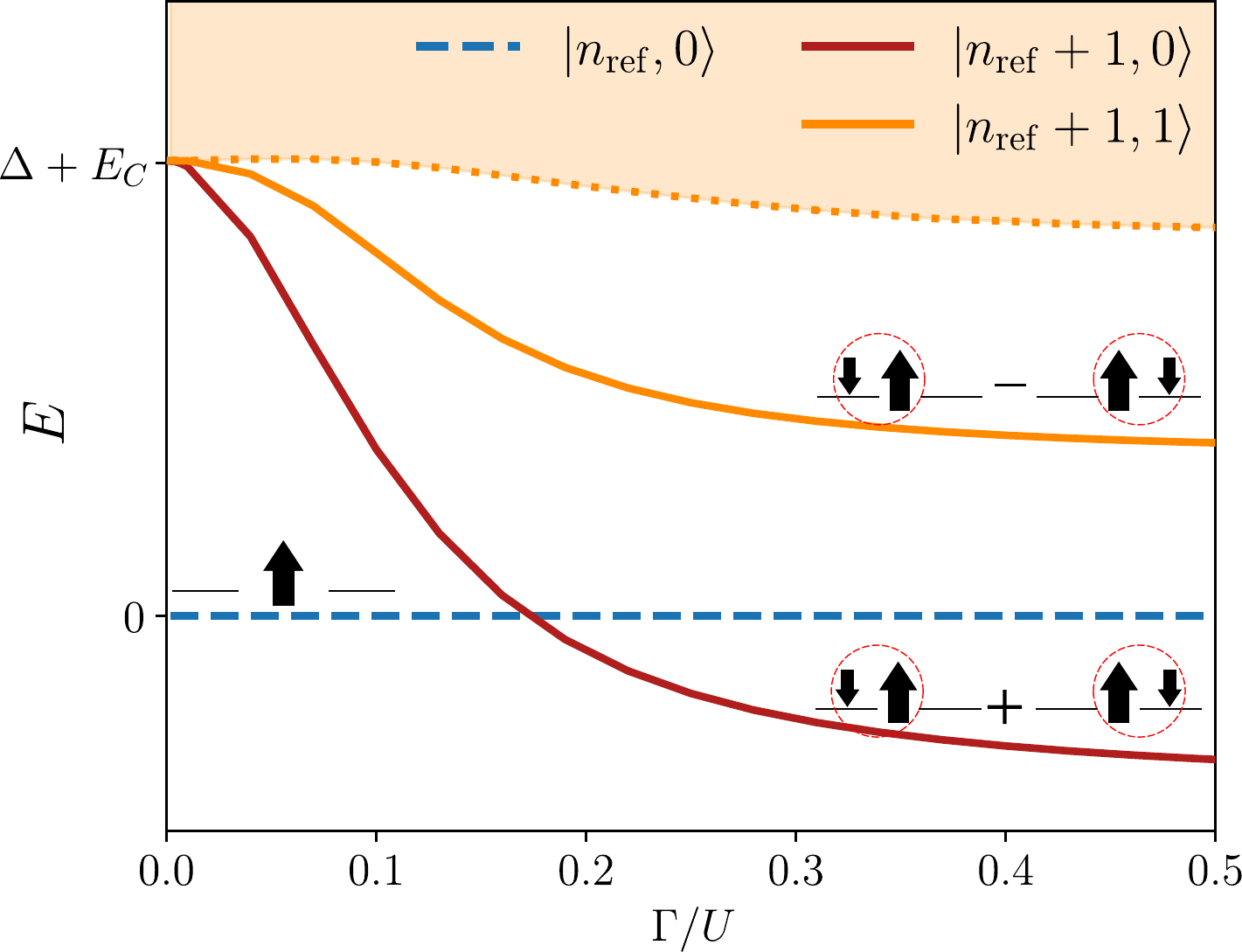}
    \caption{Sub-gap spectrum for $\Gamma_L = \Gamma_R \equiv \Gamma$ and equal charging energies $\ECL=\ECR=0.4\Delta$. 
    At each $\Gamma$, the energy of the lowest doublet state is taken as the reference value for energies, i.e., $E(n=\nref, i=0)\equiv 0$. 
    The sub-gap states are accompanied by sketches indicating their nature:
    the large spin represents the impurity electron, the small arrows denote the Bogoliubov quasiparticles, and the red circles represent a singlet configuration of the encircled levels, $\vert \uparrow \downarrow \rangle - \vert \downarrow \uparrow \rangle$. The orange shaded area above the orange dotted line denotes the quasi continuum of states in the $\nref+1$ sector.}
    \label{fig:symmetricGamma}
\end{figure}

We first examine the case of symmetric coupling of the impurity to both baths, $\Gamma_L = \Gamma_R \equiv \Gamma$. The system then has mirror symmetry and we denote the corresponding parity as gerade/ungerade in the following. The low-lying states with an odd number of electrons $\nref$  (spin-doublets) and an even number $\nref+1$ (spin-singlets) are shown in the form of an energy-level diagram in Fig. \ref{fig:symmetricGamma}. For $\Gamma=0$, the GS is a product state consisting of a BCS-like state of Cooper pairs in each channel and a decoupled electron sitting at the impurity site. 
The lowest ES with $\nref+1$ electrons is the Bogoliubov state at the bottom of the quasi continuum (represented using orange shading) with energy $\Delta+E_C$; the presence of the quasi-continuum of Bogoliubov states is confirmed by calculating a number of higher excited states.

For $\Gamma>0$, two states detach from the continuum to become the spin-singlet sub-gap states. This is clearly different from the situation in the $E_C=0$ limit, where only the gerade state is inside the gap, while the ungerade combination is fully decoupled from the impurity and remains inside the continuum (in the absence of flux bias \cite{Kirsanskas_2015, Meden2019}; see Appendix~\ref{appA}). With increasing $\Gamma$ both states descend deeper into the gap, with the gerade state always having lower energy.  This is different from the situation in the two-channel Kondo model where the singlet subgap states in a symmetric device are exactly degenerate \cite{zitko_2ch}. This is due to finite charge fluctuations in this model (and in real QD devices) and the different YSR wavefunctions, with an anti-node and a node at the impurity site for the gerade and ungerade combinations of Bogoliubov states, respectively.
The continuum of excitations in the $\nref+1$ sector begins at $\Delta + E_C$. At $\Gamma=0$, it consists of product states of the BCS-like state with $\nref$ electrons and a Bogoliubov quasiparticle in one of the channels. The lowest lying continuum states are $\ketref \otimes (\mathrm{QP}_L \pm \mathrm{QP}_R)/\sqrt{2}$, where $\mathrm{QP}_\beta$ denotes a Bogoliubov quasiparticle in channel $\beta$. The small deviation from the $\Gamma=0$ gap value of $\Delta + E_C$ is due to the $\propto \Gamma^2$ virtual tunneling of quasiparticles between the channels; these processes are suppressed if $E_C$ and/or $U$ are increased.

\begin{figure}
    \centering
    \includegraphics[width=0.49 \textwidth]{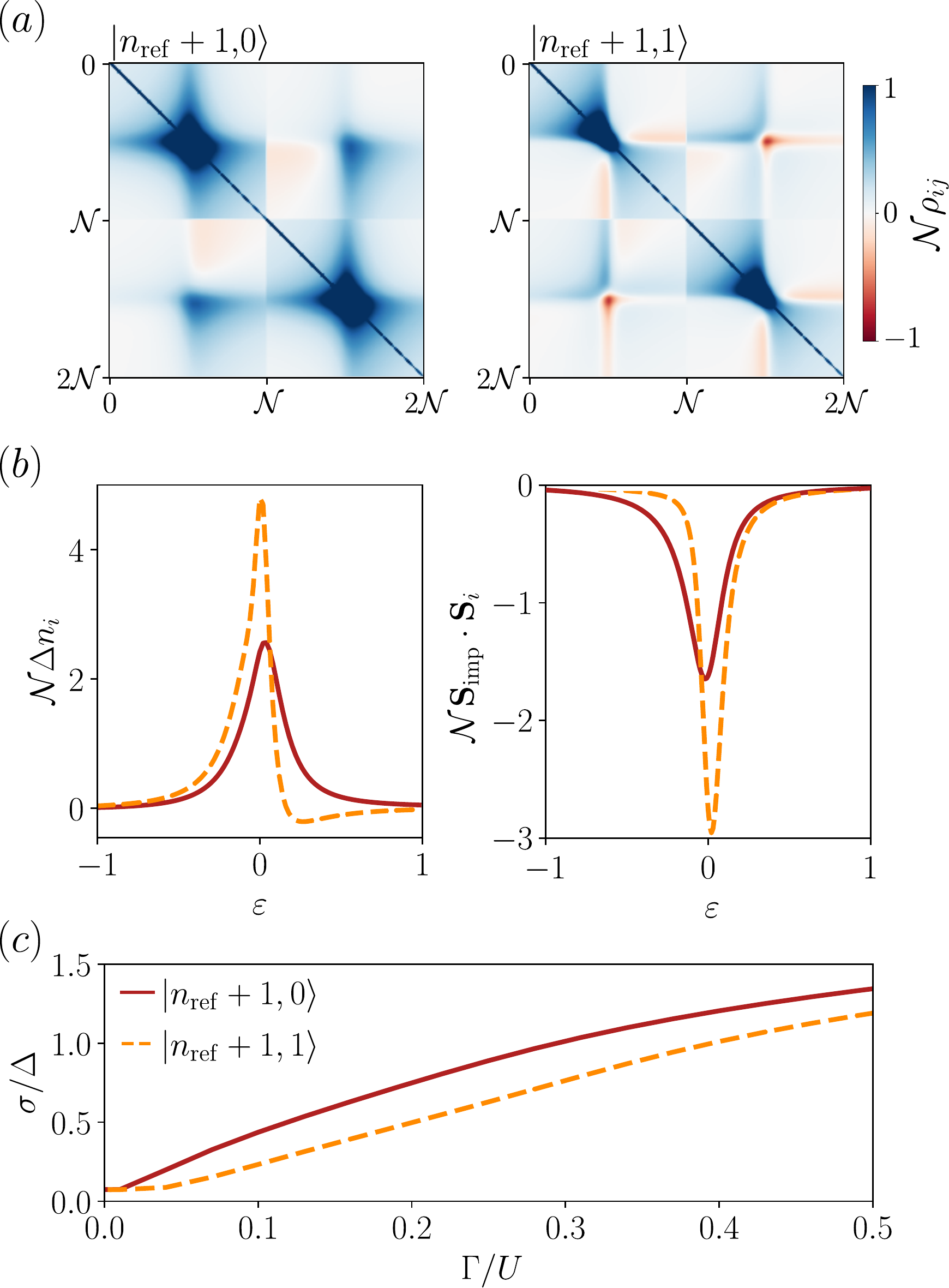}
    \caption{ Singlet subgap states. 
    (a) Channel density matrix $\rho_{ij} = \langle c_i^\dagger c_j \rangle$ for the two singlet subgap states. $\Gamma/U=0.2$, $E_C = 0.4\Delta$. The diagonal blocks represent intra-channel density matrices, while the off-diagonal parts represent the inter-channel correlations.
    (b) Occupation and spin-spin correlations between the impurity and the baths. $\Delta n_i$ is the excess charge compared to the reference doublet state at $\Gamma \rightarrow 0$. $\Gamma/U=0.2$, $E_C = 0.4\Delta$.
    (c) $\Gamma$ dependence of the width of the spin-spin correlation peak. $E_C = 0.4\Delta$.
    }
    \label{fig:symmetricGamma_nature}
\end{figure}

The nature of the sub-gap singlets is investigated in Fig. \ref{fig:symmetricGamma_nature}. 
The first subgap state is an even (gerade) linear combination of spin-singlet YSR states from each channel
\begin{equation} 
\begin{split}
&\vert \nref+1, 0 \rangle = \ketg \approx (\phiL + \phiR)/\sqrt{2} \\
&= \frac{1}{2}\left[
\left( \vert \downarrow,\uparrow,0\rangle -  \vert \uparrow,\downarrow,0\rangle \right)
+
\left( \vert 0,\uparrow,\downarrow\rangle -  \vert 0,\downarrow,\uparrow\rangle \right)
\right] \\
&= \frac{1}{2}\left[
\left( \vert \downarrow,\uparrow,0\rangle +  \vert 0,\uparrow,\downarrow\rangle
 \right)
-
\left( \vert \uparrow,\downarrow,0\rangle + \vert 0,\downarrow,\uparrow\rangle \right)
\right], \\
\end{split}
\end{equation} 
where the final line shows that this state may also be interpreted as a spin-singlet formed between the impurity spin and an even-parity combination of Bogoliubov excitations from both electrodes, while the second is an odd (ungerade) combination of the singlets, 
\begin{equation}
\begin{split}
&\vert \nref+1, 1 \rangle = \ketu \approx (\phiL - \phiR)/\sqrt{2} \\
&= \frac{1}{2}\left[
\left( \vert \downarrow,\uparrow,0\rangle -  \vert \uparrow,\downarrow,0\rangle \right)
-
\left( \vert 0,\uparrow,\downarrow\rangle -  \vert 0,\downarrow,\uparrow\rangle \right)
\right] \\
&= \frac{1}{2}\left[
\left( \vert \downarrow,\uparrow,0\rangle -  \vert 0,\uparrow,\downarrow\rangle
 \right)
-
\left( \vert \uparrow,\downarrow,0\rangle - \vert 0,\downarrow,\uparrow\rangle \right)
\right], \\
\end{split}
\end{equation}
with an alternative interpretation as a spin-singlet formed between the impurity spin and an odd-parity combination of Bogoliubov electrons from both electrodes. The parity of the states is confirmed by the calculation of the one-particle density matrices in the baths, $\rho_{ij} = \langle c_i^\dagger c_j \rangle$, at finite $\Gamma$, shown in Fig.~\ref{fig:symmetricGamma_nature}(a). The intra-channel correlations are in large part the result of superconducting pairing: $\rho_{ij}$ is large for the levels which contribute to the creation of Cooper pairs and small between levels that are fully occupied or empty. The inter-channel correlations uncover the nature of the states. 
Positive $\rho_{ij}$ in the center of the inter-channel correlation (off-diagonal blocks) in the left panel is a characteristic feature of a gerade configuration, while the negative values (right panel) are a feature of an ungerade state. 

In Fig.~\ref{fig:symmetricGamma_nature}(b), we illustrate the different composition of the two YSR singlets despite $\Gamma_L=\Gamma_R$, thus further revealing the differences compared to the pure two-channel Kondo model. 
We plot excess charge compared to a reference doublet state (left) and the spin-spin correlations $\expv{\mathbf{S}_\mathrm{imp} \cdot \mathbf{S}_i}$ between the impurity local moment and the SC single-particle energy levels $i=1,\ldots,N$, with energies between $-1$ and $1$ (right). 
Both plots indicate that the singlet is formed between the impurity and the excess electron, which is located close to the Fermi level, but the charge distribution is not identical in the two states.
The width of the spin-spin correlation peak with increasing $\Gamma$ is shown in Fig.~\ref{fig:symmetricGamma_nature}(c) and is in qualitative agreement with the well known Kondo limits. At large coupling, the impurity is screened by the electron localized at $r=0$, with a broad distribution in energy, while at small coupling the screening is performed by the electrons with energies close to Fermi energy.  

\subsection{Asymmetric QD coupling}

\begin{figure}
\centering
\includegraphics[width=\linewidth]{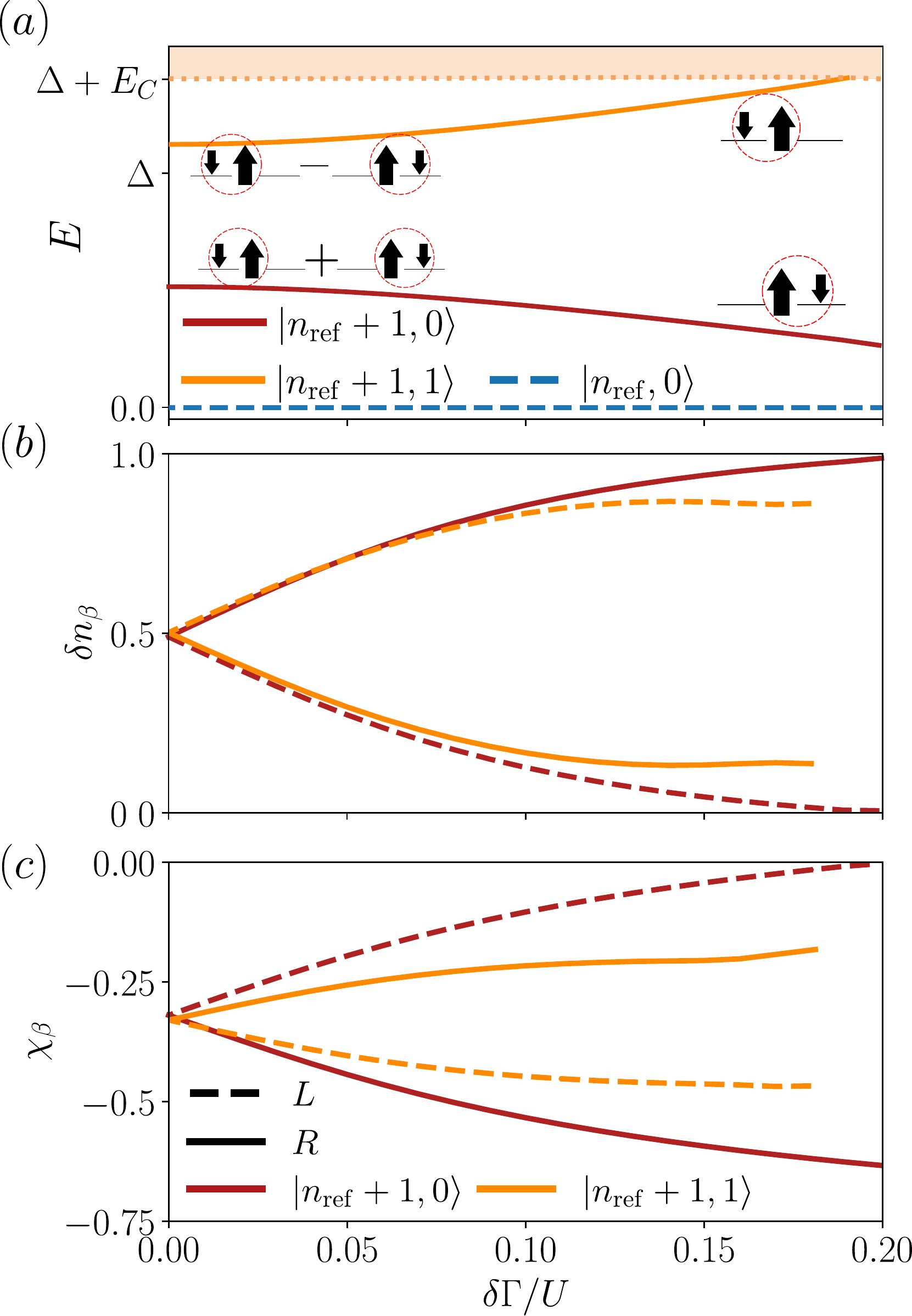}
\caption{Equivalent superconductors, asymmetric QD coupling. The depenence of the subgap states on $\dg = \Gamma_R - \Gamma_L$ with the sum $\Gamma_L + \Gamma_R = 0.2 U$ kept constant.  
(a) Subgap spectrum. Subgap states are accompanied by sketches indicating their nature. The orange shaded area represents the continuum of states in the $\nref+1$ sector. 
(b) Deviation from half-filling in the superconductors for the singlet subgap states.
(c) Spin-spin correlations between the impurity and the baths for the singlet subgap states. 
The value of $-3/4$ corresponds to a pure (saturated) singlet configuration.
}
\label{fig:asymmetricGamma}
\end{figure}

We now examine the case of $\Gamma_L \neq \Gamma_R$, see Fig. \ref{fig:asymmetricGamma}. We fix $\Gamma_L + \Gamma_R = 0.1U$ and study the effect of the deviation from the equal-$\Gamma$ tuning by simultaneously increasing $\Gamma_R$ and decreasing $\Gamma_L$. 
The energy diagram in panel (a) is accompanied by panels (b) and (c) illustrating the nature of the singlet states. In panel (b), the deviation from half-filling (even integer $\NN = (\nref-1)/2$) in each SC island is plotted. 
As we are in the limit of a very large $U/\Delta$, the impurity occupation always remains very close to 1. Therefore, this plot uncovers the position of the unpaired quasiparticle in the SC channels. The impurity-channel spin-spin correlations, defined by $\chi_\beta = \sum_{i=1}^N \expv{\mathbf{S}_\mathrm{imp} \cdot  \mathbf{S}^{(\beta)}_i }$ are presented in panel (c) and contain information on the screening of the impurity local moment. 

With increasing $\dg$ the nature of the sub-gap states gradually changes from the inversion-symmetry eigenstates $\ketg$ and $\ketu$ with equal $\chi$ in both channels to states that are closer to simple YSR singlets $\phiL$ and $\phiR$.
The state with the lowest energy is the one corresponding to the larger hybridisation ($\GR$ in our case, so $\phiR$), while the excited subgap state transforms into $\phiL$. The energy of the excited subgap state increases as $\GL$ is decreased, and it finally merges with the continuum when the left channel decouples from the system at $\dg \rightarrow 0.2$ (therefore $\GL \rightarrow 0$). This is different from the superconducting two-channel Kondo model, where breaking the coupling symmetry quickly pushes the second excited state beyond the gap \cite{zitko_2ch}. We note also that in the normal-state two-channel Kondo problem a small difference in couplings immediately leads to a cross-over to a Fermi-liquid fixed point strictly at zero temperature, with one of the channels completely decoupled. In problems with a superconducting gap the renormalization is terminated at the scale of the gap, thus a complete decoupling of a second channel does not occur except in the cases of extreme asymmetry.

\section{Subgap states: case of non-equivalent superconductors}
\label{sec4}

The experimental interest in devices where the QD is embedded between one macroscopic SC contact and one SC island provides the motivation for considering the case of strongly differing charging energies. The asymmetry in the charging energies induces a preference for even occupation in the channel with larger $E_C$. In the singlet sector with an even total number of particles, this implicitly favours the spin-singlet YSR state formed between the impurity spin with a decoupled Bogoliubov quasiparticle in the channel with smaller $E_C$.

In this section, the results will be presented by plotting the evolution of the subgap states with changing $\Gamma_R$ at fixed $\Gamma_L$ for several different cases of non-equivalent superconductor parameters.

\subsection{General considerations}

Unequal charging energies influence the subgap states in one important aspect. 
The transition from the ungerade/gerade regime at small $\dg$ towards $\phiL$/$\phiR$ subgap states at large $\dg$ is  decelerated (i.e., requires larger $\dg$) if the strongly coupled channel has larger charging energy. The hybridisation favours configurations with the unpaired particle in the strongly coupled channel, while its charging energy enforces even occupation. The crossover occurs only when the larger $\Gamma$ is big enough to overcome the charging-energy penalty. This is demonstrated in Appendix \ref{appC} using a zero-bandwidth calculation.

\subsection{Case of $E_C^{(L)} \gg E_C^{(R)}$}

\begin{figure}
\centering
\includegraphics[width=\linewidth]{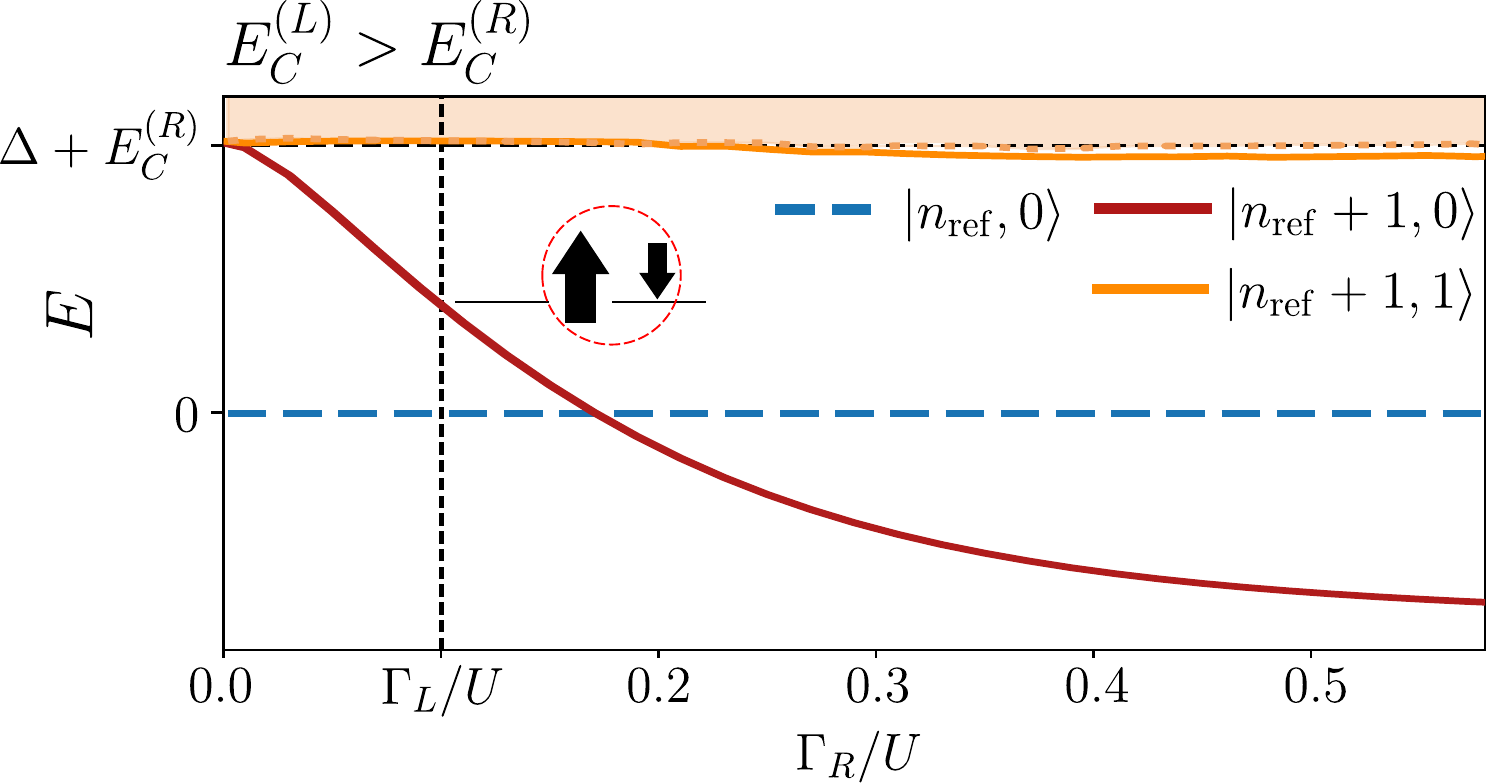}
\caption{
Non-equivalent superconductors, $E_C^{(L)} \gg E_C^{(R)}$. Subgap spectra dependence on $\GR$ for constant $\GL/U$ with $\ECL=0.1\Delta$, $\ECR=1.5\Delta$. The value of $\GL/U$ is indicated by the vertical dashed line.
}
\label{fig:EcL>>EcR}
\end{figure}

If the difference in $E_C$ is very large, this  mechanism may cause the weakly coupled channel with large charging energy to almost completely decouple from the system. Such decoupling is illustrated in Fig.~\ref{fig:EcL>>EcR} for the case of $\ECL \gg \ECR$. The second subgap state does not descend deeper into the gap even for very large $\GR$, and we effectively obtain a single channel YSR singlet in the right channel, with a barely visible contribution from the left one. 

\subsection{Case of $E_C^{(L)} \ll E_C^{(R)}$}
\label{caseECL}

\begin{figure}
\centering
\includegraphics[width=\linewidth]{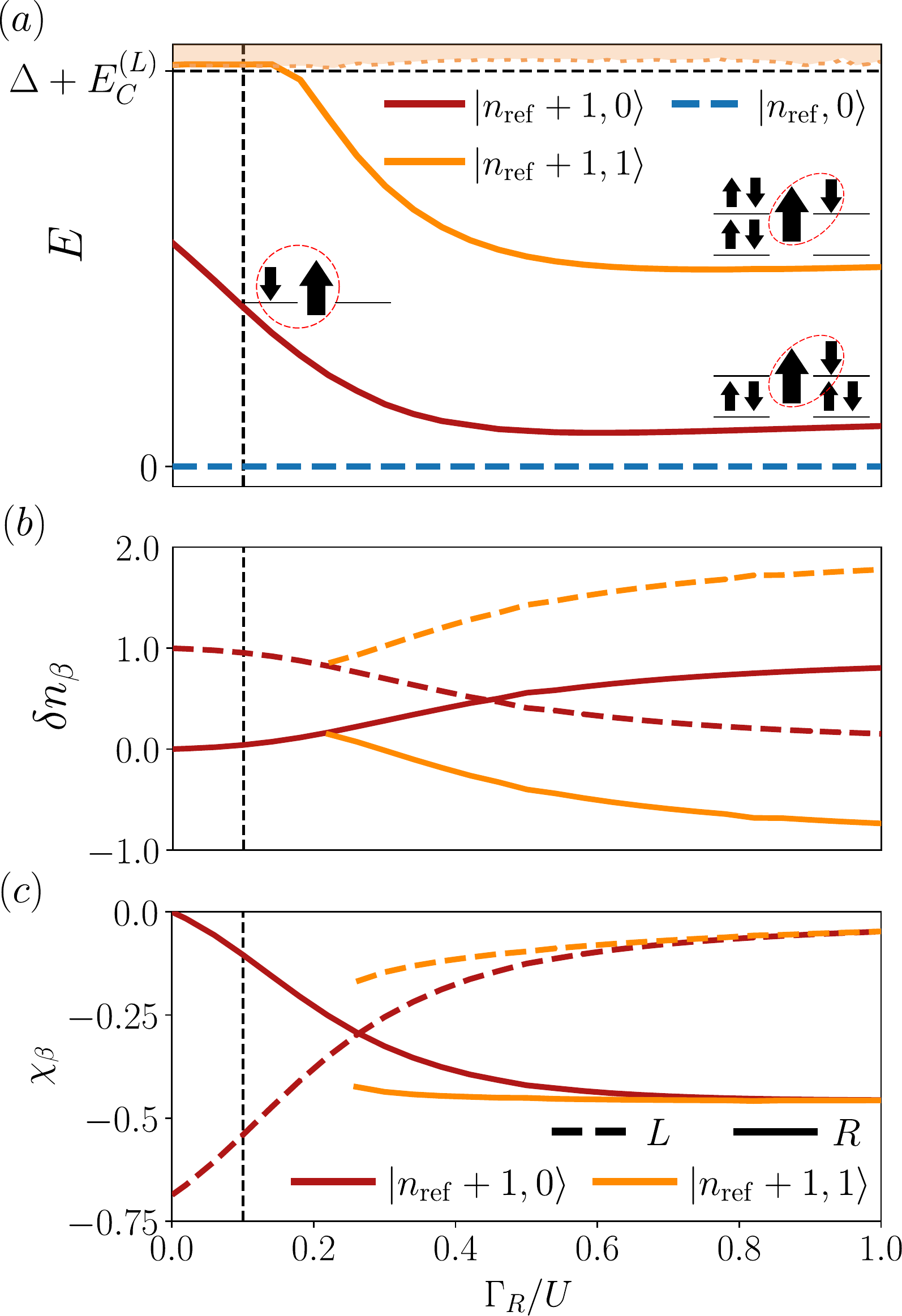}
\caption{
Non-equivalent superconductors, $\ECL \ll \ECR$.
(a) Subgap spectrum dependence on $\GR$ for constant $\GL/U=0.1$ ($\GL=\GR$ tuning is indicated by the vertical dashed line). The states are accompanied by sketches of the charge configurations. The blue ovals represent Cooper pairs. 
(b) Deviation from half-filling in the superconducting islands of the singlet subgap states.
(c) The impurity-channel spin-spin correlation of the singlet subgap states.
$\ECL = 0.1\Delta$, $\ECR = 1.5\Delta$
}
\label{fig:EcL<<EcR}
\end{figure}

The sub-gap spectrum is most intriguing in the regime of $\ECL \ll \ECR$.
The charging terms favour even occupation in the right channel, therefore  singlet screening in the left. For low $\GR < \GL$ we observe the decoupled situation with a single subgap singlet $\phiL$ and a decoupled right channel,
Fig.~\ref{fig:EcL<<EcR}.
When $\GR$ increases beyond $\GL$, the system transitions into the regime where the hybridisation terms favor $\phiR$, which is at the same time strongly disfavored by the charging energies.
The lower subgap state transforms into a right-channel singlet state only for $\GR \gg \GL$, when the strong hybridisation eventually overcomes the energy penalty due to $\ECR$. The excess charge and impurity-channel spin correlations shown in Fig.~\ref{fig:EcL<<EcR}(b,c) illustrate the smooth transition of the lower singlet state between the $\phiL$ and $\phiR$ limits. The spin correlations do not reach the saturated singlet value of $-3/4$, indicating that even for very large $\GR$ this state never quite obtains the nature of a pure singlet. 

A second subgap state descends into the gap at some finite $\GR$.
Curiously, this $i=1$ subgap state is not a left-channel YSR singlet as one might expect. Instead, the spin correlations show a very close relation to the $i=0$ subgap state, increasingly so as $\Gamma_R$ grows large.
It transpires that the $i=1$ is obtained from the $i=0$ state by a transfer of a Cooper pair between the superconductors, i.e., while the $i=0$ state has charge configuration $(\NN, 1, \NN + 1)$, the $i=1$ state corresponds to $(\NN + 2, 1, \NN - 1)$. This allows the excited state to also obtain the energetically optimal nature of a right-channel YSR singlet. The transfer of two additional electrons only costs $4 \ECL$, corresponding to approximately $0.4 \Delta$ in the example shown (plus $2d$, a finite size effect). 

\begin{figure}
\centering
\includegraphics[width=\linewidth]{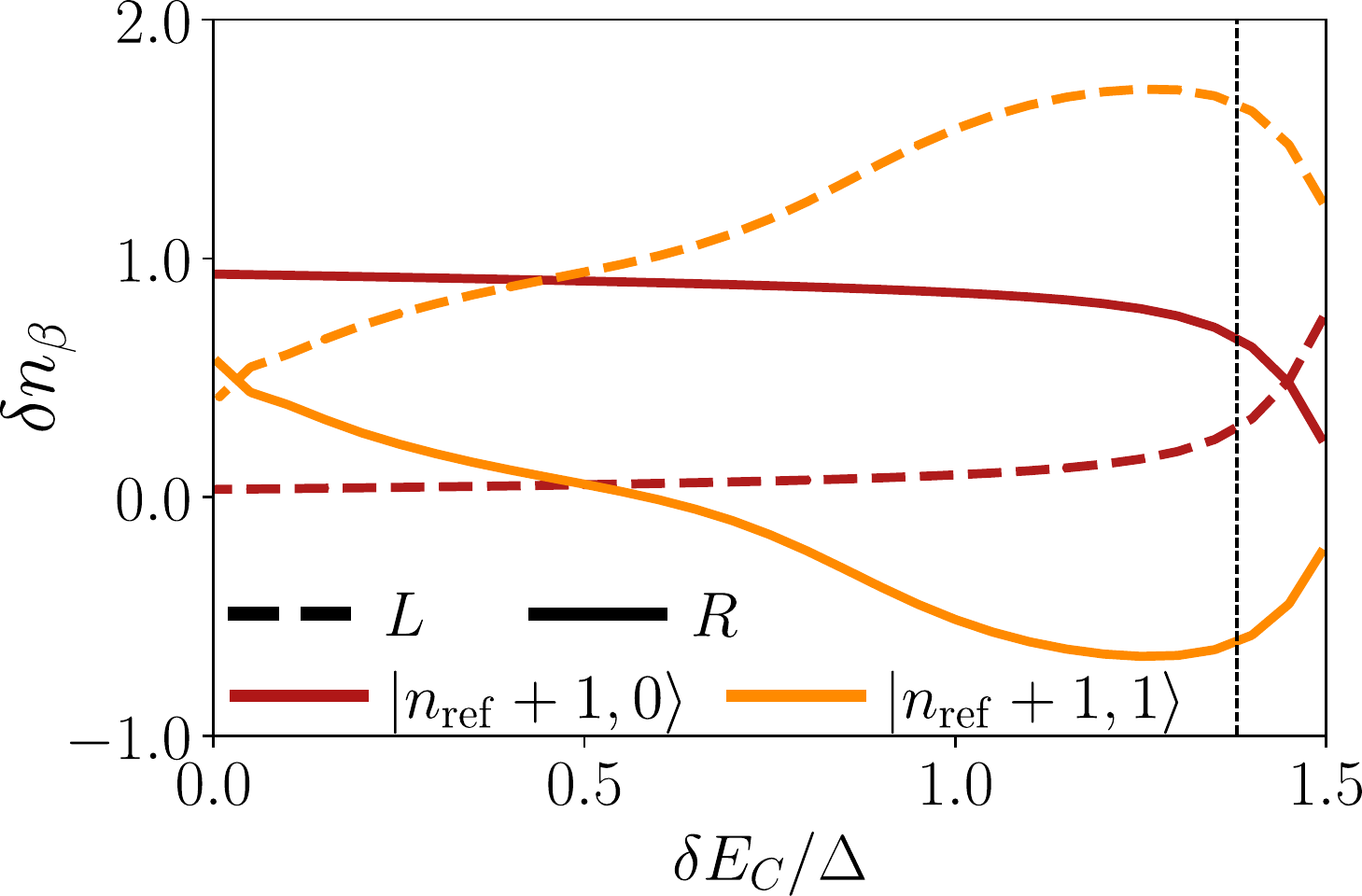}
\caption{
Evolution of the deviation from half-filling in the superconducting islands  with varying charging energy asymmetry. The sum $\ECL + \ECR = 1.5\Delta$ is kept constant, with $\delta E_C = \ECR - \ECL$. $\GL=0.1U$, $\GR=U$. The vertical dashed line roughly corresponds to the rightmost point of Fig. \ref{fig:EcL<<EcR}(b).
}
\label{fig:dEc}
\end{figure}

The evolution of the channel occupation with increasing $E_C$ asymmetry is shown in Fig. \ref{fig:dEc}. By increasing  $\delta E_C$ (decreasing $\ECL$ and increasing $\ECR$) the second subgap state approaches the $(\NN + 2, 1, \NN - 1)$ limit. For the parameters chosen here the charging energy penalty overcomes the $\Gamma$ coupling just as $\ECL\rightarrow 0$ and the occupation of the subgap states tends towards even occupation in the right channel, following the dominant charging energy effect.
It should be stressed that the existence of such excited state is not a numerical artefact, but rather a signature of the physics of small SC islands with large charging energies.

\subsection{YSR spin-singlet qubit}

The regime of $\ECL < \ECR$ with $\Gamma_L < \Gamma_R$ seems to be the most appropriate for the investigation of the occurrence of two singlet subgap states, both well separated from the quasicontinuum at higher energies. In particular, it appears possible to achieve energy separation between the singlets that is lower than the transition energy from the upper level to the continuum. This is required for the implementation of qubits based on a linear superposition of the YSR singlet states, a {\it YSR qubit}. The relative position of the states is highly tunable: the energy difference from the gap edge is increased by increasing $\GR$. The energy difference between states depends on $\ECL$, which is typically a device property. If $\ECL$ is small, there will be further singlet states with two, three, etc., Cooper pairs transferred between the superconductors, but the energy cost for those increase quadratically, resulting in non-monotonic spacing.

We further investigate this parameter regime in Sec. \ref{secgate} by considering the effects of tuning of the gate voltages in the system. These determine the favourable occupation of the system components, which can strongly influence the nature of the subgap states.

\section{Realistic QD}
\label{secsmallU}

\begin{figure}
\centering
\includegraphics[width=1\linewidth]{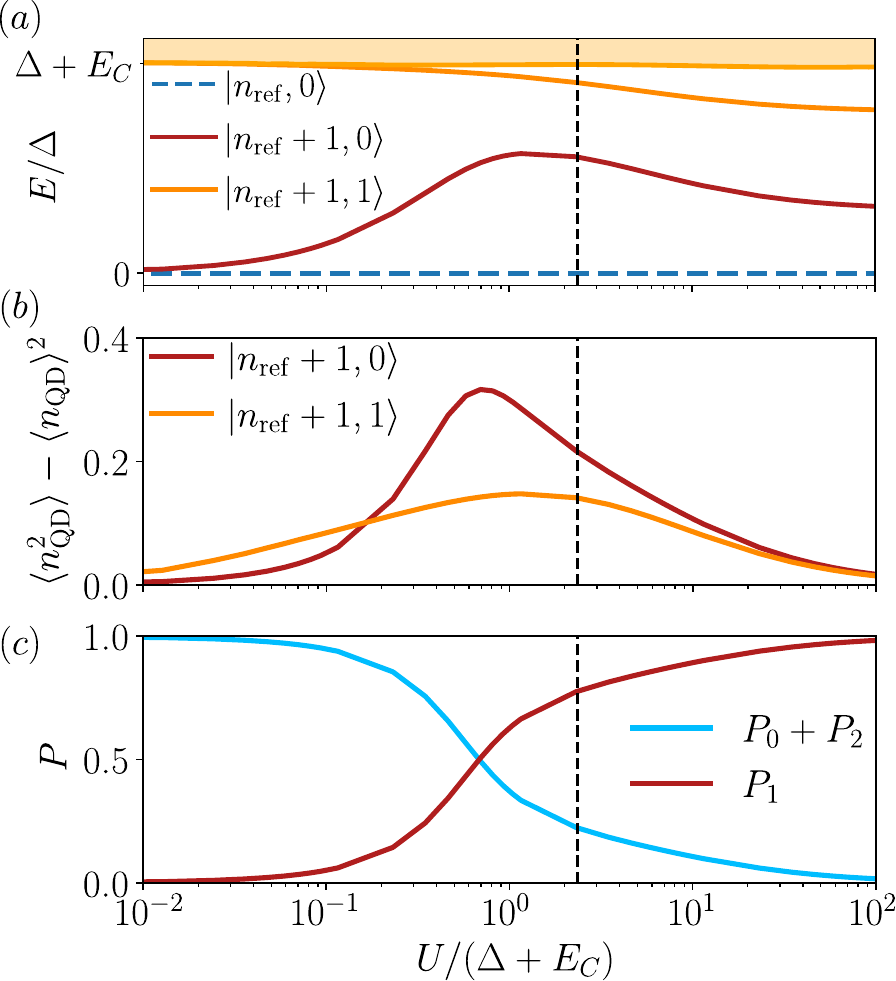}
\caption{
$U$ dependence of the subgap states.
(a) Subgap spectrum as a funciton of $U$.
(b) QD charge fluctuations in the subgap singlet states. 
(c) Probabilities $P_n$ that in the lowest singlet state the QD is occupied by $n$ electrons.
Here we show the symmetric case with constant $\Gamma = \GL = \GR = 0.1 U$, $E_C = \ECL = \ECR = 0.7\Delta$. $U$ is scaled by $\Delta + E_C$, the first being the charging energy scale of the QD and the latter of the SC islands. The vertical line at $U/\Delta=4$ correspons to the regime in which realistic devices operate.
    }
\label{fig:U_sweep}
\end{figure}

While the Kondo limit of large electron-electron repulsion on the QD, $U \gg \Delta$, is useful for studying the nature of the Yu-Shiba-Rusinov sub-gap states in an ideal setting, a large class of real devices in current use operate in the regime of comparable parameter values, $U \approx \Delta \approx E_C$. By decreasing $U$ we move away from the Kondo limit where the QD behaves almost as a pure magnetic impurity, thus changing the nature of the subgap states from the YSR singlets (due to Kondo exchange interaction) by increasing the admixture of wavefunctions with Andreev-bound-state character (due to proximity effect) and the role of Coulomb interaction. 
The changing nature of the subgap singlets is demonstrated in Fig.~\ref{fig:U_sweep}. We consider a symmetric case with constant $\Gamma /U = 0.1$ and $E_C = 0.7\Delta$.
The level diagram in panel (a) shows that the energy difference between the sub-gap singlets is constant at large $U/\Delta$, determined mostly by $\Gamma/U$. With decreasing $U$, the upper singlet disappears into the continuum, while the lower one approaches the energy of the doublet ground state. 
In the absence of charging energy a pair of singlet ABS states is expected, with energy symmetrically above and below the doublet state, the difference between them determined by $\Gamma$. As $\Gamma/U$ is constant in our plot, $U \to 0$ means $\Gamma \to 0$ as well. This is why the lowest singlet and doublet states appear degenerate in the limit of small $U$. Furthermore, the charging energy breaks the degeneracy between the states with a different number of Cooper pairs in the SC islands, thus creating an energy difference between the $\ket{0}$ and $\ket{2}$ states on the QD. The ground singlet state is therefore predominantly of $P_2$ nature, while the excited state exhibits a larger $P_0$ contribution. Its energy cost is $4 E_C$ (for an additional Cooper pair in the SC), so this state lives far beyond the continuum of decoupled excitations.

The QD charge fluctuations are presented in panel (b). They are small in the Kondo limit at $U \gg \Delta$, where the QD is a pure magnetic impurity and the only charge fluctuations correspond to virtual processes that generate the Kondo exchange scattering $J_K \propto \Gamma/U$. The charge fluctuation are also small for $U \ll \Delta + E_C$, because the charging effects in the SC islands strongly determine the QD occupation.
They are hence the strongest in the cross-over regime, which is where the experimental devices operate. Finally, panel (c) shows the diagonal elements of the QD density matrix for the ground singlet state $\ket{ \nref + 1, 0}$. $P_n$ is the probability that the QD is in a state with occupation $n$. A crossover from a large magnetic moment in the Kondo regime (large $P_1$) towards Andreev bound states (large $P_0+P_2$) is evident.

\begin{figure}
\centering
\includegraphics[width=1\linewidth]{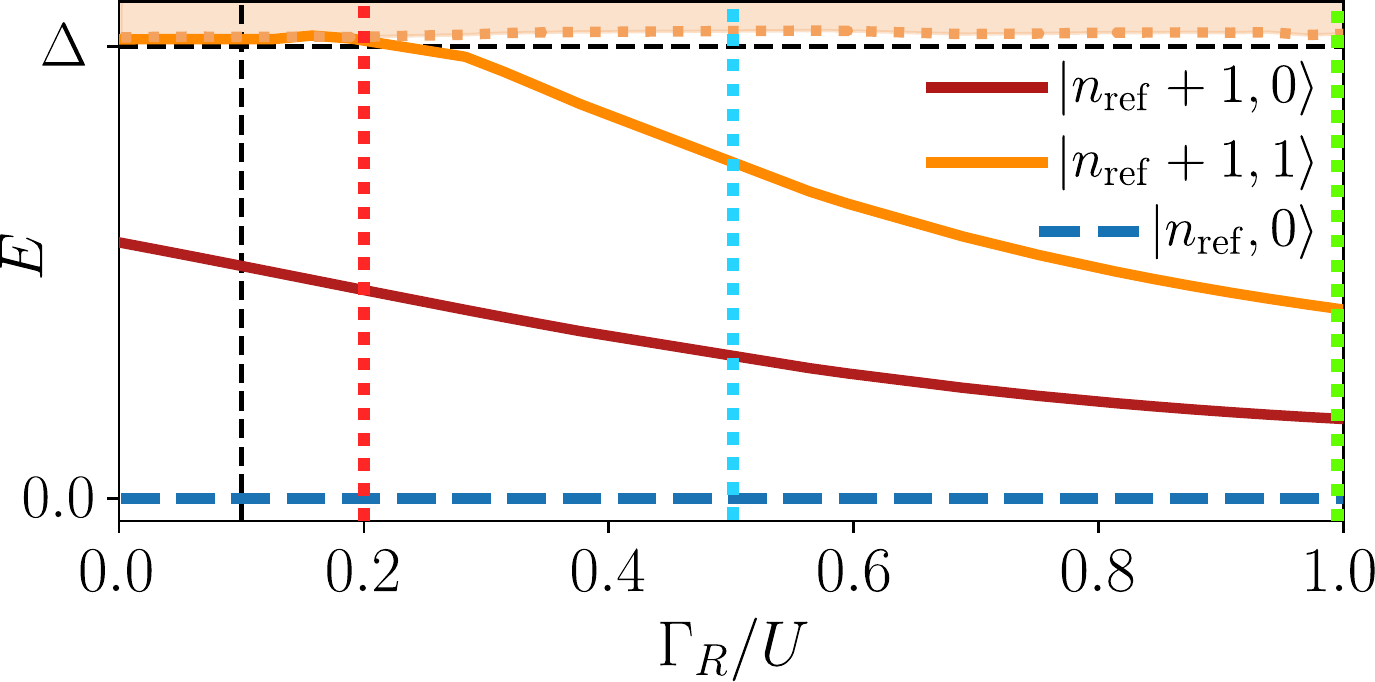}
\caption{Realistic $U = 4 \Delta$ and $\ECL \ll \ECR$. Subgap spectra dependence on $\GR$ at constant $\GL/U = 0.1$, indicated by the black vertical dashed line. $\ECL = 0.1\Delta$, $\ECR = 1.5\Delta$. The vertical dotted lines correspond to $\nR$ sweeps presented
in Fig.~\ref{fig:nR_sweeps}.}
\label{fig:smallU_Gsweep}
\end{figure}

In order to obtain experimentally relevant results, from this point on we set $U$ to a value appropriate for contemporary InAs nanowire QD devices with top gates \cite{JuanCarlos2021}, $U/\Delta = 4$.
The vertical dashed line at $U=4\Delta$ demonstrates that this situation is far away from either limit, i.e., in the regime where charge fluctuations are important and the singlet states are neither YSR nor ABS. This competition between $U$, $\Delta$ and $E_C$ thus opens a new degree of complexity in the formation of the eigenstates \cite{paper1, JuanCarlos2021}. The subgap states are not simple combinations of singlets that we had previously examined, but rather some more complicated entities that generally exhibit non-integer occupation in each system component. 
The competition reduces the local moment on the QD, which diminishes the importance of hybridisation. We demonstrate the effect in Fig. \ref{fig:smallU_Gsweep}, where the $\GR$ dependence of the sub-gap spectra is shown for $U = 4 \Delta$. The decreased importance of $\Gamma$ becomes obvious when one compares this plot to Fig. \ref{fig:EcL<<EcR}a, which is equal in all parameters, but at much larger $U$.

\section{Gate voltage tuning}
\label{secgate}

In the following section we investigate the gate voltage dependencies on the SC islands and in the QD. We mainly focus on the regime with two nearby subgap states presented in Sec. \ref{caseECL}, but with $U = 4\Delta$. 
The goal in this section is twofold. Firstly, we presented results in the experimentally relevant regime.
Second, we point out the optimal points for the operation of a proposed YSR qubit, and show that they occur when the energy difference between the sub-gap states is minimal.

\subsection{SC island gate voltage tuning}
\label{sec5}

\begin{figure}
\centering
\includegraphics[width=1\linewidth]{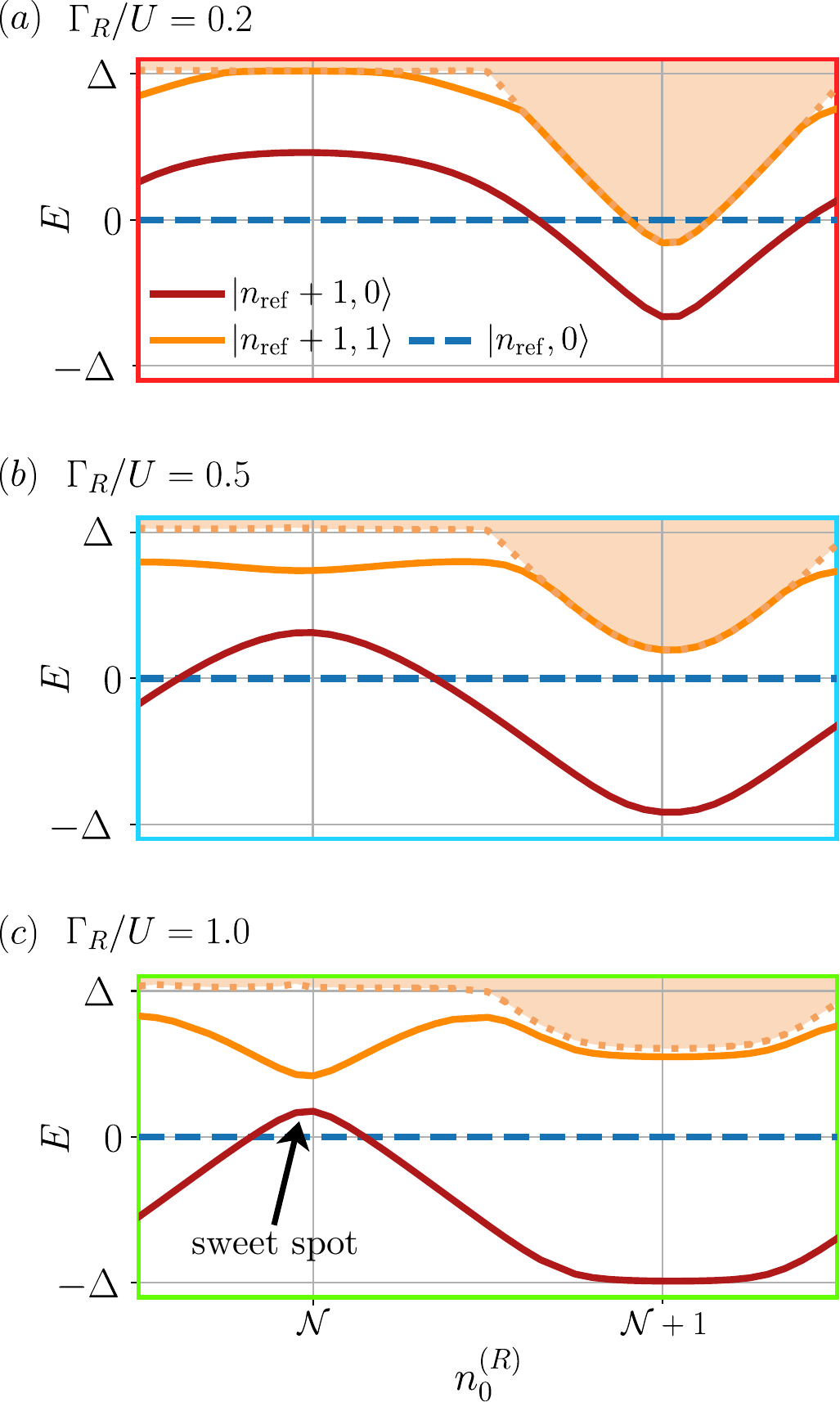}
\caption{
Subgap spectra with a realistic value of $U = 4 \Delta$, $\ECL = 0.1 \Delta$ and $\ECR = 1.5 \Delta$. Varying $\nR$ and (top to bottom) increasing $\GR$. The colors of the panel frames correspond to vertical dotted lines in Fig. \ref{fig:smallU_Gsweep} indicating the position of the cut.   
}
\label{fig:nR_sweeps}
\end{figure}

Fig.~\ref{fig:nR_sweeps} illustrates the
transformation of the subgap spectra with varying $\nR$ (the gate
voltage setting the occupancy of the right SC island) for a range of
$\GR$. By tuning $\nR$ towards an odd integer value $\NN +1$, the
large $\ECR$ enforces odd occupation in the right channel. This
results in the decoupling of the left channel in a large interval
around odd $\nR$ and a single YSR-like subgap singlet. This is the
same decoupling effect as seen in Fig. \ref{fig:EcL>>EcR}, where odd
occupation of the right channel is enforced by a large $\ECL$ term at
even $\nL$. As $\ECR > \Delta$ we observe the closure of the transport
superconducting gap (the difference between the BCS-like doublet state
$\ketref$ and the $\nref +1$ quasi continuum) around
odd $\nR$, as predicted in the single-channel model \cite{paper1}.
Increasing $\GR$ summons a second bound state into the gap and
increases the interval at which it is present. At very large $\GR$ and
close to $\nR=\NN$ the excited subgap state exhibits a further
decrease in energy, as exchange of the additional Cooper pair between
the channels becomes favourable.  The energy difference between the
subgap states is minimal at  a value of $\nR$ which is somewhat lower than $\NN$.
This is due to the fact that we are considering the states in
the $\nref +1$ sector. These have an additional particle compared to
half filling, thus having lower energy (compared to the reference
$\ketref$ state) in the $(\nL, \nR) = (\NN, \NN - \delta)$ parameter
point compared to the $(\NN, \NN + \delta)$ one, where $\delta$
represents a small deviation from the half-filling point.  If the
states in the $\nref - 1$ sector were shown instead, the
situation would be symmetrically mirrored over the $\nR = \NN$ point,
with the minimal energy difference at $\nR = \NN + \delta$.

\begin{figure}
\centering
\includegraphics[width=1\linewidth]{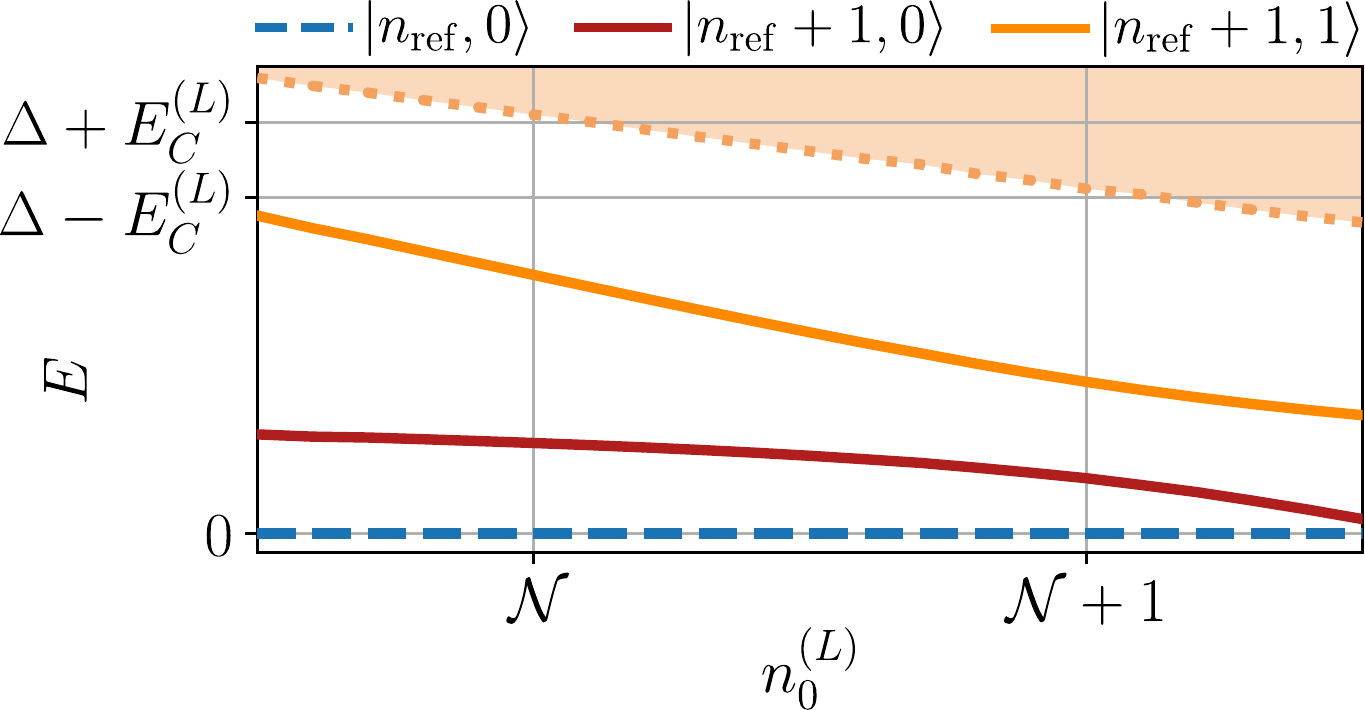}
\caption{
Subgap spectra with varying $\nL$. Realistic value of $U = 4 \Delta$, at the strongly asymmetric point with $\ECL = 0.1 \Delta$ and $\ECR = 1.5 \Delta$ and $\GL /U = 0.1 U$ $\GR /U = 1$. 
}
\label{fig:nL_sweeps}
\end{figure}

Fig. \ref{fig:nL_sweeps} shows the $\nL$ dependence of the energy spectrum, corresponding to the case with large hybridisation asymmetry shown in Fig. \ref{fig:nR_sweeps}c. 
The energy of the decoupled continuum states depends linearly on $\nL$, with the slope given by $\ECL$. The gap width is $\Delta + \ECL$ when $\nL = \NN$ and $\Delta - \ECL$ when $\nL = \NN + 1$. 
While the energies of the states themselves exhibit a finite slope throughout the $\nL$ range, the derivative of the energy difference becomes zero close to $\nL = \NN + 1$. The system thus has an operational sweet spot in both island gate voltage tunings.

\subsection{QD gate voltage tuning}
\label{sec6}

In typical experimental setups, the QD gate voltage is easily tunable as well. In our model, this is represented by the term $U/2 (\hat{n}_\mathrm{imp} - \nu)^2$, where $U/2$ plays the role of the QD charging energy and $\nu$ is the favourable occupation of the QD set by the gate. Here we investigate the properties of the spectra for the system in the symmetrical and the strongly asymmetrical cases.
Tuning the QD gate voltage away from the particle-hole (p-h) symmetric point $\nu=1$ decreases the local magnetic moment of the dot. 
A completely filled (or empty) QD thus does not have a local magnetic moment, suppressing the spin physics entirely. This means that the subgap states asymptotically approach the gap edge for extreme values of $\nu$, resulting in the well known eye-shaped dispersion for the YSR subgap states. 

\subsubsection{Symmetric case}

First we revisit the symmetrical system with $\GL = \GR$, $\ECL = \ECR$ and $\nL = \nR = \NN$. We show the energy diagram as a function of $\nu$ in Fig.~\ref{fig:nu_sweeps}a. 
In the p-h symmetric point at $\nu=1$, the singlet subgap states are $\ketg$ and $\ketu$. Deviation from the p-h symmetric point does not break space-inversion symmetry, so the states retain their parity throughout.
While increasing $\nu$ towards $\nu=2$, the QD levels fills up, transforming the first subgap state into a $(\NN, 2, \NN)$ configuration. The roles of the singlet and doublet sectors switch, as now the singlet states can be formed without mustering the Bogoliubov quasiparticles, while the doublet states are forced to have odd occupation in the SC islands.
The ground state in the doublet $\nref$ sector has a full QD level, which is required by $\nu=2$ and enforced by $U$. The resulting odd occupation in one of the SC islands results in the energy of this state at $\Delta + E_C$ above the lowest lying singlet state. In the singlet sector, the first excitation consists of a broken Cooper pair in one of the channels, which does not change the occupation of the channels and therefore costs $2\Delta$. 
Decreasing $\nu$ has the opposite effect of emptying the QD level. The subgap states in the singlet $\nref+1$ sector are superpositions of the strongly coupled singlet states and states with an empty quantum dot and an additional Cooper pair in the SC: $a \big( \ket{(\NN+2, 0, \NN)} \pm \ket{(\NN, 0, \NN+2)} \big) + b \big(\phiL \pm \phiR \big)$, where the plus sign applies to the first (gerade) and the minus sign is for the second (ungerade) subgap state.
With decreasing $\nu$ the amplitude $a$ continuously increases at the cost of $b$, as the YSR singlets become less energetically favourable. Despite this, both states remain in the gap throughout the $\nu$ range, keeping their gerade/ungerade nature. 
It should be noted that when $\nu=0$, the true ground state is actually the singlet state in the $\nref - 1$ sector, its energy being equal to that of the $\nref + 1$ ground state at $\nu=2$. 
The procedure of tuning the device to this point would therefore require two steps: First tuning the device
into the ground state at $\nu=2$ would fill it with $\nref+1$
electrons. Next, one should completely pinch off the device from the source and
drain electrodes by strongly raising the tunnel barriers, and finally tune the QD gate voltage to $\nu=0$. The pair of singlet subgap states could then be probed by microwave spectroscopy\cite{Lange_2015, Larsen_2015, Woerkom_2017, Tosi_2019}.

\subsubsection{Non-symmetric case}

\begin{figure}
\centering
\includegraphics[width=1\linewidth]{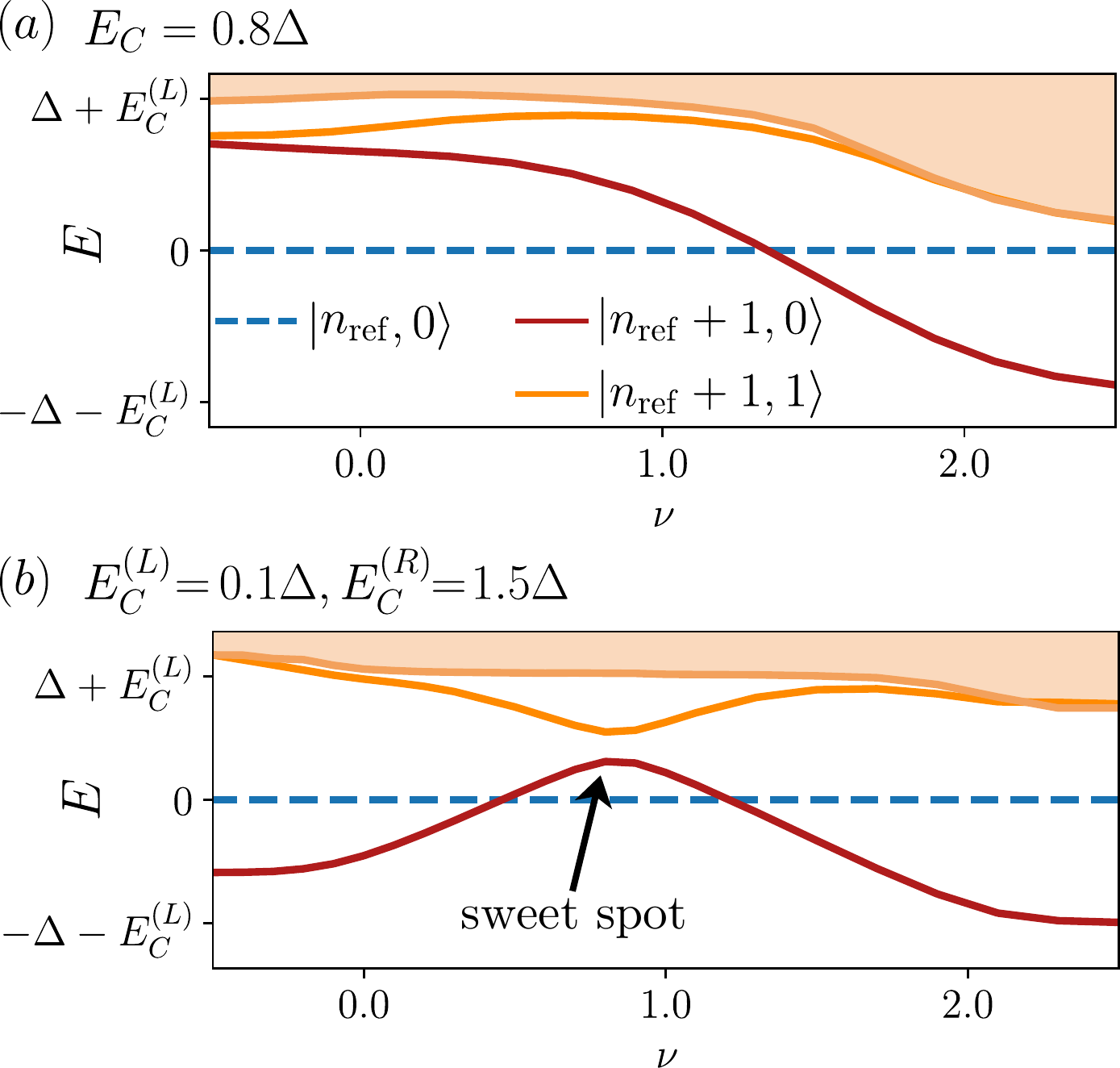}
\caption{ QD gate voltage tuning with a realistic value of $U \approx 4 \Delta$.
a) $\ECL=\ECR=0.8\Delta$, $\GL=\GR = 0.2 U$,
b) $\ECL = 0.1\Delta$, $\ECR = 1.5\Delta$, $\GL = 0.1U$, $\GR = 1.0 U$.
In both panels $U=0.64=4\Delta$.}
\label{fig:nu_sweeps}
\end{figure}

Next, consider the asymmetrical situation with $\ECL \ll \ECR$ and $\GL \ll \GR$. The energy diagram is shown in Fig.~\ref{fig:nu_sweeps}(b).
The charging energy of the QD and the right channel are comparable and much larger than the charging energy of the left channel. 
The energy cost of the unfavourable occupation of the left channel is the relatively small factor of $\ECL$.
Thus the left channel tends to act as a reservoir for the rest of the system, primarily optimizing the charge configuration of the QD and the right channel. 
The subgap states have $(\NN, 1, \NN+1)$ and $(\NN+2, 1, \NN-1)$ configurations at $\nu=1$. 
By tuning the gate voltage away from the particle hole symmetric point, the first subgap state assumes the $(\NN+2, 0, \NN)$ configuration for $\nu=0$ and $(\NN, 2, \NN)$ for $\nu = 2$. Such configurations do not allow for a second subgap state as there is no local moment on the QD. The first excitation is a broken Cooper pair in the SC - there is a continuum of such excitations $2\Delta$ above the ground state.  
The value of $\nu$ where the energy difference between the two subgap states is minimal is again a bit below half-filling, for the same reasons as when tuning $\nR$.  

This is an optimal operation point for a YSR qubit, pointed to by black arrows. The derivative of energy difference with respect to all gate voltages has a root at this triple sweet-spot point, making the system relatively insensitive to electric noise, while the energy difference between the two subgap states itself is minimal. For further comments on the use of the system as a qubit, refer to the Discussion section.

\subsection{Capacitive coupling}
Capacitive coupling between component parts of the system is typically present in experimental devices, including in InAs quantum dots \cite{Galpin_2006, Nishikawa_2012, JuanCarlos2021}. 
We include it in our model by augmenting it with terms that couple charge in parts of the system: 
\begin{equation}
    V_\beta \big( \hat{n}_\mathrm{imp} - \nu \big) \big(\hat{n}_\mathrm{SC}^{(\beta)} - \hat{n}_\mathrm{0}^{(\beta)} \big),
\end{equation}
for $\beta = \mathrm{L, R}$. The strength of the coupling $V_\beta$ is bounded from above by $\mathrm{max}(U/2, E_C^{(\beta)})$, otherwise the system enters a charge-ordered state \cite{Nishikawa_2012}. An experimentally relevant value is on the order of $E_C^{(\beta)} / 5$ \cite{JuanCarlos2021}. 
We find that as long as the ratio of $V_\beta / E_C^{(\beta)}$ is equal for both channels and within reasonable strength, the effect of capacitive coupling on the nature of the subgap singlet states is minute, especially in the proposed qubit regime.     

\section{Electric transition moments}
\label{sec7}

Manipulation of the subgap states in the same charge sector in the SC-QD-SC devices is possible by inducing microwave transitions \cite{Oosterkamp1998}. 
The electric field of a linear resonator typically couples to the electric dipole moment of the device \cite{Wallraff2004, Stockklauser2017}, but coupling to the quadrupole moment is also possible in a triple quantum dot architecture \cite{Koski2020}. 
In order to demonstrate that manipulation of subgap states in the proposed device is possible, we calculate both quantities for the transitions between the two subgap single states.

\begin{figure}
\centering
\includegraphics[width=1\linewidth]{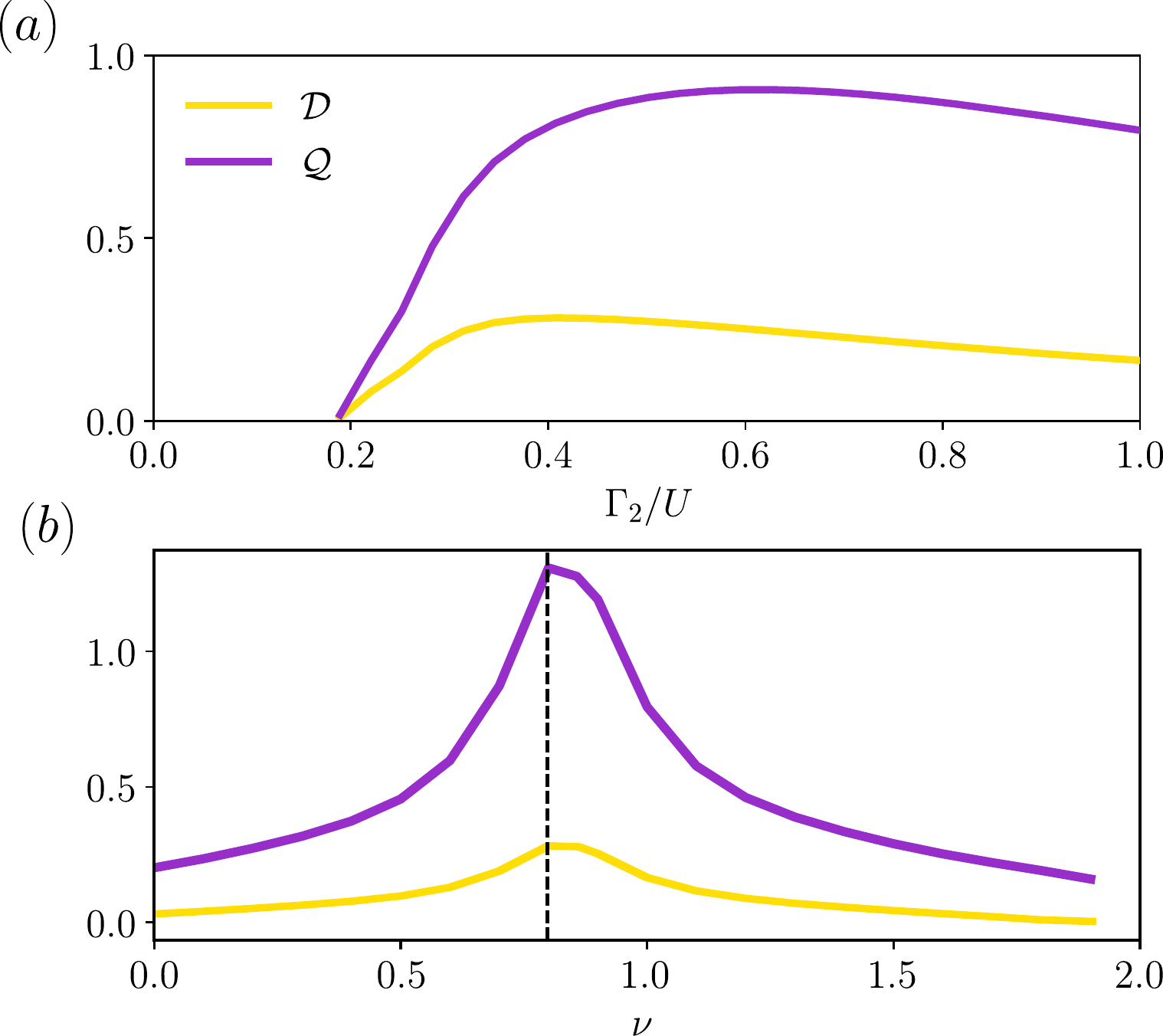}
\caption{ 
Electric transition dipole moment $\mathcal{D}$ and quadrupole moment $\mathcal{Q}$.
(a) Hybridisation strength $\Gamma_R$ dependence for parameters as in Fig.~\ref{fig:nR_sweeps}(a).
(b) QD gate voltage dependence for parameters as in Fig.~\ref{fig:nu_sweeps}(b). The horizontal dashed line corresponds to the position of the sweet spot in Fig.~\ref{fig:nu_sweeps}(b).
}
\label{fig:transition_dipole_moment}
\end{figure}

The matrix element for the transition dipole moment is $\bra{1} q\hat{\mathbf{r}} \ket{0}$, where $\hat{\mathbf{r}}$ is the position operator. In the language of our model with two point-like SC islands at unit distance from the QD, this is (up to a constant prefactor) equivalent to: $\bra{1} (\hat{n}_L - \nL) - (\hat{n}_R - \nR) \ket{0}$. The $n_0$ factors represent the charge accumulated on the gate electrodes, but their contribution is nullified by the orthogonality of the subgap states. This gives the expression for the transition dipole moment $\mathcal{D} = \bra{1} \hat{n}_L - \hat{n}_R \ket{0}$. Following the same arguments, the quadrupole moment $\bra{1} q\hat{\mathbf{r}}^2 \ket{0}$ is written as $\mathcal{Q} =  \bra{1} \hat{n}_L + \hat{n}_R \ket{0}$.
In Fig. \ref{fig:transition_dipole_moment} the transition dipole and quadrupole moments for singlet subgap states are shown. Panel (a) corresponds to states shown in Fig.~\ref{fig:nR_sweeps}(a), while panel (b) corresponds to Fig.~\ref{fig:nu_sweeps}(b). 
Both transition moments roughly follow the energy difference between the two singlet states, exhibiting a peak when the singlets are close in energy and the system is close to its sweet spot. Both quickly decrease to zero when one of the states merges with the continuum.

\section{Discussion}

Practical quantum computing, especially one that would be cryptographically relevant, requires systems with a large number of
qubits \cite{Loss_1998}. This need will drive increasing integration
and scaling down of all constituent parts. Miniaturization of devices
based on superconducting regions will ultimately lead to the use of SC
islands, which will then become a ubiquitous element. It thus appears
pertinant to explore the possible advantages that these will bring. In
this work we have studied one possible implementation that makes use
of two separate SC islands that bring about two separate spin-singlet
YSR subgap states which may be used to store one qubit of information.
We now comment on the relation of this class of qubits to other known 
architectures \cite{nielsen_chuang_2010, Kjaergaard2020}, and on its relative merits
and downsides.

The YSR spin-singlet qubit may at first appear related to the Andreev
qubit in a Josephson junction \cite{Janvier_2015, Hays_2018}. The Andreev qubit makes use
of the Andreev bound states (ABS) in the even parity subsector: the
basis states consist of zero or double occupancy of the ABS. This
concept does not directly apply to the YSR spin-singlet states: these
cannot be ``doubly occupied'' since only a single quasiparticle from a
given SC island can bind antiferromagnetically to the impurity to
form a singlet state. An Andreev qubit can be approximately modelled using a
non-interacting QD which correspond to the $U\to0$ limit of our model.
The higher-energy singlet subgap state of the Andreev qubit at $U\to0$
is not connected to the higher-energy singlet YSR state at $U \gg
\Delta$. 
Starting from the $U=0$, $E_c=0$ limit, increasing $U$ or $E_C$ will rapidly drive the higher-energy spin-singlet Andreev state into the continuum. Conversely, the higher-energy YSR singlet will disappear in the continuum when moving away from the $U\gg \Delta$, $E_c \neq 0$ regime by decreasing $U$ or $E_c$.

The YSR singlet qubit is actually more closely related to the charge
qubit in double QD systems \cite{gorman2005}. The basis states for the
charge qubit correspond to the position of the single electron in
either of the two QDs. This is similar to YSR singlet qubit states
which likewise have different distribution of the wavefunction in
space. In this sense, the SI-QD-SI device is a variant of a charge
qubit in a non-homogeneous triple QD system, however with the qubit
state encoded in a pair of complex many-body states, rather than in a
single-particle position basis. Depending on the operating point of
the device (parameter choice), the spatial structure of the YSR
singlets differs. For example, the regime $E_C^{(L)}<E_C^{(R)}$ and
$\GL<\GR$, discussed in subsection~\ref{caseECL}, corresponds to a
pair of singlet states which mainly {\it differ in the position in
space of a single Cooper pair}. In this respect, at this operating
point the YSR qubit is actually very similar to the Cooper-pair box
charge qubit \cite{Shnirman1997,Bouchiat1998,Makhlin2001, Koch_2007}.

The YSR singlet qubits are relatively insensitive to magnetic field
noise and do not require any magnetic flux tuning through
superconducting loops and are in this aspect similar to spin singlet qubits \cite{Sala_2017, Sala_2020}. The robustness with respect to magnetic field
weakly depends on the $g$-factors of the subsystems (QD vs. SC
islands). The Zeeman shifts of spin-singlet states are namely
proportional to the differences in the $g$-factors. Nevertheless, even
in the case of different $g$-factors there exist operating points (such as the
one discussed above) which are largely insensitive to the field,
because both states are Zeeman shifted in the same way, since they
differ mostly in the position of a Cooper pair (two-particle singlet
confined within a single SC island, insensitive to the magnetic field
to first order).

Finally, we mention the YSR qubits recently proposed in
Ref.~\cite{mishra2021}. Those are based on classical magnetic moments
(e.g. on high-spin magnetic adsorbates with large magnetic anisotropy)
which have spontaneously broken spin symmetry and are sensitive to
external magnetic field. Another key difference is that our platform
requires a single local moment (QD) and two superconductors (SC
islands), while their proposal requires two local moments (adatoms)
and a singlet superconductor (surface of a bulk SC). Beside making use
of YSR subgap states, the two approaches are thus quite different. 

\section{Conclusion}

We have explored the properties of the subgap states in the system of an interacting
quantum dot coupled to two small superconducting islands, focusing on the spin-singlet
subspace. We have uncovered regimes where two sub-gap singlets are present deep in the gap
with good separation from the continuum states. Their energies can be tuned to obtain an energy difference 
suitable for manipulation with microwaves pulses. The transitions are facilitated by large
electric dipole (and quadrupole) transition moments. Such devices could be built using
the present-day technology. Using a third superconducting island, the scheme could
be generalized to qutrits.

This class of devices has also interesting properties in the doublet subspace: two Bogoliubov
quasiparticles, one from each superconducting island, could overscreen the impurity moment to produce
an overscreened Yu-Shiba-Rusinov state, a state clearly distinct from the ``decoupled'' spin-doublet
state.

\begin{acknowledgments}
R\v{Z} and LP acknowledge the support of the Slovenian Research Agency (ARRS)
under P1-0044.
We acknowledge discussions with Daniel Bauernfeind who participated in the early
stages of this project.
Calculations were performed with the ITensor library \cite{itensor}.
\end{acknowledgments}

\bibliography{bibliography.bib}

\begin{thebibliography}{89}%
\makeatletter
\providecommand \@ifxundefined [1]{%
 \@ifx{#1\undefined}
}%
\providecommand \@ifnum [1]{%
 \ifnum #1\expandafter \@firstoftwo
 \else \expandafter \@secondoftwo
 \fi
}%
\providecommand \@ifx [1]{%
 \ifx #1\expandafter \@firstoftwo
 \else \expandafter \@secondoftwo
 \fi
}%
\providecommand \natexlab [1]{#1}%
\providecommand \enquote  [1]{``#1''}%
\providecommand \bibnamefont  [1]{#1}%
\providecommand \bibfnamefont [1]{#1}%
\providecommand \citenamefont [1]{#1}%
\providecommand \href@noop [0]{\@secondoftwo}%
\providecommand \href [0]{\begingroup \@sanitize@url \@href}%
\providecommand \@href[1]{\@@startlink{#1}\@@href}%
\providecommand \@@href[1]{\endgroup#1\@@endlink}%
\providecommand \@sanitize@url [0]{\catcode `\\12\catcode `\$12\catcode
  `\&12\catcode `\#12\catcode `\^12\catcode `\_12\catcode `\%12\relax}%
\providecommand \@@startlink[1]{}%
\providecommand \@@endlink[0]{}%
\providecommand \url  [0]{\begingroup\@sanitize@url \@url }%
\providecommand \@url [1]{\endgroup\@href {#1}{\urlprefix }}%
\providecommand \urlprefix  [0]{URL }%
\providecommand \Eprint [0]{\href }%
\providecommand \doibase [0]{http://dx.doi.org/}%
\providecommand \selectlanguage [0]{\@gobble}%
\providecommand \bibinfo  [0]{\@secondoftwo}%
\providecommand \bibfield  [0]{\@secondoftwo}%
\providecommand \translation [1]{[#1]}%
\providecommand \BibitemOpen [0]{}%
\providecommand \bibitemStop [0]{}%
\providecommand \bibitemNoStop [0]{.\EOS\space}%
\providecommand \EOS [0]{\spacefactor3000\relax}%
\providecommand \BibitemShut  [1]{\csname bibitem#1\endcsname}%
\let\auto@bib@innerbib\@empty
\bibitem [{\citenamefont {Yu}(1965)}]{Yu}%
  \BibitemOpen
  \bibfield  {author} {\bibinfo {author} {\bibfnamefont {L.}~\bibnamefont
  {Yu}},\ }\bibfield  {title} {\enquote {\bibinfo {title} {Bound state in
  superconductors with paramagnetic impurities},}\ }\href@noop {} {\bibfield
  {journal} {\bibinfo  {journal} {Acta Phys. Sin.}\ }\textbf {\bibinfo {volume}
  {21}} (\bibinfo {year} {1965})}\BibitemShut {NoStop}%
\bibitem [{\citenamefont {Shiba}(1968)}]{Shiba}%
  \BibitemOpen
  \bibfield  {author} {\bibinfo {author} {\bibfnamefont {Hiroyuki}\
  \bibnamefont {Shiba}},\ }\bibfield  {title} {\enquote {\bibinfo {title}
  {Classical spins in superconductors},}\ }\href {\doibase 10.1143/PTP.40.435}
  {\bibfield  {journal} {\bibinfo  {journal} {Progress of Theoretical Physics}\
  }\textbf {\bibinfo {volume} {40}},\ \bibinfo {pages} {435--451} (\bibinfo
  {year} {1968})},\ \Eprint
  {http://arxiv.org/abs/https://academic.oup.com/ptp/article-pdf/40/3/435/5185550/40-3-435.pdf}
  {https://academic.oup.com/ptp/article-pdf/40/3/435/5185550/40-3-435.pdf}
  \BibitemShut {NoStop}%
\bibitem [{\citenamefont {{Rusinov}}(1969)}]{Rusinov}%
  \BibitemOpen
  \bibfield  {author} {\bibinfo {author} {\bibfnamefont {A.~I.}\ \bibnamefont
  {{Rusinov}}},\ }\bibfield  {title} {\enquote {\bibinfo {title}
  {Superconductivity near a paramagnetic impurity},}\ }\href@noop {} {\bibfield
   {journal} {\bibinfo  {journal} {Soviet Journal of Experimental and
  Theoretical Physics Letters}\ }\textbf {\bibinfo {volume} {9}},\ \bibinfo
  {pages} {85} (\bibinfo {year} {1969})}\BibitemShut {NoStop}%
\bibitem [{\citenamefont {Balatsky}\ \emph {et~al.}(2006)\citenamefont
  {Balatsky}, \citenamefont {Vekhter},\ and\ \citenamefont
  {Zhu}}]{Balatasky2006_review}%
  \BibitemOpen
  \bibfield  {author} {\bibinfo {author} {\bibfnamefont {A.~V.}\ \bibnamefont
  {Balatsky}}, \bibinfo {author} {\bibfnamefont {I.}~\bibnamefont {Vekhter}}, \
  and\ \bibinfo {author} {\bibfnamefont {Jian-Xin}\ \bibnamefont {Zhu}},\
  }\bibfield  {title} {\enquote {\bibinfo {title} {Impurity-induced states in
  conventional and unconventional superconductors},}\ }\href {\doibase
  10.1103/RevModPhys.78.373} {\bibfield  {journal} {\bibinfo  {journal} {Rev.
  Mod. Phys.}\ }\textbf {\bibinfo {volume} {78}},\ \bibinfo {pages} {373--433}
  (\bibinfo {year} {2006})}\BibitemShut {NoStop}%
\bibitem [{\citenamefont {Mart{\'{\i}}n-Rodero}\ and\ \citenamefont
  {Yeyati}(2011)}]{MartinRodero_2011}%
  \BibitemOpen
  \bibfield  {author} {\bibinfo {author} {\bibfnamefont {A.}~\bibnamefont
  {Mart{\'{\i}}n-Rodero}}\ and\ \bibinfo {author} {\bibfnamefont {A.~Levy}\
  \bibnamefont {Yeyati}},\ }\bibfield  {title} {\enquote {\bibinfo {title}
  {{Josephson and Andreev transport through quantum dots}},}\ }\href {\doibase
  10.1080/00018732.2011.624266} {\bibfield  {journal} {\bibinfo  {journal}
  {Advances in Physics}\ }\textbf {\bibinfo {volume} {60}},\ \bibinfo {pages}
  {899--958} (\bibinfo {year} {2011})}\BibitemShut {NoStop}%
\bibitem [{\citenamefont {Heinrich}\ \emph {et~al.}(2018)\citenamefont
  {Heinrich}, \citenamefont {Pascual},\ and\ \citenamefont
  {Franke}}]{Heinrich_2018}%
  \BibitemOpen
  \bibfield  {author} {\bibinfo {author} {\bibfnamefont {Benjamin~W.}\
  \bibnamefont {Heinrich}}, \bibinfo {author} {\bibfnamefont {Jose~I.}\
  \bibnamefont {Pascual}}, \ and\ \bibinfo {author} {\bibfnamefont
  {Katharina~J.}\ \bibnamefont {Franke}},\ }\bibfield  {title} {\enquote
  {\bibinfo {title} {Single magnetic adsorbates on s -wave superconductors},}\
  }\href {\doibase 10.1016/j.progsurf.2018.01.001} {\bibfield  {journal}
  {\bibinfo  {journal} {Prog. Surf. Sci.}\ }\textbf {\bibinfo {volume} {93}},\
  \bibinfo {pages} {1--19} (\bibinfo {year} {2018})}\BibitemShut {NoStop}%
\bibitem [{\citenamefont {Lutchyn}\ \emph {et~al.}(2018)\citenamefont
  {Lutchyn}, \citenamefont {Bakkers}, \citenamefont {Kouwenhoven},
  \citenamefont {Krogstrup}, \citenamefont {Marcus},\ and\ \citenamefont
  {Oreg}}]{Lutchyn_2018}%
  \BibitemOpen
  \bibfield  {author} {\bibinfo {author} {\bibfnamefont {R.~M.}\ \bibnamefont
  {Lutchyn}}, \bibinfo {author} {\bibfnamefont {E.~P. A.~M.}\ \bibnamefont
  {Bakkers}}, \bibinfo {author} {\bibfnamefont {L.~P.}\ \bibnamefont
  {Kouwenhoven}}, \bibinfo {author} {\bibfnamefont {P.}~\bibnamefont
  {Krogstrup}}, \bibinfo {author} {\bibfnamefont {C.~M.}\ \bibnamefont
  {Marcus}}, \ and\ \bibinfo {author} {\bibfnamefont {Y.}~\bibnamefont
  {Oreg}},\ }\bibfield  {title} {\enquote {\bibinfo {title} {Majorana zero
  modes in superconductor{\textendash}semiconductor heterostructures},}\ }\href
  {\doibase 10.1038/s41578-018-0003-1} {\bibfield  {journal} {\bibinfo
  {journal} {Nat. Rev. Mater.}\ }\textbf {\bibinfo {volume} {3}},\ \bibinfo
  {pages} {52--68} (\bibinfo {year} {2018})}\BibitemShut {NoStop}%
\bibitem [{\citenamefont {Meden}(2019)}]{Meden2019}%
  \BibitemOpen
  \bibfield  {author} {\bibinfo {author} {\bibfnamefont {V}~\bibnamefont
  {Meden}},\ }\bibfield  {title} {\enquote {\bibinfo {title} {The
  {Anderson{\textendash}Josephson} quantum dot{\textemdash}a theory
  perspective},}\ }\href {\doibase 10.1088/1361-648x/aafd6a} {\bibfield
  {journal} {\bibinfo  {journal} {Journal of Physics: Condensed Matter}\
  }\textbf {\bibinfo {volume} {31}},\ \bibinfo {pages} {163001} (\bibinfo
  {year} {2019})}\BibitemShut {NoStop}%
\bibitem [{\citenamefont {Prada}\ \emph {et~al.}(2020)\citenamefont {Prada},
  \citenamefont {San-Jose}, \citenamefont {de~Moor}, \citenamefont {Geresdi},
  \citenamefont {Lee}, \citenamefont {Klinovaja}, \citenamefont {Loss},
  \citenamefont {Nyg{\aa}rd}, \citenamefont {Aguado},\ and\ \citenamefont
  {Kouwenhoven}}]{Prada_2020}%
  \BibitemOpen
  \bibfield  {author} {\bibinfo {author} {\bibfnamefont {Elsa}\ \bibnamefont
  {Prada}}, \bibinfo {author} {\bibfnamefont {Pablo}\ \bibnamefont {San-Jose}},
  \bibinfo {author} {\bibfnamefont {Michiel W.~A.}\ \bibnamefont {de~Moor}},
  \bibinfo {author} {\bibfnamefont {Attila}\ \bibnamefont {Geresdi}}, \bibinfo
  {author} {\bibfnamefont {Eduardo J.~H.}\ \bibnamefont {Lee}}, \bibinfo
  {author} {\bibfnamefont {Jelena}\ \bibnamefont {Klinovaja}}, \bibinfo
  {author} {\bibfnamefont {Daniel}\ \bibnamefont {Loss}}, \bibinfo {author}
  {\bibfnamefont {Jesper}\ \bibnamefont {Nyg{\aa}rd}}, \bibinfo {author}
  {\bibfnamefont {Ram{\'{o}}n}\ \bibnamefont {Aguado}}, \ and\ \bibinfo
  {author} {\bibfnamefont {Leo~P.}\ \bibnamefont {Kouwenhoven}},\ }\bibfield
  {title} {\enquote {\bibinfo {title} {{From Andreev to Majorana bound states
  in hybrid superconductor{\textendash}semiconductor nanowires}},}\ }\href
  {\doibase 10.1038/s42254-020-0228-y} {\bibfield  {journal} {\bibinfo
  {journal} {Nat Rev Phys}\ }\textbf {\bibinfo {volume} {2}},\ \bibinfo {pages}
  {575--594} (\bibinfo {year} {2020})}\BibitemShut {NoStop}%
\bibitem [{\citenamefont {Frolov}\ \emph {et~al.}(2020)\citenamefont {Frolov},
  \citenamefont {Manfra},\ and\ \citenamefont {Sau}}]{Frolov_2020}%
  \BibitemOpen
  \bibfield  {author} {\bibinfo {author} {\bibfnamefont {S.~M.}\ \bibnamefont
  {Frolov}}, \bibinfo {author} {\bibfnamefont {M.~J.}\ \bibnamefont {Manfra}},
  \ and\ \bibinfo {author} {\bibfnamefont {J.~D.}\ \bibnamefont {Sau}},\
  }\bibfield  {title} {\enquote {\bibinfo {title} {Topological
  superconductivity in hybrid devices},}\ }\href {\doibase
  10.1038/s41567-020-0925-6} {\bibfield  {journal} {\bibinfo  {journal} {Nat.
  Phys.}\ }\textbf {\bibinfo {volume} {16}},\ \bibinfo {pages} {718--724}
  (\bibinfo {year} {2020})}\BibitemShut {NoStop}%
\bibitem [{\citenamefont {Anderson}(1961)}]{SIAM}%
  \BibitemOpen
  \bibfield  {author} {\bibinfo {author} {\bibfnamefont {P.~W.}\ \bibnamefont
  {Anderson}},\ }\bibfield  {title} {\enquote {\bibinfo {title} {Localized
  magnetic states in metals},}\ }\href {\doibase 10.1103/PhysRev.124.41}
  {\bibfield  {journal} {\bibinfo  {journal} {Phys. Rev.}\ }\textbf {\bibinfo
  {volume} {124}},\ \bibinfo {pages} {41--53} (\bibinfo {year}
  {1961})}\BibitemShut {NoStop}%
\bibitem [{\citenamefont {Bardeen}\ \emph {et~al.}(1957)\citenamefont
  {Bardeen}, \citenamefont {Cooper},\ and\ \citenamefont {Schrieffer}}]{BCS}%
  \BibitemOpen
  \bibfield  {author} {\bibinfo {author} {\bibfnamefont {J.}~\bibnamefont
  {Bardeen}}, \bibinfo {author} {\bibfnamefont {L.~N.}\ \bibnamefont {Cooper}},
  \ and\ \bibinfo {author} {\bibfnamefont {J.~R.}\ \bibnamefont {Schrieffer}},\
  }\bibfield  {title} {\enquote {\bibinfo {title} {Microscopic theory of
  superconductivity},}\ }\href {\doibase 10.1103/PhysRev.106.162} {\bibfield
  {journal} {\bibinfo  {journal} {Phys. Rev.}\ }\textbf {\bibinfo {volume}
  {106}},\ \bibinfo {pages} {162--164} (\bibinfo {year} {1957})}\BibitemShut
  {NoStop}%
\bibitem [{\citenamefont {Wilson}(1975)}]{Wilson_NRG}%
  \BibitemOpen
  \bibfield  {author} {\bibinfo {author} {\bibfnamefont {Kenneth~G.}\
  \bibnamefont {Wilson}},\ }\bibfield  {title} {\enquote {\bibinfo {title}
  {{The renormalization group: Critical phenomena and the Kondo problem}},}\
  }\href {\doibase 10.1103/RevModPhys.47.773} {\bibfield  {journal} {\bibinfo
  {journal} {Rev. Mod. Phys.}\ }\textbf {\bibinfo {volume} {47}},\ \bibinfo
  {pages} {773--840} (\bibinfo {year} {1975})}\BibitemShut {NoStop}%
\bibitem [{\citenamefont {Krishna-murthy}\ \emph {et~al.}(1980)\citenamefont
  {Krishna-murthy}, \citenamefont {Wilkins},\ and\ \citenamefont
  {Wilson}}]{Krishnamurthy_NRG}%
  \BibitemOpen
  \bibfield  {author} {\bibinfo {author} {\bibfnamefont {H.~R.}\ \bibnamefont
  {Krishna-murthy}}, \bibinfo {author} {\bibfnamefont {J.~W.}\ \bibnamefont
  {Wilkins}}, \ and\ \bibinfo {author} {\bibfnamefont {K.~G.}\ \bibnamefont
  {Wilson}},\ }\bibfield  {title} {\enquote {\bibinfo {title}
  {Renormalization-group approach to the {Anderson} model of dilute magnetic
  alloys. ii. static properties for the asymmetric case},}\ }\href {\doibase
  10.1103/PhysRevB.21.1044} {\bibfield  {journal} {\bibinfo  {journal} {Phys.
  Rev. B}\ }\textbf {\bibinfo {volume} {21}},\ \bibinfo {pages} {1044--1083}
  (\bibinfo {year} {1980})}\BibitemShut {NoStop}%
\bibitem [{\citenamefont {Bulla}\ \emph {et~al.}(2008)\citenamefont {Bulla},
  \citenamefont {Costi},\ and\ \citenamefont {Pruschke}}]{Bulla2008}%
  \BibitemOpen
  \bibfield  {author} {\bibinfo {author} {\bibfnamefont {Ralf}\ \bibnamefont
  {Bulla}}, \bibinfo {author} {\bibfnamefont {Theo~A.}\ \bibnamefont {Costi}},
  \ and\ \bibinfo {author} {\bibfnamefont {Thomas}\ \bibnamefont {Pruschke}},\
  }\bibfield  {title} {\enquote {\bibinfo {title} {Numerical renormalization
  group method for quantum impurity systems},}\ }\href {\doibase
  10.1103/revmodphys.80.395} {\bibfield  {journal} {\bibinfo  {journal}
  {Reviews of Modern Physics}\ }\textbf {\bibinfo {volume} {80}},\ \bibinfo
  {pages} {395--450} (\bibinfo {year} {2008})}\BibitemShut {NoStop}%
\bibitem [{\citenamefont {Sakai}\ \emph {et~al.}(1993)\citenamefont {Sakai},
  \citenamefont {Shimizu}, \citenamefont {Shiba},\ and\ \citenamefont
  {Satori}}]{Sakai_1993}%
  \BibitemOpen
  \bibfield  {author} {\bibinfo {author} {\bibfnamefont {Osamu}\ \bibnamefont
  {Sakai}}, \bibinfo {author} {\bibfnamefont {Yukihiro}\ \bibnamefont
  {Shimizu}}, \bibinfo {author} {\bibfnamefont {Hiroyuki}\ \bibnamefont
  {Shiba}}, \ and\ \bibinfo {author} {\bibfnamefont {Koji}\ \bibnamefont
  {Satori}},\ }\bibfield  {title} {\enquote {\bibinfo {title} {Numerical
  renormalization group study of magnetic impurities in superconductors. {II}.
  dynamical excitation spectra and spatial variation of the order parameter},}\
  }\href {\doibase 10.1143/jpsj.62.3181} {\bibfield  {journal} {\bibinfo
  {journal} {J. Phys. Soc. Jpn.}\ }\textbf {\bibinfo {volume} {62}},\ \bibinfo
  {pages} {3181--3197} (\bibinfo {year} {1993})}\BibitemShut {NoStop}%
\bibitem [{\citenamefont {Yoshioka}\ and\ \citenamefont
  {Ohashi}(2000)}]{Yoshioka_2000}%
  \BibitemOpen
  \bibfield  {author} {\bibinfo {author} {\bibfnamefont {Tomoki}\ \bibnamefont
  {Yoshioka}}\ and\ \bibinfo {author} {\bibfnamefont {Yoji}\ \bibnamefont
  {Ohashi}},\ }\bibfield  {title} {\enquote {\bibinfo {title} {{Numerical
  renormalization group studies on single impurity Anderson model in
  superconductivity: A unified treatment of magnetic, nonmagnetic impurities,
  and resonance scattering}},}\ }\href {\doibase 10.1143/jpsj.69.1812}
  {\bibfield  {journal} {\bibinfo  {journal} {J. Phys. Soc. Jpn.}\ }\textbf
  {\bibinfo {volume} {69}},\ \bibinfo {pages} {1812--1823} (\bibinfo {year}
  {2000})}\BibitemShut {NoStop}%
\bibitem [{\citenamefont {Satori}\ \emph {et~al.}(1992)\citenamefont {Satori},
  \citenamefont {Shiba}, \citenamefont {Sakai},\ and\ \citenamefont
  {Shimizu}}]{Satori_1992}%
  \BibitemOpen
  \bibfield  {author} {\bibinfo {author} {\bibfnamefont {Koji}\ \bibnamefont
  {Satori}}, \bibinfo {author} {\bibfnamefont {Hiroyuki}\ \bibnamefont
  {Shiba}}, \bibinfo {author} {\bibfnamefont {Osamu}\ \bibnamefont {Sakai}}, \
  and\ \bibinfo {author} {\bibfnamefont {Yukihiro}\ \bibnamefont {Shimizu}},\
  }\bibfield  {title} {\enquote {\bibinfo {title} {Numerical renormalization
  group study of magnetic impurities in superconductors},}\ }\href {\doibase
  10.1143/jpsj.61.3239} {\bibfield  {journal} {\bibinfo  {journal} {J. Phys.
  Soc. Jpn.}\ }\textbf {\bibinfo {volume} {61}},\ \bibinfo {pages} {3239--3254}
  (\bibinfo {year} {1992})}\BibitemShut {NoStop}%
\bibitem [{\citenamefont {Krogstrup}\ \emph {et~al.}(2015)\citenamefont
  {Krogstrup}, \citenamefont {Ziino}, \citenamefont {Chang}, \citenamefont
  {Albrecht}, \citenamefont {Madsen}, \citenamefont {Johnson}, \citenamefont
  {Nyg{\aa}rd}, \citenamefont {Marcus},\ and\ \citenamefont
  {Jespersen}}]{Krogstrup_2015}%
  \BibitemOpen
  \bibfield  {author} {\bibinfo {author} {\bibfnamefont {P.}~\bibnamefont
  {Krogstrup}}, \bibinfo {author} {\bibfnamefont {N.~L.~B.}\ \bibnamefont
  {Ziino}}, \bibinfo {author} {\bibfnamefont {W.}~\bibnamefont {Chang}},
  \bibinfo {author} {\bibfnamefont {S.~M.}\ \bibnamefont {Albrecht}}, \bibinfo
  {author} {\bibfnamefont {M.~H.}\ \bibnamefont {Madsen}}, \bibinfo {author}
  {\bibfnamefont {E.}~\bibnamefont {Johnson}}, \bibinfo {author} {\bibfnamefont
  {J.}~\bibnamefont {Nyg{\aa}rd}}, \bibinfo {author} {\bibfnamefont {C.~M.}\
  \bibnamefont {Marcus}}, \ and\ \bibinfo {author} {\bibfnamefont {T.~S.}\
  \bibnamefont {Jespersen}},\ }\bibfield  {title} {\enquote {\bibinfo {title}
  {Epitaxy of semiconductor{\textendash}superconductor nanowires},}\ }\href
  {\doibase 10.1038/nmat4176} {\bibfield  {journal} {\bibinfo  {journal}
  {Nature Mater}\ }\textbf {\bibinfo {volume} {14}},\ \bibinfo {pages}
  {400--406} (\bibinfo {year} {2015})}\BibitemShut {NoStop}%
\bibitem [{\citenamefont {Chang}\ \emph {et~al.}(2015)\citenamefont {Chang},
  \citenamefont {Albrecht}, \citenamefont {Jespersen}, \citenamefont
  {Kuemmeth}, \citenamefont {Krogstrup}, \citenamefont {Nyg{\aa}rd},\ and\
  \citenamefont {Marcus}}]{Chang_2015}%
  \BibitemOpen
  \bibfield  {author} {\bibinfo {author} {\bibfnamefont {W.}~\bibnamefont
  {Chang}}, \bibinfo {author} {\bibfnamefont {S.~M.}\ \bibnamefont {Albrecht}},
  \bibinfo {author} {\bibfnamefont {T.~S.}\ \bibnamefont {Jespersen}}, \bibinfo
  {author} {\bibfnamefont {F.}~\bibnamefont {Kuemmeth}}, \bibinfo {author}
  {\bibfnamefont {P.}~\bibnamefont {Krogstrup}}, \bibinfo {author}
  {\bibfnamefont {J.}~\bibnamefont {Nyg{\aa}rd}}, \ and\ \bibinfo {author}
  {\bibfnamefont {C.~M.}\ \bibnamefont {Marcus}},\ }\bibfield  {title}
  {\enquote {\bibinfo {title} {Hard gap in epitaxial
  semiconductor{\textendash}superconductor nanowires},}\ }\href {\doibase
  10.1038/nnano.2014.306} {\bibfield  {journal} {\bibinfo  {journal} {Nature
  Nanotech}\ }\textbf {\bibinfo {volume} {10}},\ \bibinfo {pages} {232--236}
  (\bibinfo {year} {2015})}\BibitemShut {NoStop}%
\bibitem [{\citenamefont {Flensberg}(1993)}]{Flensberg_1993}%
  \BibitemOpen
  \bibfield  {author} {\bibinfo {author} {\bibfnamefont {Karsten}\ \bibnamefont
  {Flensberg}},\ }\bibfield  {title} {\enquote {\bibinfo {title} {Capacitance
  and conductance of mesoscopic systems connected by quantum point contacts},}\
  }\href {\doibase 10.1103/PhysRevB.48.11156} {\bibfield  {journal} {\bibinfo
  {journal} {Phys. Rev. B}\ }\textbf {\bibinfo {volume} {48}},\ \bibinfo
  {pages} {11156--11166} (\bibinfo {year} {1993})}\BibitemShut {NoStop}%
\bibitem [{\citenamefont {Matveev}(1995)}]{Matveev_1995}%
  \BibitemOpen
  \bibfield  {author} {\bibinfo {author} {\bibfnamefont {K.~A.}\ \bibnamefont
  {Matveev}},\ }\bibfield  {title} {\enquote {\bibinfo {title} {Coulomb
  blockade at almost perfect transmission},}\ }\href {\doibase
  10.1103/PhysRevB.51.1743} {\bibfield  {journal} {\bibinfo  {journal} {Phys.
  Rev. B}\ }\textbf {\bibinfo {volume} {51}},\ \bibinfo {pages} {1743--1751}
  (\bibinfo {year} {1995})}\BibitemShut {NoStop}%
\bibitem [{\citenamefont {von Delft}\ and\ \citenamefont
  {Ralph}(2001)}]{vonDelft2001}%
  \BibitemOpen
  \bibfield  {author} {\bibinfo {author} {\bibfnamefont {Jan}\ \bibnamefont
  {von Delft}}\ and\ \bibinfo {author} {\bibfnamefont {D.C.}\ \bibnamefont
  {Ralph}},\ }\bibfield  {title} {\enquote {\bibinfo {title} {Spectroscopy of
  discrete energy levels in ultrasmall metallic grains},}\ }\href {\doibase
  10.1016/s0370-1573(00)00099-5} {\bibfield  {journal} {\bibinfo  {journal}
  {Physics Reports}\ }\textbf {\bibinfo {volume} {345}},\ \bibinfo {pages}
  {61--173} (\bibinfo {year} {2001})}\BibitemShut {NoStop}%
\bibitem [{\citenamefont {van~der Wiel}\ \emph {et~al.}(2002)\citenamefont
  {van~der Wiel}, \citenamefont {De~Franceschi}, \citenamefont {Elzerman},
  \citenamefont {Fujisawa}, \citenamefont {Tarucha},\ and\ \citenamefont
  {Kouwenhoven}}]{Wiel_2002}%
  \BibitemOpen
  \bibfield  {author} {\bibinfo {author} {\bibfnamefont {W.~G.}\ \bibnamefont
  {van~der Wiel}}, \bibinfo {author} {\bibfnamefont {S.}~\bibnamefont
  {De~Franceschi}}, \bibinfo {author} {\bibfnamefont {J.~M.}\ \bibnamefont
  {Elzerman}}, \bibinfo {author} {\bibfnamefont {T.}~\bibnamefont {Fujisawa}},
  \bibinfo {author} {\bibfnamefont {S.}~\bibnamefont {Tarucha}}, \ and\
  \bibinfo {author} {\bibfnamefont {L.~P.}\ \bibnamefont {Kouwenhoven}},\
  }\bibfield  {title} {\enquote {\bibinfo {title} {Electron transport through
  double quantum dots},}\ }\href {\doibase 10.1103/RevModPhys.75.1} {\bibfield
  {journal} {\bibinfo  {journal} {Rev. Mod. Phys.}\ }\textbf {\bibinfo {volume}
  {75}},\ \bibinfo {pages} {1--22} (\bibinfo {year} {2002})}\BibitemShut
  {NoStop}%
\bibitem [{\citenamefont {Oreg}\ and\ \citenamefont
  {Goldhaber-Gordon}(2003)}]{Yuval_2003}%
  \BibitemOpen
  \bibfield  {author} {\bibinfo {author} {\bibfnamefont {Yuval}\ \bibnamefont
  {Oreg}}\ and\ \bibinfo {author} {\bibfnamefont {David}\ \bibnamefont
  {Goldhaber-Gordon}},\ }\bibfield  {title} {\enquote {\bibinfo {title}
  {Two-channel kondo effect in a modified single electron transistor},}\ }\href
  {\doibase 10.1103/PhysRevLett.90.136602} {\bibfield  {journal} {\bibinfo
  {journal} {Phys. Rev. Lett.}\ }\textbf {\bibinfo {volume} {90}},\ \bibinfo
  {pages} {136602} (\bibinfo {year} {2003})}\BibitemShut {NoStop}%
\bibitem [{\citenamefont {Anders}\ \emph {et~al.}(2004)\citenamefont {Anders},
  \citenamefont {Lebanon},\ and\ \citenamefont {Schiller}}]{Frithjof_2004}%
  \BibitemOpen
  \bibfield  {author} {\bibinfo {author} {\bibfnamefont {Frithjof~B.}\
  \bibnamefont {Anders}}, \bibinfo {author} {\bibfnamefont {Eran}\ \bibnamefont
  {Lebanon}}, \ and\ \bibinfo {author} {\bibfnamefont {Avraham}\ \bibnamefont
  {Schiller}},\ }\bibfield  {title} {\enquote {\bibinfo {title} {Coulomb
  blockade and {non-Fermi-liquid} behavior in quantum dots},}\ }\href {\doibase
  10.1103/PhysRevB.70.201306} {\bibfield  {journal} {\bibinfo  {journal} {Phys.
  Rev. B}\ }\textbf {\bibinfo {volume} {70}},\ \bibinfo {pages} {201306}
  (\bibinfo {year} {2004})}\BibitemShut {NoStop}%
\bibitem [{\citenamefont {Anders}\ \emph {et~al.}(2005)\citenamefont {Anders},
  \citenamefont {Lebanon},\ and\ \citenamefont {Schiller}}]{Frithjof_2005}%
  \BibitemOpen
  \bibfield  {author} {\bibinfo {author} {\bibfnamefont {Frithjof~B.}\
  \bibnamefont {Anders}}, \bibinfo {author} {\bibfnamefont {Eran}\ \bibnamefont
  {Lebanon}}, \ and\ \bibinfo {author} {\bibfnamefont {Avraham}\ \bibnamefont
  {Schiller}},\ }\bibfield  {title} {\enquote {\bibinfo {title} {Coulomb
  blockade and quantum critical points in quantum dots},}\ }\href {\doibase
  10.1016/j.physb.2005.01.427} {\bibfield  {journal} {\bibinfo  {journal}
  {Physica B: Condensed Matter}\ }\textbf {\bibinfo {volume} {359-361}},\
  \bibinfo {pages} {1381--1383} (\bibinfo {year} {2005})}\BibitemShut {NoStop}%
\bibitem [{\citenamefont {Zhang}\ \emph {et~al.}(2019)\citenamefont {Zhang},
  \citenamefont {Liu}, \citenamefont {Wimmer},\ and\ \citenamefont
  {Kouwenhoven}}]{Zhang_2019}%
  \BibitemOpen
  \bibfield  {author} {\bibinfo {author} {\bibfnamefont {Hao}\ \bibnamefont
  {Zhang}}, \bibinfo {author} {\bibfnamefont {Dong~E.}\ \bibnamefont {Liu}},
  \bibinfo {author} {\bibfnamefont {Michael}\ \bibnamefont {Wimmer}}, \ and\
  \bibinfo {author} {\bibfnamefont {Leo~P.}\ \bibnamefont {Kouwenhoven}},\
  }\bibfield  {title} {\enquote {\bibinfo {title} {Next steps of quantum
  transport in majorana nanowire devices},}\ }\href {\doibase
  10.1038/s41467-019-13133-1} {\bibfield  {journal} {\bibinfo  {journal} {Nat
  Commun}\ }\textbf {\bibinfo {volume} {10}} (\bibinfo {year} {2019}),\
  10.1038/s41467-019-13133-1}\BibitemShut {NoStop}%
\bibitem [{\citenamefont {Mitchell}\ \emph {et~al.}(2021)\citenamefont
  {Mitchell}, \citenamefont {Liberman}, \citenamefont {Sela},\ and\
  \citenamefont {Affleck}}]{Mitchell_2021}%
  \BibitemOpen
  \bibfield  {author} {\bibinfo {author} {\bibfnamefont {Andrew~K.}\
  \bibnamefont {Mitchell}}, \bibinfo {author} {\bibfnamefont {Alon}\
  \bibnamefont {Liberman}}, \bibinfo {author} {\bibfnamefont {Eran}\
  \bibnamefont {Sela}}, \ and\ \bibinfo {author} {\bibfnamefont {Ian}\
  \bibnamefont {Affleck}},\ }\bibfield  {title} {\enquote {\bibinfo {title}
  {{SO(5) Non-Fermi Liquid in a Coulomb box device}},}\ }\href {\doibase
  10.1103/PhysRevLett.126.147702} {\bibfield  {journal} {\bibinfo  {journal}
  {Phys. Rev. Lett.}\ }\textbf {\bibinfo {volume} {126}},\ \bibinfo {pages}
  {147702} (\bibinfo {year} {2021})}\BibitemShut {NoStop}%
\bibitem [{\citenamefont {Matveev}(1991)}]{Matveev1991}%
  \BibitemOpen
  \bibfield  {author} {\bibinfo {author} {\bibfnamefont {K.~A.}\ \bibnamefont
  {Matveev}},\ }\bibfield  {title} {\enquote {\bibinfo {title} {Quantum
  fluctuations of the charge of a metal particle under the {Coulomb} blockade
  conditions},}\ }\href@noop {} {\bibfield  {journal} {\bibinfo  {journal} {Zh.
  Eksp. Theor. Fiz.}\ }\textbf {\bibinfo {volume} {99}},\ \bibinfo {pages}
  {1598} (\bibinfo {year} {1991})},\ \bibinfo {note} {{Sov.} Phys. JETP 72, 892
  (1991)}\BibitemShut {NoStop}%
\bibitem [{\citenamefont {Pavešić}\ \emph {et~al.}(2021)\citenamefont
  {Pavešić}, \citenamefont {Bauernfeind},\ and\ \citenamefont
  {Žitko}}]{paper1}%
  \BibitemOpen
  \bibfield  {author} {\bibinfo {author} {\bibfnamefont {Luka}\ \bibnamefont
  {Pavešić}}, \bibinfo {author} {\bibfnamefont {Daniel}\ \bibnamefont
  {Bauernfeind}}, \ and\ \bibinfo {author} {\bibfnamefont {Rok}\ \bibnamefont
  {Žitko}},\ }\bibfield  {title} {\enquote {\bibinfo {title}
  {{Yu-Shiba-Rusinov states in superconducting islands with finite charging
  energy}},}\ }\href@noop {} {\  (\bibinfo {year} {2021})},\ \Eprint
  {http://arxiv.org/abs/2101.10168} {arXiv:2101.10168 [cond-mat.mes-hall]}
  \BibitemShut {NoStop}%
\bibitem [{\citenamefont {Ralph}\ \emph {et~al.}(1995)\citenamefont {Ralph},
  \citenamefont {Black},\ and\ \citenamefont {Tinkham}}]{RBT1995}%
  \BibitemOpen
  \bibfield  {author} {\bibinfo {author} {\bibfnamefont {D.~C.}\ \bibnamefont
  {Ralph}}, \bibinfo {author} {\bibfnamefont {C.~T.}\ \bibnamefont {Black}}, \
  and\ \bibinfo {author} {\bibfnamefont {M.}~\bibnamefont {Tinkham}},\
  }\bibfield  {title} {\enquote {\bibinfo {title} {Spectroscopic measurements
  of discrete electronic states in single metal particles},}\ }\href {\doibase
  10.1103/PhysRevLett.74.3241} {\bibfield  {journal} {\bibinfo  {journal}
  {Phys. Rev. Lett.}\ }\textbf {\bibinfo {volume} {74}},\ \bibinfo {pages}
  {3241--3244} (\bibinfo {year} {1995})}\BibitemShut {NoStop}%
\bibitem [{\citenamefont {Ralph}\ \emph {et~al.}(1997)\citenamefont {Ralph},
  \citenamefont {Black},\ and\ \citenamefont {Tinkham}}]{RBT1997}%
  \BibitemOpen
  \bibfield  {author} {\bibinfo {author} {\bibfnamefont {D.~C.}\ \bibnamefont
  {Ralph}}, \bibinfo {author} {\bibfnamefont {C.~T.}\ \bibnamefont {Black}}, \
  and\ \bibinfo {author} {\bibfnamefont {M.}~\bibnamefont {Tinkham}},\
  }\bibfield  {title} {\enquote {\bibinfo {title} {Gate-voltage studies of
  discrete electronic states in aluminum nanoparticles},}\ }\href {\doibase
  10.1103/PhysRevLett.78.4087} {\bibfield  {journal} {\bibinfo  {journal}
  {Phys. Rev. Lett.}\ }\textbf {\bibinfo {volume} {78}},\ \bibinfo {pages}
  {4087--4090} (\bibinfo {year} {1997})}\BibitemShut {NoStop}%
\bibitem [{\citenamefont {Gobert}\ \emph {et~al.}(2004)\citenamefont {Gobert},
  \citenamefont {Schollw\"{o}ck},\ and\ \citenamefont {von
  Delft}}]{Gobert_2004}%
  \BibitemOpen
  \bibfield  {author} {\bibinfo {author} {\bibfnamefont {D.}~\bibnamefont
  {Gobert}}, \bibinfo {author} {\bibfnamefont {U.}~\bibnamefont
  {Schollw\"{o}ck}}, \ and\ \bibinfo {author} {\bibfnamefont {J.}~\bibnamefont
  {von Delft}},\ }\bibfield  {title} {\enquote {\bibinfo {title} {Josephson
  effect between superconducting nanograins with discrete energy levels},}\
  }\href {\doibase 10.1140/epjb/e2004-00145-6} {\bibfield  {journal} {\bibinfo
  {journal} {Eur. Phys. J. B}\ }\textbf {\bibinfo {volume} {38}},\ \bibinfo
  {pages} {501--513} (\bibinfo {year} {2004})}\BibitemShut {NoStop}%
\bibitem [{\citenamefont {Richardson}(1963)}]{Richardson1963}%
  \BibitemOpen
  \bibfield  {author} {\bibinfo {author} {\bibfnamefont {R.W.}\ \bibnamefont
  {Richardson}},\ }\bibfield  {title} {\enquote {\bibinfo {title} {A restricted
  class of exact eigenstates of the pairing-force {Hamiltonian}},}\ }\href
  {\doibase 10.1016/0031-9163(63)90259-2} {\bibfield  {journal} {\bibinfo
  {journal} {Physics Letters}\ }\textbf {\bibinfo {volume} {3}},\ \bibinfo
  {pages} {277--279} (\bibinfo {year} {1963})}\BibitemShut {NoStop}%
\bibitem [{\citenamefont {Richardson}\ and\ \citenamefont
  {Sherman}(1964)}]{Richardson1964}%
  \BibitemOpen
  \bibfield  {author} {\bibinfo {author} {\bibfnamefont {R.W.}\ \bibnamefont
  {Richardson}}\ and\ \bibinfo {author} {\bibfnamefont {N.}~\bibnamefont
  {Sherman}},\ }\bibfield  {title} {\enquote {\bibinfo {title} {Exact
  eigenstates of the pairing-force {Hamiltonian}},}\ }\href {\doibase
  10.1016/0029-5582(64)90687-x} {\bibfield  {journal} {\bibinfo  {journal}
  {Nuclear Physics}\ }\textbf {\bibinfo {volume} {52}},\ \bibinfo {pages}
  {221--238} (\bibinfo {year} {1964})}\BibitemShut {NoStop}%
\bibitem [{\citenamefont {von Delft}\ and\ \citenamefont
  {Braun}(1999)}]{vonDelft1999}%
  \BibitemOpen
  \bibfield  {author} {\bibinfo {author} {\bibfnamefont {Jan}\ \bibnamefont
  {von Delft}}\ and\ \bibinfo {author} {\bibfnamefont {Fabian}\ \bibnamefont
  {Braun}},\ }\bibfield  {title} {\enquote {\bibinfo {title} {Superconductivity
  in ultrasmall grains: Introduction to {Richardson}'s exact solution},}\
  }\href@noop {} {\  (\bibinfo {year} {1999})},\ \Eprint
  {http://arxiv.org/abs/9911058v1} {arXiv:9911058v1 [cond-mat]} \BibitemShut
  {NoStop}%
\bibitem [{\citenamefont {Schollw\"ock}(2005)}]{DMRG}%
  \BibitemOpen
  \bibfield  {author} {\bibinfo {author} {\bibfnamefont {U.}~\bibnamefont
  {Schollw\"ock}},\ }\bibfield  {title} {\enquote {\bibinfo {title} {The
  density-matrix renormalization group},}\ }\href {\doibase
  10.1103/RevModPhys.77.259} {\bibfield  {journal} {\bibinfo  {journal} {Rev.
  Mod. Phys.}\ }\textbf {\bibinfo {volume} {77}},\ \bibinfo {pages} {259--315}
  (\bibinfo {year} {2005})}\BibitemShut {NoStop}%
\bibitem [{\citenamefont {White}(1992)}]{White_1992}%
  \BibitemOpen
  \bibfield  {author} {\bibinfo {author} {\bibfnamefont {Steven~R.}\
  \bibnamefont {White}},\ }\bibfield  {title} {\enquote {\bibinfo {title}
  {Density matrix formulation for quantum renormalization groups},}\ }\href
  {\doibase 10.1103/PhysRevLett.69.2863} {\bibfield  {journal} {\bibinfo
  {journal} {Phys. Rev. Lett.}\ }\textbf {\bibinfo {volume} {69}},\ \bibinfo
  {pages} {2863--2866} (\bibinfo {year} {1992})}\BibitemShut {NoStop}%
\bibitem [{\citenamefont {White}(1993)}]{White_1993}%
  \BibitemOpen
  \bibfield  {author} {\bibinfo {author} {\bibfnamefont {Steven~R.}\
  \bibnamefont {White}},\ }\bibfield  {title} {\enquote {\bibinfo {title}
  {Density-matrix algorithms for quantum renormalization groups},}\ }\href
  {\doibase 10.1103/PhysRevB.48.10345} {\bibfield  {journal} {\bibinfo
  {journal} {Phys. Rev. B}\ }\textbf {\bibinfo {volume} {48}},\ \bibinfo
  {pages} {10345--10356} (\bibinfo {year} {1993})}\BibitemShut {NoStop}%
\bibitem [{\citenamefont {Saldaña}\ \emph {et~al.}(2021)\citenamefont
  {Saldaña}, \citenamefont {Vekris}, \citenamefont {Pavešič}, \citenamefont
  {Krogstrup}, \citenamefont {Žitko}, \citenamefont {Grove-Rasmussen},\ and\
  \citenamefont {Nygård}}]{JuanCarlos2021}%
  \BibitemOpen
  \bibfield  {author} {\bibinfo {author} {\bibfnamefont {Juan Carlos~Estrada}\
  \bibnamefont {Saldaña}}, \bibinfo {author} {\bibfnamefont {Alexandros}\
  \bibnamefont {Vekris}}, \bibinfo {author} {\bibfnamefont {Luka}\ \bibnamefont
  {Pavešič}}, \bibinfo {author} {\bibfnamefont {Peter}\ \bibnamefont
  {Krogstrup}}, \bibinfo {author} {\bibfnamefont {Rok}\ \bibnamefont {Žitko}},
  \bibinfo {author} {\bibfnamefont {Kasper}\ \bibnamefont {Grove-Rasmussen}}, \
  and\ \bibinfo {author} {\bibfnamefont {Jesper}\ \bibnamefont {Nygård}},\
  }\href@noop {} {\enquote {\bibinfo {title} {Coulombic subgap states},}\ }
  (\bibinfo {year} {2021}),\ \Eprint {http://arxiv.org/abs/2101.10794}
  {arXiv:2101.10794 [cond-mat.mes-hall]} \BibitemShut {NoStop}%
\bibitem [{\citenamefont {Franceschi}\ \emph {et~al.}(2010)\citenamefont
  {Franceschi}, \citenamefont {Kouwenhoven}, \citenamefont
  {Sch\"{o}nenberger},\ and\ \citenamefont {Wernsdorfer}}]{DeFranceschi2010}%
  \BibitemOpen
  \bibfield  {author} {\bibinfo {author} {\bibfnamefont {Silvano~De}\
  \bibnamefont {Franceschi}}, \bibinfo {author} {\bibfnamefont {Leo}\
  \bibnamefont {Kouwenhoven}}, \bibinfo {author} {\bibfnamefont {Christian}\
  \bibnamefont {Sch\"{o}nenberger}}, \ and\ \bibinfo {author} {\bibfnamefont
  {Wolfgang}\ \bibnamefont {Wernsdorfer}},\ }\bibfield  {title} {\enquote
  {\bibinfo {title} {Hybrid superconductor{\textendash}quantum dot devices},}\
  }\href {\doibase 10.1038/nnano.2010.173} {\bibfield  {journal} {\bibinfo
  {journal} {Nature Nanotechnology}\ }\textbf {\bibinfo {volume} {5}},\
  \bibinfo {pages} {703--711} (\bibinfo {year} {2010})}\BibitemShut {NoStop}%
\bibitem [{\citenamefont {Aguado}(2020)}]{Aguado_2020}%
  \BibitemOpen
  \bibfield  {author} {\bibinfo {author} {\bibfnamefont {Ram{\'{o}}n}\
  \bibnamefont {Aguado}},\ }\bibfield  {title} {\enquote {\bibinfo {title} {A
  perspective on semiconductor-based superconducting qubits},}\ }\href
  {\doibase 10.1063/5.0024124} {\bibfield  {journal} {\bibinfo  {journal}
  {Appl. Phys. Lett.}\ }\textbf {\bibinfo {volume} {117}},\ \bibinfo {pages}
  {240501} (\bibinfo {year} {2020})}\BibitemShut {NoStop}%
\bibitem [{\citenamefont {Pillet}\ \emph {et~al.}(2010)\citenamefont {Pillet},
  \citenamefont {Quay}, \citenamefont {Morfin}, \citenamefont {Bena},
  \citenamefont {Yeyati},\ and\ \citenamefont {Joyez}}]{Pillet_2010}%
  \BibitemOpen
  \bibfield  {author} {\bibinfo {author} {\bibfnamefont {J-D.}\ \bibnamefont
  {Pillet}}, \bibinfo {author} {\bibfnamefont {C.~H.~L.}\ \bibnamefont {Quay}},
  \bibinfo {author} {\bibfnamefont {P.}~\bibnamefont {Morfin}}, \bibinfo
  {author} {\bibfnamefont {C.}~\bibnamefont {Bena}}, \bibinfo {author}
  {\bibfnamefont {A.~Levy}\ \bibnamefont {Yeyati}}, \ and\ \bibinfo {author}
  {\bibfnamefont {P.}~\bibnamefont {Joyez}},\ }\bibfield  {title} {\enquote
  {\bibinfo {title} {Andreev bound states in supercurrent-carrying carbon
  nanotubes revealed},}\ }\href {\doibase 10.1038/nphys1811} {\bibfield
  {journal} {\bibinfo  {journal} {Nature Physics}\ }\textbf {\bibinfo {volume}
  {6}},\ \bibinfo {pages} {965--969} (\bibinfo {year} {2010})}\BibitemShut
  {NoStop}%
\bibitem [{\citenamefont {Chang}\ \emph {et~al.}(2013)\citenamefont {Chang},
  \citenamefont {Manucharyan}, \citenamefont {Jespersen}, \citenamefont
  {Nyg\aa{}rd},\ and\ \citenamefont {Marcus}}]{Chang_2013}%
  \BibitemOpen
  \bibfield  {author} {\bibinfo {author} {\bibfnamefont {W.}~\bibnamefont
  {Chang}}, \bibinfo {author} {\bibfnamefont {V.~E.}\ \bibnamefont
  {Manucharyan}}, \bibinfo {author} {\bibfnamefont {T.~S.}\ \bibnamefont
  {Jespersen}}, \bibinfo {author} {\bibfnamefont {J.}~\bibnamefont
  {Nyg\aa{}rd}}, \ and\ \bibinfo {author} {\bibfnamefont {C.~M.}\ \bibnamefont
  {Marcus}},\ }\bibfield  {title} {\enquote {\bibinfo {title} {Tunneling
  spectroscopy of quasiparticle bound states in a spinful {Josephson}
  junction},}\ }\href {\doibase 10.1103/PhysRevLett.110.217005} {\bibfield
  {journal} {\bibinfo  {journal} {Phys. Rev. Lett.}\ }\textbf {\bibinfo
  {volume} {110}},\ \bibinfo {pages} {217005} (\bibinfo {year}
  {2013})}\BibitemShut {NoStop}%
\bibitem [{\citenamefont {Casparis}\ \emph {et~al.}(2016)\citenamefont
  {Casparis}, \citenamefont {Larsen}, \citenamefont {Olsen}, \citenamefont
  {Kuemmeth}, \citenamefont {Krogstrup}, \citenamefont {Nyg\aa{}rd},
  \citenamefont {Petersson},\ and\ \citenamefont {Marcus}}]{Casparis_2016}%
  \BibitemOpen
  \bibfield  {author} {\bibinfo {author} {\bibfnamefont {L.}~\bibnamefont
  {Casparis}}, \bibinfo {author} {\bibfnamefont {T.~W.}\ \bibnamefont
  {Larsen}}, \bibinfo {author} {\bibfnamefont {M.~S.}\ \bibnamefont {Olsen}},
  \bibinfo {author} {\bibfnamefont {F.}~\bibnamefont {Kuemmeth}}, \bibinfo
  {author} {\bibfnamefont {P.}~\bibnamefont {Krogstrup}}, \bibinfo {author}
  {\bibfnamefont {J.}~\bibnamefont {Nyg\aa{}rd}}, \bibinfo {author}
  {\bibfnamefont {K.~D.}\ \bibnamefont {Petersson}}, \ and\ \bibinfo {author}
  {\bibfnamefont {C.~M.}\ \bibnamefont {Marcus}},\ }\bibfield  {title}
  {\enquote {\bibinfo {title} {Gatemon benchmarking and two-qubit
  operations},}\ }\href {\doibase 10.1103/PhysRevLett.116.150505} {\bibfield
  {journal} {\bibinfo  {journal} {Phys. Rev. Lett.}\ }\textbf {\bibinfo
  {volume} {116}},\ \bibinfo {pages} {150505} (\bibinfo {year}
  {2016})}\BibitemShut {NoStop}%
\bibitem [{\citenamefont {Casparis}\ \emph {et~al.}(2018)\citenamefont
  {Casparis}, \citenamefont {Connolly}, \citenamefont {Kjaergaard},
  \citenamefont {Pearson}, \citenamefont {Kringh{\o}j}, \citenamefont {Larsen},
  \citenamefont {Kuemmeth}, \citenamefont {Wang}, \citenamefont {Thomas},
  \citenamefont {Gronin}, \citenamefont {Gardner}, \citenamefont {Manfra},
  \citenamefont {Marcus},\ and\ \citenamefont {Petersson}}]{Casparis_2018}%
  \BibitemOpen
  \bibfield  {author} {\bibinfo {author} {\bibfnamefont {Lucas}\ \bibnamefont
  {Casparis}}, \bibinfo {author} {\bibfnamefont {Malcolm~R.}\ \bibnamefont
  {Connolly}}, \bibinfo {author} {\bibfnamefont {Morten}\ \bibnamefont
  {Kjaergaard}}, \bibinfo {author} {\bibfnamefont {Natalie~J.}\ \bibnamefont
  {Pearson}}, \bibinfo {author} {\bibfnamefont {Anders}\ \bibnamefont
  {Kringh{\o}j}}, \bibinfo {author} {\bibfnamefont {Thorvald~W.}\ \bibnamefont
  {Larsen}}, \bibinfo {author} {\bibfnamefont {Ferdinand}\ \bibnamefont
  {Kuemmeth}}, \bibinfo {author} {\bibfnamefont {Tiantian}\ \bibnamefont
  {Wang}}, \bibinfo {author} {\bibfnamefont {Candice}\ \bibnamefont {Thomas}},
  \bibinfo {author} {\bibfnamefont {Sergei}\ \bibnamefont {Gronin}}, \bibinfo
  {author} {\bibfnamefont {Geoffrey~C.}\ \bibnamefont {Gardner}}, \bibinfo
  {author} {\bibfnamefont {Michael~J.}\ \bibnamefont {Manfra}}, \bibinfo
  {author} {\bibfnamefont {Charles~M.}\ \bibnamefont {Marcus}}, \ and\ \bibinfo
  {author} {\bibfnamefont {Karl~D.}\ \bibnamefont {Petersson}},\ }\bibfield
  {title} {\enquote {\bibinfo {title} {Superconducting gatemon qubit based on a
  proximitized two-dimensional electron gas},}\ }\href {\doibase
  10.1038/s41565-018-0207-y} {\bibfield  {journal} {\bibinfo  {journal} {Nature
  Nanotech}\ }\textbf {\bibinfo {volume} {13}},\ \bibinfo {pages} {915--919}
  (\bibinfo {year} {2018})}\BibitemShut {NoStop}%
\bibitem [{\citenamefont {Kurilovich}\ \emph {et~al.}(2021)\citenamefont
  {Kurilovich}, \citenamefont {Kurilovich}, \citenamefont {Fatemi},
  \citenamefont {Devoret},\ and\ \citenamefont {Glazman}}]{Kurilovich_2021}%
  \BibitemOpen
  \bibfield  {author} {\bibinfo {author} {\bibfnamefont {Pavel~D.}\
  \bibnamefont {Kurilovich}}, \bibinfo {author} {\bibfnamefont {Vladislav~D.}\
  \bibnamefont {Kurilovich}}, \bibinfo {author} {\bibfnamefont {Valla}\
  \bibnamefont {Fatemi}}, \bibinfo {author} {\bibfnamefont {Michel~H.}\
  \bibnamefont {Devoret}}, \ and\ \bibinfo {author} {\bibfnamefont {Leonid~I.}\
  \bibnamefont {Glazman}},\ }\bibfield  {title} {\enquote {\bibinfo {title}
  {Microwave response of an {Andreev} bound state},}\ }\href@noop {} {\
  (\bibinfo {year} {2021})},\ \Eprint {http://arxiv.org/abs/2106.00028}
  {arXiv:2106.00028 [cond-mat.mes-hall]} \BibitemShut {NoStop}%
\bibitem [{\citenamefont {Affleck}(2005)}]{Affleck_2005}%
  \BibitemOpen
  \bibfield  {author} {\bibinfo {author} {\bibfnamefont {Ian}\ \bibnamefont
  {Affleck}},\ }\bibfield  {title} {\enquote {\bibinfo {title} {Non-fermi
  liquid behavior in kondo models},}\ }\href {\doibase 10.1143/jpsj.74.59}
  {\bibfield  {journal} {\bibinfo  {journal} {J. Phys. Soc. Jpn.}\ }\textbf
  {\bibinfo {volume} {74}},\ \bibinfo {pages} {59--66} (\bibinfo {year}
  {2005})}\BibitemShut {NoStop}%
\bibitem [{\citenamefont {Potok}\ \emph {et~al.}(2007)\citenamefont {Potok},
  \citenamefont {Rau}, \citenamefont {Shtrikman}, \citenamefont {Oreg},\ and\
  \citenamefont {Goldhaber-Gordon}}]{Potok2007}%
  \BibitemOpen
  \bibfield  {author} {\bibinfo {author} {\bibfnamefont {R.~M.}\ \bibnamefont
  {Potok}}, \bibinfo {author} {\bibfnamefont {I.~G.}\ \bibnamefont {Rau}},
  \bibinfo {author} {\bibfnamefont {Hadas}\ \bibnamefont {Shtrikman}}, \bibinfo
  {author} {\bibfnamefont {Yuval}\ \bibnamefont {Oreg}}, \ and\ \bibinfo
  {author} {\bibfnamefont {D.}~\bibnamefont {Goldhaber-Gordon}},\ }\bibfield
  {title} {\enquote {\bibinfo {title} {Observation of the two-channel {Kondo}
  effect},}\ }\href {\doibase 10.1038/nature05556} {\bibfield  {journal}
  {\bibinfo  {journal} {Nature}\ }\textbf {\bibinfo {volume} {446}},\ \bibinfo
  {pages} {167--171} (\bibinfo {year} {2007})}\BibitemShut {NoStop}%
\bibitem [{\citenamefont {Iftikhar}\ \emph {et~al.}(2015)\citenamefont
  {Iftikhar}, \citenamefont {Jezouin}, \citenamefont {Anthore}, \citenamefont
  {Gennser}, \citenamefont {Parmentier}, \citenamefont {Cavanna},\ and\
  \citenamefont {Pierre}}]{Iftikhar_2015}%
  \BibitemOpen
  \bibfield  {author} {\bibinfo {author} {\bibfnamefont {Z.}~\bibnamefont
  {Iftikhar}}, \bibinfo {author} {\bibfnamefont {S.}~\bibnamefont {Jezouin}},
  \bibinfo {author} {\bibfnamefont {A.}~\bibnamefont {Anthore}}, \bibinfo
  {author} {\bibfnamefont {U.}~\bibnamefont {Gennser}}, \bibinfo {author}
  {\bibfnamefont {F.~D.}\ \bibnamefont {Parmentier}}, \bibinfo {author}
  {\bibfnamefont {A.}~\bibnamefont {Cavanna}}, \ and\ \bibinfo {author}
  {\bibfnamefont {F.}~\bibnamefont {Pierre}},\ }\bibfield  {title} {\enquote
  {\bibinfo {title} {Two-channel kondo effect and renormalization flow with
  macroscopic quantum charge states},}\ }\href {\doibase 10.1038/nature15384}
  {\bibfield  {journal} {\bibinfo  {journal} {Nature}\ }\textbf {\bibinfo
  {volume} {526}},\ \bibinfo {pages} {233--236} (\bibinfo {year}
  {2015})}\BibitemShut {NoStop}%
\bibitem [{\citenamefont {Kirchner}(2020)}]{Kirchner_2020}%
  \BibitemOpen
  \bibfield  {author} {\bibinfo {author} {\bibfnamefont {Stefan}\ \bibnamefont
  {Kirchner}},\ }\bibfield  {title} {\enquote {\bibinfo {title} {Two-channel
  kondo physics: From engineered structures to quantum materials
  realizations},}\ }\href {\doibase 10.1002/qute.201900128} {\bibfield
  {journal} {\bibinfo  {journal} {Advanced Quantum Technologies}\ }\textbf
  {\bibinfo {volume} {3}},\ \bibinfo {pages} {1900128} (\bibinfo {year}
  {2020})}\BibitemShut {NoStop}%
\bibitem [{\citenamefont {\ifmmode~\check{Z}\else \v{Z}\fi{}itko}\ and\
  \citenamefont {Fabrizio}(2017)}]{zitko_2ch}%
  \BibitemOpen
  \bibfield  {author} {\bibinfo {author} {\bibfnamefont {Rok}\ \bibnamefont
  {\ifmmode~\check{Z}\else \v{Z}\fi{}itko}}\ and\ \bibinfo {author}
  {\bibfnamefont {Michele}\ \bibnamefont {Fabrizio}},\ }\bibfield  {title}
  {\enquote {\bibinfo {title} {Non-{Fermi}-liquid behavior in quantum impurity
  models with superconducting channels},}\ }\href {\doibase
  10.1103/PhysRevB.95.085121} {\bibfield  {journal} {\bibinfo  {journal} {Phys.
  Rev. B}\ }\textbf {\bibinfo {volume} {95}},\ \bibinfo {pages} {085121}
  (\bibinfo {year} {2017})}\BibitemShut {NoStop}%
\bibitem [{\citenamefont {Braun}\ and\ \citenamefont {von
  Delft}(1998)}]{Braun_vonDelft_1}%
  \BibitemOpen
  \bibfield  {author} {\bibinfo {author} {\bibfnamefont {Fabian}\ \bibnamefont
  {Braun}}\ and\ \bibinfo {author} {\bibfnamefont {Jan}\ \bibnamefont {von
  Delft}},\ }\bibfield  {title} {\enquote {\bibinfo {title} {Fixed-{N}
  superconductivity: The crossover from the bulk to the few-electron limit},}\
  }\href {\doibase 10.1103/PhysRevLett.81.4712} {\bibfield  {journal} {\bibinfo
   {journal} {Phys. Rev. Lett.}\ }\textbf {\bibinfo {volume} {81}},\ \bibinfo
  {pages} {4712--4715} (\bibinfo {year} {1998})}\BibitemShut {NoStop}%
\bibitem [{\citenamefont {Braun}\ and\ \citenamefont {von
  Delft}(1999)}]{Braun_vonDelft_2}%
  \BibitemOpen
  \bibfield  {author} {\bibinfo {author} {\bibfnamefont {Fabian}\ \bibnamefont
  {Braun}}\ and\ \bibinfo {author} {\bibfnamefont {Jan}\ \bibnamefont {von
  Delft}},\ }\bibfield  {title} {\enquote {\bibinfo {title} {Superconductivity
  in ultrasmall metallic grains},}\ }\href {\doibase 10.1103/PhysRevB.59.9527}
  {\bibfield  {journal} {\bibinfo  {journal} {Phys. Rev. B}\ }\textbf {\bibinfo
  {volume} {59}},\ \bibinfo {pages} {9527--9544} (\bibinfo {year}
  {1999})}\BibitemShut {NoStop}%
\bibitem [{\citenamefont {Averin}\ and\ \citenamefont
  {Nazarov}(1992)}]{Averin_1992}%
  \BibitemOpen
  \bibfield  {author} {\bibinfo {author} {\bibfnamefont {D.~V.}\ \bibnamefont
  {Averin}}\ and\ \bibinfo {author} {\bibfnamefont {Yu.~V.}\ \bibnamefont
  {Nazarov}},\ }\bibfield  {title} {\enquote {\bibinfo {title} {Single-electron
  charging of a superconducting island},}\ }\href {\doibase
  10.1103/PhysRevLett.69.1993} {\bibfield  {journal} {\bibinfo  {journal}
  {Phys. Rev. Lett.}\ }\textbf {\bibinfo {volume} {69}},\ \bibinfo {pages}
  {1993--1996} (\bibinfo {year} {1992})}\BibitemShut {NoStop}%
\bibitem [{\citenamefont {Lafarge}\ \emph {et~al.}(1993)\citenamefont
  {Lafarge}, \citenamefont {Joyez}, \citenamefont {Esteve}, \citenamefont
  {Urbina},\ and\ \citenamefont {Devoret}}]{Lafarge_1993}%
  \BibitemOpen
  \bibfield  {author} {\bibinfo {author} {\bibfnamefont {P.}~\bibnamefont
  {Lafarge}}, \bibinfo {author} {\bibfnamefont {P.}~\bibnamefont {Joyez}},
  \bibinfo {author} {\bibfnamefont {D.}~\bibnamefont {Esteve}}, \bibinfo
  {author} {\bibfnamefont {C.}~\bibnamefont {Urbina}}, \ and\ \bibinfo {author}
  {\bibfnamefont {M.~H.}\ \bibnamefont {Devoret}},\ }\bibfield  {title}
  {\enquote {\bibinfo {title} {Measurement of the even-odd free-energy
  difference of an isolated superconductor},}\ }\href {\doibase
  10.1103/PhysRevLett.70.994} {\bibfield  {journal} {\bibinfo  {journal} {Phys.
  Rev. Lett.}\ }\textbf {\bibinfo {volume} {70}},\ \bibinfo {pages} {994--997}
  (\bibinfo {year} {1993})}\BibitemShut {NoStop}%
\bibitem [{\citenamefont {von Delft}\ \emph {et~al.}(1996)\citenamefont {von
  Delft}, \citenamefont {Zaikin}, \citenamefont {Golubev},\ and\ \citenamefont
  {Tichy}}]{vonDelft_1996}%
  \BibitemOpen
  \bibfield  {author} {\bibinfo {author} {\bibfnamefont {Jan}\ \bibnamefont
  {von Delft}}, \bibinfo {author} {\bibfnamefont {Andrei~D.}\ \bibnamefont
  {Zaikin}}, \bibinfo {author} {\bibfnamefont {Dmitrii~S.}\ \bibnamefont
  {Golubev}}, \ and\ \bibinfo {author} {\bibfnamefont {Wolfgang}\ \bibnamefont
  {Tichy}},\ }\bibfield  {title} {\enquote {\bibinfo {title} {Parity-affected
  superconductivity in ultrasmall metallic grains},}\ }\href {\doibase
  10.1103/PhysRevLett.77.3189} {\bibfield  {journal} {\bibinfo  {journal}
  {Phys. Rev. Lett.}\ }\textbf {\bibinfo {volume} {77}},\ \bibinfo {pages}
  {3189--3192} (\bibinfo {year} {1996})}\BibitemShut {NoStop}%
\bibitem [{\citenamefont {Matveev}\ and\ \citenamefont
  {Larkin}(1997)}]{Matveev_1997}%
  \BibitemOpen
  \bibfield  {author} {\bibinfo {author} {\bibfnamefont {K.~A.}\ \bibnamefont
  {Matveev}}\ and\ \bibinfo {author} {\bibfnamefont {A.~I.}\ \bibnamefont
  {Larkin}},\ }\bibfield  {title} {\enquote {\bibinfo {title} {Parity effect in
  ground state energies of ultrasmall superconducting grains},}\ }\href
  {\doibase 10.1103/PhysRevLett.78.3749} {\bibfield  {journal} {\bibinfo
  {journal} {Phys. Rev. Lett.}\ }\textbf {\bibinfo {volume} {78}},\ \bibinfo
  {pages} {3749--3752} (\bibinfo {year} {1997})}\BibitemShut {NoStop}%
\bibitem [{\citenamefont {Mastellone}\ \emph {et~al.}(1998)\citenamefont
  {Mastellone}, \citenamefont {Falci},\ and\ \citenamefont
  {Fazio}}]{Mastellone_1998}%
  \BibitemOpen
  \bibfield  {author} {\bibinfo {author} {\bibfnamefont {A.}~\bibnamefont
  {Mastellone}}, \bibinfo {author} {\bibfnamefont {G.}~\bibnamefont {Falci}}, \
  and\ \bibinfo {author} {\bibfnamefont {Rosario}\ \bibnamefont {Fazio}},\
  }\bibfield  {title} {\enquote {\bibinfo {title} {Small superconducting grain
  in the canonical ensemble},}\ }\href {\doibase 10.1103/PhysRevLett.80.4542}
  {\bibfield  {journal} {\bibinfo  {journal} {Phys. Rev. Lett.}\ }\textbf
  {\bibinfo {volume} {80}},\ \bibinfo {pages} {4542--4545} (\bibinfo {year}
  {1998})}\BibitemShut {NoStop}%
\bibitem [{\citenamefont {Tuominen}\ \emph {et~al.}(1992)\citenamefont
  {Tuominen}, \citenamefont {Hergenrother}, \citenamefont {Tighe},\ and\
  \citenamefont {Tinkham}}]{Tuominen_1997}%
  \BibitemOpen
  \bibfield  {author} {\bibinfo {author} {\bibfnamefont {M.~T.}\ \bibnamefont
  {Tuominen}}, \bibinfo {author} {\bibfnamefont {J.~M.}\ \bibnamefont
  {Hergenrother}}, \bibinfo {author} {\bibfnamefont {T.~S.}\ \bibnamefont
  {Tighe}}, \ and\ \bibinfo {author} {\bibfnamefont {M.}~\bibnamefont
  {Tinkham}},\ }\bibfield  {title} {\enquote {\bibinfo {title} {Experimental
  evidence for parity-based 2e periodicity in a superconducting single-electron
  tunneling transistor},}\ }\href {\doibase 10.1103/PhysRevLett.69.1997}
  {\bibfield  {journal} {\bibinfo  {journal} {Phys. Rev. Lett.}\ }\textbf
  {\bibinfo {volume} {69}},\ \bibinfo {pages} {1997--2000} (\bibinfo {year}
  {1992})}\BibitemShut {NoStop}%
\bibitem [{\citenamefont {Tinkham}(2004)}]{tinkham_book}%
  \BibitemOpen
  \bibfield  {author} {\bibinfo {author} {\bibfnamefont {Michael}\ \bibnamefont
  {Tinkham}},\ }\href {http://www.worldcat.org/isbn/0486435032} {\emph
  {\bibinfo {title} {Introduction to Superconductivity}}},\ \bibinfo {edition}
  {2nd}\ ed.\ (\bibinfo  {publisher} {Dover Publications},\ \bibinfo {year}
  {2004})\BibitemShut {NoStop}%
\bibitem [{\citenamefont {Kir\ifmmode~\check{s}\else \v{s}\fi{}anskas}\ \emph
  {et~al.}(2015)\citenamefont {Kir\ifmmode~\check{s}\else \v{s}\fi{}anskas},
  \citenamefont {Goldstein}, \citenamefont {Flensberg}, \citenamefont
  {Glazman},\ and\ \citenamefont {Paaske}}]{Kirsanskas_2015}%
  \BibitemOpen
  \bibfield  {author} {\bibinfo {author} {\bibfnamefont {Gediminas}\
  \bibnamefont {Kir\ifmmode~\check{s}\else \v{s}\fi{}anskas}}, \bibinfo
  {author} {\bibfnamefont {Moshe}\ \bibnamefont {Goldstein}}, \bibinfo {author}
  {\bibfnamefont {Karsten}\ \bibnamefont {Flensberg}}, \bibinfo {author}
  {\bibfnamefont {Leonid~I.}\ \bibnamefont {Glazman}}, \ and\ \bibinfo {author}
  {\bibfnamefont {Jens}\ \bibnamefont {Paaske}},\ }\bibfield  {title} {\enquote
  {\bibinfo {title} {{Yu-Shiba-Rusinov} states in phase-biased
  superconductor--quantum dot--superconductor junctions},}\ }\href {\doibase
  10.1103/PhysRevB.92.235422} {\bibfield  {journal} {\bibinfo  {journal} {Phys.
  Rev. B}\ }\textbf {\bibinfo {volume} {92}},\ \bibinfo {pages} {235422}
  (\bibinfo {year} {2015})}\BibitemShut {NoStop}%
\bibitem [{\citenamefont {de~Lange}\ \emph {et~al.}(2015)\citenamefont
  {de~Lange}, \citenamefont {van Heck}, \citenamefont {Bruno}, \citenamefont
  {van Woerkom}, \citenamefont {Geresdi}, \citenamefont {Plissard},
  \citenamefont {Bakkers}, \citenamefont {Akhmerov},\ and\ \citenamefont
  {DiCarlo}}]{Lange_2015}%
  \BibitemOpen
  \bibfield  {author} {\bibinfo {author} {\bibfnamefont {G.}~\bibnamefont
  {de~Lange}}, \bibinfo {author} {\bibfnamefont {B.}~\bibnamefont {van Heck}},
  \bibinfo {author} {\bibfnamefont {A.}~\bibnamefont {Bruno}}, \bibinfo
  {author} {\bibfnamefont {D.~J.}\ \bibnamefont {van Woerkom}}, \bibinfo
  {author} {\bibfnamefont {A.}~\bibnamefont {Geresdi}}, \bibinfo {author}
  {\bibfnamefont {S.~R.}\ \bibnamefont {Plissard}}, \bibinfo {author}
  {\bibfnamefont {E.~P. A.~M.}\ \bibnamefont {Bakkers}}, \bibinfo {author}
  {\bibfnamefont {A.~R.}\ \bibnamefont {Akhmerov}}, \ and\ \bibinfo {author}
  {\bibfnamefont {L.}~\bibnamefont {DiCarlo}},\ }\bibfield  {title} {\enquote
  {\bibinfo {title} {Realization of microwave quantum circuits using hybrid
  superconducting-semiconducting nanowire {Josephson} elements},}\ }\href
  {\doibase 10.1103/PhysRevLett.115.127002} {\bibfield  {journal} {\bibinfo
  {journal} {Phys. Rev. Lett.}\ }\textbf {\bibinfo {volume} {115}},\ \bibinfo
  {pages} {127002} (\bibinfo {year} {2015})}\BibitemShut {NoStop}%
\bibitem [{\citenamefont {Larsen}\ \emph {et~al.}(2015)\citenamefont {Larsen},
  \citenamefont {Petersson}, \citenamefont {Kuemmeth}, \citenamefont
  {Jespersen}, \citenamefont {Krogstrup}, \citenamefont {Nyg\aa{}rd},\ and\
  \citenamefont {Marcus}}]{Larsen_2015}%
  \BibitemOpen
  \bibfield  {author} {\bibinfo {author} {\bibfnamefont {T.~W.}\ \bibnamefont
  {Larsen}}, \bibinfo {author} {\bibfnamefont {K.~D.}\ \bibnamefont
  {Petersson}}, \bibinfo {author} {\bibfnamefont {F.}~\bibnamefont {Kuemmeth}},
  \bibinfo {author} {\bibfnamefont {T.~S.}\ \bibnamefont {Jespersen}}, \bibinfo
  {author} {\bibfnamefont {P.}~\bibnamefont {Krogstrup}}, \bibinfo {author}
  {\bibfnamefont {J.}~\bibnamefont {Nyg\aa{}rd}}, \ and\ \bibinfo {author}
  {\bibfnamefont {C.~M.}\ \bibnamefont {Marcus}},\ }\bibfield  {title}
  {\enquote {\bibinfo {title} {Semiconductor-nanowire-based superconducting
  qubit},}\ }\href {\doibase 10.1103/PhysRevLett.115.127001} {\bibfield
  {journal} {\bibinfo  {journal} {Phys. Rev. Lett.}\ }\textbf {\bibinfo
  {volume} {115}},\ \bibinfo {pages} {127001} (\bibinfo {year}
  {2015})}\BibitemShut {NoStop}%
\bibitem [{\citenamefont {van Woerkom}\ \emph {et~al.}(2017)\citenamefont {van
  Woerkom}, \citenamefont {Proutski}, \citenamefont {van Heck}, \citenamefont
  {Bouman}, \citenamefont {V\"{a}yrynen}, \citenamefont {Glazman},
  \citenamefont {Krogstrup}, \citenamefont {Nyg{\aa}rd}, \citenamefont
  {Kouwenhoven},\ and\ \citenamefont {Geresdi}}]{Woerkom_2017}%
  \BibitemOpen
  \bibfield  {author} {\bibinfo {author} {\bibfnamefont {David~J.}\
  \bibnamefont {van Woerkom}}, \bibinfo {author} {\bibfnamefont {Alex}\
  \bibnamefont {Proutski}}, \bibinfo {author} {\bibfnamefont {Bernard}\
  \bibnamefont {van Heck}}, \bibinfo {author} {\bibfnamefont {Daniël}\
  \bibnamefont {Bouman}}, \bibinfo {author} {\bibfnamefont {Jukka~I.}\
  \bibnamefont {V\"{a}yrynen}}, \bibinfo {author} {\bibfnamefont {Leonid~I.}\
  \bibnamefont {Glazman}}, \bibinfo {author} {\bibfnamefont {Peter}\
  \bibnamefont {Krogstrup}}, \bibinfo {author} {\bibfnamefont {Jesper}\
  \bibnamefont {Nyg{\aa}rd}}, \bibinfo {author} {\bibfnamefont {Leo~P.}\
  \bibnamefont {Kouwenhoven}}, \ and\ \bibinfo {author} {\bibfnamefont
  {Attila}\ \bibnamefont {Geresdi}},\ }\bibfield  {title} {\enquote {\bibinfo
  {title} {Microwave spectroscopy of spinful {Andreev} bound states in
  ballistic semiconductor {Josephson}~junctions},}\ }\href {\doibase
  10.1038/nphys4150} {\bibfield  {journal} {\bibinfo  {journal} {Nature Phys}\
  }\textbf {\bibinfo {volume} {13}},\ \bibinfo {pages} {876--881} (\bibinfo
  {year} {2017})}\BibitemShut {NoStop}%
\bibitem [{\citenamefont {Tosi}\ \emph {et~al.}(2019)\citenamefont {Tosi},
  \citenamefont {Metzger}, \citenamefont {Goffman}, \citenamefont {Urbina},
  \citenamefont {Pothier}, \citenamefont {Park}, \citenamefont {Yeyati},
  \citenamefont {Nyg\aa{}rd},\ and\ \citenamefont {Krogstrup}}]{Tosi_2019}%
  \BibitemOpen
  \bibfield  {author} {\bibinfo {author} {\bibfnamefont {L.}~\bibnamefont
  {Tosi}}, \bibinfo {author} {\bibfnamefont {C.}~\bibnamefont {Metzger}},
  \bibinfo {author} {\bibfnamefont {M.~F.}\ \bibnamefont {Goffman}}, \bibinfo
  {author} {\bibfnamefont {C.}~\bibnamefont {Urbina}}, \bibinfo {author}
  {\bibfnamefont {H.}~\bibnamefont {Pothier}}, \bibinfo {author} {\bibfnamefont
  {Sunghun}\ \bibnamefont {Park}}, \bibinfo {author} {\bibfnamefont {A.~Levy}\
  \bibnamefont {Yeyati}}, \bibinfo {author} {\bibfnamefont {J.}~\bibnamefont
  {Nyg\aa{}rd}}, \ and\ \bibinfo {author} {\bibfnamefont {P.}~\bibnamefont
  {Krogstrup}},\ }\bibfield  {title} {\enquote {\bibinfo {title} {Spin-orbit
  splitting of {Andreev} states revealed by microwave spectroscopy},}\ }\href
  {\doibase 10.1103/PhysRevX.9.011010} {\bibfield  {journal} {\bibinfo
  {journal} {Phys. Rev. X}\ }\textbf {\bibinfo {volume} {9}},\ \bibinfo {pages}
  {011010} (\bibinfo {year} {2019})}\BibitemShut {NoStop}%
\bibitem [{\citenamefont {Galpin}\ \emph {et~al.}(2006)\citenamefont {Galpin},
  \citenamefont {Logan},\ and\ \citenamefont {Krishnamurthy}}]{Galpin_2006}%
  \BibitemOpen
  \bibfield  {author} {\bibinfo {author} {\bibfnamefont {Martin~R}\
  \bibnamefont {Galpin}}, \bibinfo {author} {\bibfnamefont {David~E}\
  \bibnamefont {Logan}}, \ and\ \bibinfo {author} {\bibfnamefont {H~R}\
  \bibnamefont {Krishnamurthy}},\ }\bibfield  {title} {\enquote {\bibinfo
  {title} {Renormalization group study of capacitively coupled double quantum
  dots},}\ }\href {\doibase 10.1088/0953-8984/18/29/001} {\ \textbf {\bibinfo
  {volume} {18}},\ \bibinfo {pages} {6545--6570} (\bibinfo {year}
  {2006})}\BibitemShut {NoStop}%
\bibitem [{\citenamefont {Nishikawa}\ \emph {et~al.}(2012)\citenamefont
  {Nishikawa}, \citenamefont {Crow},\ and\ \citenamefont
  {Hewson}}]{Nishikawa_2012}%
  \BibitemOpen
  \bibfield  {author} {\bibinfo {author} {\bibfnamefont {Y.}~\bibnamefont
  {Nishikawa}}, \bibinfo {author} {\bibfnamefont {D.~J.~G.}\ \bibnamefont
  {Crow}}, \ and\ \bibinfo {author} {\bibfnamefont {A.~C.}\ \bibnamefont
  {Hewson}},\ }\bibfield  {title} {\enquote {\bibinfo {title} {Phase diagram
  and critical points of a double quantum dot},}\ }\href {\doibase
  10.1103/physrevb.86.125134} {\ \textbf {\bibinfo {volume} {86}} (\bibinfo
  {year} {2012}),\ 10.1103/physrevb.86.125134}\BibitemShut {NoStop}%
\bibitem [{\citenamefont {Oosterkamp}\ \emph {et~al.}(1998)\citenamefont
  {Oosterkamp}, \citenamefont {Fujisawa}, \citenamefont {van~der Wiel},
  \citenamefont {Ishibashi}, \citenamefont {Hijman}, \citenamefont {Tarucha},\
  and\ \citenamefont {Kouwenhoven}}]{Oosterkamp1998}%
  \BibitemOpen
  \bibfield  {author} {\bibinfo {author} {\bibfnamefont {T.~H.}\ \bibnamefont
  {Oosterkamp}}, \bibinfo {author} {\bibfnamefont {T.}~\bibnamefont
  {Fujisawa}}, \bibinfo {author} {\bibfnamefont {W.~G.}\ \bibnamefont {van~der
  Wiel}}, \bibinfo {author} {\bibfnamefont {K.}~\bibnamefont {Ishibashi}},
  \bibinfo {author} {\bibfnamefont {R.~V.}\ \bibnamefont {Hijman}}, \bibinfo
  {author} {\bibfnamefont {S.}~\bibnamefont {Tarucha}}, \ and\ \bibinfo
  {author} {\bibfnamefont {L.~P.}\ \bibnamefont {Kouwenhoven}},\ }\bibfield
  {title} {\enquote {\bibinfo {title} {Microwave spectroscopy of a quantum-dot
  molecule},}\ }\href {\doibase 10.1038/27617} {\bibfield  {journal} {\bibinfo
  {journal} {Nature}\ }\textbf {\bibinfo {volume} {395}},\ \bibinfo {pages}
  {873--876} (\bibinfo {year} {1998})}\BibitemShut {NoStop}%
\bibitem [{\citenamefont {Wallraff}\ \emph {et~al.}(2004)\citenamefont
  {Wallraff}, \citenamefont {Schuster}, \citenamefont {Blais}, \citenamefont
  {Frunzio}, \citenamefont {Huang}, \citenamefont {Majer}, \citenamefont
  {Kumar}, \citenamefont {Girvin},\ and\ \citenamefont
  {Schoelkopf}}]{Wallraff2004}%
  \BibitemOpen
  \bibfield  {author} {\bibinfo {author} {\bibfnamefont {A.}~\bibnamefont
  {Wallraff}}, \bibinfo {author} {\bibfnamefont {D.~I.}\ \bibnamefont
  {Schuster}}, \bibinfo {author} {\bibfnamefont {A.}~\bibnamefont {Blais}},
  \bibinfo {author} {\bibfnamefont {L.}~\bibnamefont {Frunzio}}, \bibinfo
  {author} {\bibfnamefont {R.-S.}\ \bibnamefont {Huang}}, \bibinfo {author}
  {\bibfnamefont {J.}~\bibnamefont {Majer}}, \bibinfo {author} {\bibfnamefont
  {S.}~\bibnamefont {Kumar}}, \bibinfo {author} {\bibfnamefont {S.~M.}\
  \bibnamefont {Girvin}}, \ and\ \bibinfo {author} {\bibfnamefont {R.~J.}\
  \bibnamefont {Schoelkopf}},\ }\bibfield  {title} {\enquote {\bibinfo {title}
  {Strong coupling of a single photon to a superconducting qubit using circuit
  quantum electrodynamics},}\ }\href {\doibase 10.1038/nature02851} {\bibfield
  {journal} {\bibinfo  {journal} {Nature}\ }\textbf {\bibinfo {volume} {431}},\
  \bibinfo {pages} {162--167} (\bibinfo {year} {2004})}\BibitemShut {NoStop}%
\bibitem [{\citenamefont {Stockklauser}\ \emph {et~al.}(2017)\citenamefont
  {Stockklauser}, \citenamefont {Scarlino}, \citenamefont {Koski},
  \citenamefont {Gasparinetti}, \citenamefont {Andersen}, \citenamefont
  {Reichl}, \citenamefont {Wegscheider}, \citenamefont {Ihn}, \citenamefont
  {Ensslin},\ and\ \citenamefont {Wallraff}}]{Stockklauser2017}%
  \BibitemOpen
  \bibfield  {author} {\bibinfo {author} {\bibfnamefont {A.}~\bibnamefont
  {Stockklauser}}, \bibinfo {author} {\bibfnamefont {P.}~\bibnamefont
  {Scarlino}}, \bibinfo {author} {\bibfnamefont {J.~V.}\ \bibnamefont {Koski}},
  \bibinfo {author} {\bibfnamefont {S.}~\bibnamefont {Gasparinetti}}, \bibinfo
  {author} {\bibfnamefont {C.~K.}\ \bibnamefont {Andersen}}, \bibinfo {author}
  {\bibfnamefont {C.}~\bibnamefont {Reichl}}, \bibinfo {author} {\bibfnamefont
  {W.}~\bibnamefont {Wegscheider}}, \bibinfo {author} {\bibfnamefont
  {T.}~\bibnamefont {Ihn}}, \bibinfo {author} {\bibfnamefont {K.}~\bibnamefont
  {Ensslin}}, \ and\ \bibinfo {author} {\bibfnamefont {A.}~\bibnamefont
  {Wallraff}},\ }\bibfield  {title} {\enquote {\bibinfo {title} {Strong
  coupling cavity qed with gate-defined double quantum dots enabled by a high
  impedance resonator},}\ }\href {\doibase 10.1103/PhysRevX.7.011030}
  {\bibfield  {journal} {\bibinfo  {journal} {Phys. Rev. X}\ }\textbf {\bibinfo
  {volume} {7}},\ \bibinfo {pages} {011030} (\bibinfo {year}
  {2017})}\BibitemShut {NoStop}%
\bibitem [{\citenamefont {Koski}\ \emph {et~al.}(2020)\citenamefont {Koski},
  \citenamefont {Landig}, \citenamefont {Russ}, \citenamefont {Abadillo-Uriel},
  \citenamefont {Scarlino}, \citenamefont {Kratochwil}, \citenamefont {Reichl},
  \citenamefont {Wegscheider}, \citenamefont {Burkard}, \citenamefont
  {Friesen}, \citenamefont {Coppersmith}, \citenamefont {Wallraff},
  \citenamefont {Ensslin},\ and\ \citenamefont {Ihn}}]{Koski2020}%
  \BibitemOpen
  \bibfield  {author} {\bibinfo {author} {\bibfnamefont {J.~V.}\ \bibnamefont
  {Koski}}, \bibinfo {author} {\bibfnamefont {A.~J.}\ \bibnamefont {Landig}},
  \bibinfo {author} {\bibfnamefont {M.}~\bibnamefont {Russ}}, \bibinfo {author}
  {\bibfnamefont {J.~C.}\ \bibnamefont {Abadillo-Uriel}}, \bibinfo {author}
  {\bibfnamefont {P.}~\bibnamefont {Scarlino}}, \bibinfo {author}
  {\bibfnamefont {B.}~\bibnamefont {Kratochwil}}, \bibinfo {author}
  {\bibfnamefont {C.}~\bibnamefont {Reichl}}, \bibinfo {author} {\bibfnamefont
  {W.}~\bibnamefont {Wegscheider}}, \bibinfo {author} {\bibfnamefont {Guido}\
  \bibnamefont {Burkard}}, \bibinfo {author} {\bibfnamefont {Mark}\
  \bibnamefont {Friesen}}, \bibinfo {author} {\bibfnamefont {S.~N.}\
  \bibnamefont {Coppersmith}}, \bibinfo {author} {\bibfnamefont
  {A.}~\bibnamefont {Wallraff}}, \bibinfo {author} {\bibfnamefont
  {K.}~\bibnamefont {Ensslin}}, \ and\ \bibinfo {author} {\bibfnamefont
  {T.}~\bibnamefont {Ihn}},\ }\bibfield  {title} {\enquote {\bibinfo {title}
  {Strong photon coupling to the quadrupole moment of an electron in a
  solid-state qubit},}\ }\href {\doibase 10.1038/s41567-020-0862-4} {\bibfield
  {journal} {\bibinfo  {journal} {Nature Physics}\ }\textbf {\bibinfo {volume}
  {16}},\ \bibinfo {pages} {642--646} (\bibinfo {year} {2020})}\BibitemShut
  {NoStop}%
\bibitem [{\citenamefont {Loss}\ and\ \citenamefont
  {DiVincenzo}(1998)}]{Loss_1998}%
  \BibitemOpen
  \bibfield  {author} {\bibinfo {author} {\bibfnamefont {Daniel}\ \bibnamefont
  {Loss}}\ and\ \bibinfo {author} {\bibfnamefont {David~P.}\ \bibnamefont
  {DiVincenzo}},\ }\bibfield  {title} {\enquote {\bibinfo {title} {Quantum
  computation with quantum dots},}\ }\href@noop {} {\bibfield  {journal}
  {\bibinfo  {journal} {Phys. Rev. A}\ }\textbf {\bibinfo {volume} {57}},\
  \bibinfo {pages} {120} (\bibinfo {year} {1998})}\BibitemShut {NoStop}%
\bibitem [{\citenamefont {Nielsen}\ and\ \citenamefont
  {Chuang}(2010)}]{nielsen_chuang_2010}%
  \BibitemOpen
  \bibfield  {author} {\bibinfo {author} {\bibfnamefont {Michael~A.}\
  \bibnamefont {Nielsen}}\ and\ \bibinfo {author} {\bibfnamefont {Isaac~L.}\
  \bibnamefont {Chuang}},\ }\href {\doibase 10.1017/CBO9780511976667} {\emph
  {\bibinfo {title} {Quantum Computation and Quantum Information: 10th
  Anniversary Edition}}}\ (\bibinfo  {publisher} {Cambridge University Press},\
  \bibinfo {year} {2010})\BibitemShut {NoStop}%
\bibitem [{\citenamefont {Kjaergaard}\ \emph {et~al.}(2020)\citenamefont
  {Kjaergaard}, \citenamefont {Schwartz}, \citenamefont {Braum\"{u}ller},
  \citenamefont {Krantz}, \citenamefont {Wang}, \citenamefont {Gustavsson},\
  and\ \citenamefont {Oliver}}]{Kjaergaard2020}%
  \BibitemOpen
  \bibfield  {author} {\bibinfo {author} {\bibfnamefont {Morten}\ \bibnamefont
  {Kjaergaard}}, \bibinfo {author} {\bibfnamefont {Mollie~E.}\ \bibnamefont
  {Schwartz}}, \bibinfo {author} {\bibfnamefont {Jochen}\ \bibnamefont
  {Braum\"{u}ller}}, \bibinfo {author} {\bibfnamefont {Philip}\ \bibnamefont
  {Krantz}}, \bibinfo {author} {\bibfnamefont {Joel I.-J.}\ \bibnamefont
  {Wang}}, \bibinfo {author} {\bibfnamefont {Simon}\ \bibnamefont
  {Gustavsson}}, \ and\ \bibinfo {author} {\bibfnamefont {William~D.}\
  \bibnamefont {Oliver}},\ }\bibfield  {title} {\enquote {\bibinfo {title}
  {Superconducting qubits: Current state of play},}\ }\href {\doibase
  10.1146/annurev-conmatphys-031119-050605} {\bibfield  {journal} {\bibinfo
  {journal} {Annual Review of Condensed Matter Physics}\ }\textbf {\bibinfo
  {volume} {11}},\ \bibinfo {pages} {369--395} (\bibinfo {year}
  {2020})}\BibitemShut {NoStop}%
\bibitem [{\citenamefont {Janvier}\ \emph {et~al.}(2015)\citenamefont
  {Janvier}, \citenamefont {Tosi}, \citenamefont {Bretheau}, \citenamefont
  {Girit}, \citenamefont {Stern}, \citenamefont {Bertet}, \citenamefont
  {Joyez}, \citenamefont {Vion}, \citenamefont {Esteve}, \citenamefont
  {Goffman}, \citenamefont {Pothier},\ and\ \citenamefont
  {Urbina}}]{Janvier_2015}%
  \BibitemOpen
  \bibfield  {author} {\bibinfo {author} {\bibfnamefont {C.}~\bibnamefont
  {Janvier}}, \bibinfo {author} {\bibfnamefont {L.}~\bibnamefont {Tosi}},
  \bibinfo {author} {\bibfnamefont {L.}~\bibnamefont {Bretheau}}, \bibinfo
  {author} {\bibfnamefont {{\c{C}}.~\"{O}.}\ \bibnamefont {Girit}}, \bibinfo
  {author} {\bibfnamefont {M.}~\bibnamefont {Stern}}, \bibinfo {author}
  {\bibfnamefont {P.}~\bibnamefont {Bertet}}, \bibinfo {author} {\bibfnamefont
  {P.}~\bibnamefont {Joyez}}, \bibinfo {author} {\bibfnamefont
  {D.}~\bibnamefont {Vion}}, \bibinfo {author} {\bibfnamefont {D.}~\bibnamefont
  {Esteve}}, \bibinfo {author} {\bibfnamefont {M.~F.}\ \bibnamefont {Goffman}},
  \bibinfo {author} {\bibfnamefont {H.}~\bibnamefont {Pothier}}, \ and\
  \bibinfo {author} {\bibfnamefont {C.}~\bibnamefont {Urbina}},\ }\bibfield
  {title} {\enquote {\bibinfo {title} {Coherent manipulation of {Andreev}
  states in superconducting atomic contacts},}\ }\href {\doibase
  10.1126/science.aab2179} {\bibfield  {journal} {\bibinfo  {journal}
  {Science}\ }\textbf {\bibinfo {volume} {349}},\ \bibinfo {pages} {1199--1202}
  (\bibinfo {year} {2015})}\BibitemShut {NoStop}%
\bibitem [{\citenamefont {Hays}\ \emph {et~al.}(2018)\citenamefont {Hays},
  \citenamefont {de~Lange}, \citenamefont {Serniak}, \citenamefont {van
  Woerkom}, \citenamefont {Bouman}, \citenamefont {Krogstrup}, \citenamefont
  {Nyg\aa{}rd}, \citenamefont {Geresdi},\ and\ \citenamefont
  {Devoret}}]{Hays_2018}%
  \BibitemOpen
  \bibfield  {author} {\bibinfo {author} {\bibfnamefont {M.}~\bibnamefont
  {Hays}}, \bibinfo {author} {\bibfnamefont {G.}~\bibnamefont {de~Lange}},
  \bibinfo {author} {\bibfnamefont {K.}~\bibnamefont {Serniak}}, \bibinfo
  {author} {\bibfnamefont {D.~J.}\ \bibnamefont {van Woerkom}}, \bibinfo
  {author} {\bibfnamefont {D.}~\bibnamefont {Bouman}}, \bibinfo {author}
  {\bibfnamefont {P.}~\bibnamefont {Krogstrup}}, \bibinfo {author}
  {\bibfnamefont {J.}~\bibnamefont {Nyg\aa{}rd}}, \bibinfo {author}
  {\bibfnamefont {A.}~\bibnamefont {Geresdi}}, \ and\ \bibinfo {author}
  {\bibfnamefont {M.~H.}\ \bibnamefont {Devoret}},\ }\bibfield  {title}
  {\enquote {\bibinfo {title} {Direct microwave measurement of
  {Andreev}-bound-state dynamics in a semiconductor-nanowire {Josephson}
  junction},}\ }\href {\doibase 10.1103/PhysRevLett.121.047001} {\bibfield
  {journal} {\bibinfo  {journal} {Phys. Rev. Lett.}\ }\textbf {\bibinfo
  {volume} {121}},\ \bibinfo {pages} {047001} (\bibinfo {year}
  {2018})}\BibitemShut {NoStop}%
\bibitem [{\citenamefont {Gorman}\ \emph {et~al.}(2005)\citenamefont {Gorman},
  \citenamefont {Hasko},\ and\ \citenamefont {Williams}}]{gorman2005}%
  \BibitemOpen
  \bibfield  {author} {\bibinfo {author} {\bibfnamefont {J.}~\bibnamefont
  {Gorman}}, \bibinfo {author} {\bibfnamefont {D.~G.}\ \bibnamefont {Hasko}}, \
  and\ \bibinfo {author} {\bibfnamefont {D.~A.}\ \bibnamefont {Williams}},\
  }\bibfield  {title} {\enquote {\bibinfo {title} {Charge-qubit operation of an
  isolated double quantum dot},}\ }\href {\doibase
  10.1103/PhysRevLett.95.090502} {\bibfield  {journal} {\bibinfo  {journal}
  {Phys. Rev. Lett.}\ }\textbf {\bibinfo {volume} {95}},\ \bibinfo {pages}
  {090502} (\bibinfo {year} {2005})}\BibitemShut {NoStop}%
\bibitem [{\citenamefont {Shnirman}\ \emph {et~al.}(1997)\citenamefont
  {Shnirman}, \citenamefont {Sch\"on},\ and\ \citenamefont
  {Hermon}}]{Shnirman1997}%
  \BibitemOpen
  \bibfield  {author} {\bibinfo {author} {\bibfnamefont {Alexander}\
  \bibnamefont {Shnirman}}, \bibinfo {author} {\bibfnamefont {Gerd}\
  \bibnamefont {Sch\"on}}, \ and\ \bibinfo {author} {\bibfnamefont {Ziv}\
  \bibnamefont {Hermon}},\ }\bibfield  {title} {\enquote {\bibinfo {title}
  {Quantum manipulations of small {Josephson} junctions},}\ }\href {\doibase
  10.1103/PhysRevLett.79.2371} {\bibfield  {journal} {\bibinfo  {journal}
  {Phys. Rev. Lett.}\ }\textbf {\bibinfo {volume} {79}},\ \bibinfo {pages}
  {2371--2374} (\bibinfo {year} {1997})}\BibitemShut {NoStop}%
\bibitem [{\citenamefont {Bouchiat}\ \emph {et~al.}(1998)\citenamefont
  {Bouchiat}, \citenamefont {Vion}, \citenamefont {Joyez}, \citenamefont
  {Esteve},\ and\ \citenamefont {Devoret}}]{Bouchiat1998}%
  \BibitemOpen
  \bibfield  {author} {\bibinfo {author} {\bibfnamefont {V.}~\bibnamefont
  {Bouchiat}}, \bibinfo {author} {\bibfnamefont {D.}~\bibnamefont {Vion}},
  \bibinfo {author} {\bibfnamefont {P.}~\bibnamefont {Joyez}}, \bibinfo
  {author} {\bibfnamefont {D.}~\bibnamefont {Esteve}}, \ and\ \bibinfo {author}
  {\bibfnamefont {M.~H.}\ \bibnamefont {Devoret}},\ }\bibfield  {title}
  {\enquote {\bibinfo {title} {Quantum coherence with a single {Cooper}
  pair},}\ }\href {\doibase 10.1238/physica.topical.076a00165} {\bibfield
  {journal} {\bibinfo  {journal} {Physica Scripta}\ }\textbf {\bibinfo {volume}
  {T76}},\ \bibinfo {pages} {165} (\bibinfo {year} {1998})}\BibitemShut
  {NoStop}%
\bibitem [{\citenamefont {Makhlin}\ \emph {et~al.}(2001)\citenamefont
  {Makhlin}, \citenamefont {Sch\"on},\ and\ \citenamefont
  {Shnirman}}]{Makhlin2001}%
  \BibitemOpen
  \bibfield  {author} {\bibinfo {author} {\bibfnamefont {Yuriy}\ \bibnamefont
  {Makhlin}}, \bibinfo {author} {\bibfnamefont {Gerd}\ \bibnamefont {Sch\"on}},
  \ and\ \bibinfo {author} {\bibfnamefont {Alexander}\ \bibnamefont
  {Shnirman}},\ }\bibfield  {title} {\enquote {\bibinfo {title} {Quantum-state
  engineering with {Josephson-junction} devices},}\ }\href {\doibase
  10.1103/RevModPhys.73.357} {\bibfield  {journal} {\bibinfo  {journal} {Rev.
  Mod. Phys.}\ }\textbf {\bibinfo {volume} {73}},\ \bibinfo {pages} {357--400}
  (\bibinfo {year} {2001})}\BibitemShut {NoStop}%
\bibitem [{\citenamefont {Koch}\ \emph {et~al.}(2007)\citenamefont {Koch},
  \citenamefont {Yu}, \citenamefont {Gambetta}, \citenamefont {Houck},
  \citenamefont {Schuster}, \citenamefont {Majer}, \citenamefont {Blais},
  \citenamefont {Devoret}, \citenamefont {Girvin},\ and\ \citenamefont
  {Schoelkopf}}]{Koch_2007}%
  \BibitemOpen
  \bibfield  {author} {\bibinfo {author} {\bibfnamefont {Jens}\ \bibnamefont
  {Koch}}, \bibinfo {author} {\bibfnamefont {Terri~M.}\ \bibnamefont {Yu}},
  \bibinfo {author} {\bibfnamefont {Jay}\ \bibnamefont {Gambetta}}, \bibinfo
  {author} {\bibfnamefont {A.~A.}\ \bibnamefont {Houck}}, \bibinfo {author}
  {\bibfnamefont {D.~I.}\ \bibnamefont {Schuster}}, \bibinfo {author}
  {\bibfnamefont {J.}~\bibnamefont {Majer}}, \bibinfo {author} {\bibfnamefont
  {Alexandre}\ \bibnamefont {Blais}}, \bibinfo {author} {\bibfnamefont {M.~H.}\
  \bibnamefont {Devoret}}, \bibinfo {author} {\bibfnamefont {S.~M.}\
  \bibnamefont {Girvin}}, \ and\ \bibinfo {author} {\bibfnamefont {R.~J.}\
  \bibnamefont {Schoelkopf}},\ }\bibfield  {title} {\enquote {\bibinfo {title}
  {Charge-insensitive qubit design derived from the {Cooper} pair box},}\
  }\href {\doibase 10.1103/PhysRevA.76.042319} {\bibfield  {journal} {\bibinfo
  {journal} {Phys. Rev. A}\ }\textbf {\bibinfo {volume} {76}},\ \bibinfo
  {pages} {042319} (\bibinfo {year} {2007})}\BibitemShut {NoStop}%
\bibitem [{\citenamefont {Sala}\ and\ \citenamefont {Danon}(2017)}]{Sala_2017}%
  \BibitemOpen
  \bibfield  {author} {\bibinfo {author} {\bibfnamefont {Arnau}\ \bibnamefont
  {Sala}}\ and\ \bibinfo {author} {\bibfnamefont {Jeroen}\ \bibnamefont
  {Danon}},\ }\bibfield  {title} {\enquote {\bibinfo {title} {Exchange-only
  singlet-only spin qubit},}\ }\href {\doibase 10.1103/PhysRevB.95.241303}
  {\bibfield  {journal} {\bibinfo  {journal} {Phys. Rev. B}\ }\textbf {\bibinfo
  {volume} {95}},\ \bibinfo {pages} {241303} (\bibinfo {year}
  {2017})}\BibitemShut {NoStop}%
\bibitem [{\citenamefont {Sala}\ \emph {et~al.}(2020)\citenamefont {Sala},
  \citenamefont {Qvist},\ and\ \citenamefont {Danon}}]{Sala_2020}%
  \BibitemOpen
  \bibfield  {author} {\bibinfo {author} {\bibfnamefont {Arnau}\ \bibnamefont
  {Sala}}, \bibinfo {author} {\bibfnamefont {J\o{}rgen~Holme}\ \bibnamefont
  {Qvist}}, \ and\ \bibinfo {author} {\bibfnamefont {Jeroen}\ \bibnamefont
  {Danon}},\ }\bibfield  {title} {\enquote {\bibinfo {title} {Highly tunable
  exchange-only singlet-only qubit in a {GaAs} triple quantum dot},}\ }\href
  {\doibase 10.1103/PhysRevResearch.2.012062} {\bibfield  {journal} {\bibinfo
  {journal} {Phys. Rev. Research}\ }\textbf {\bibinfo {volume} {2}},\ \bibinfo
  {pages} {012062} (\bibinfo {year} {2020})}\BibitemShut {NoStop}%
\bibitem [{\citenamefont {Mishra}\ \emph {et~al.}(2021)\citenamefont {Mishra},
  \citenamefont {Simon}, \citenamefont {Hyart},\ and\ \citenamefont
  {Trif}}]{mishra2021}%
  \BibitemOpen
  \bibfield  {author} {\bibinfo {author} {\bibfnamefont {Archana}\ \bibnamefont
  {Mishra}}, \bibinfo {author} {\bibfnamefont {Pascal}\ \bibnamefont {Simon}},
  \bibinfo {author} {\bibfnamefont {Timo}\ \bibnamefont {Hyart}}, \ and\
  \bibinfo {author} {\bibfnamefont {Mircea}\ \bibnamefont {Trif}},\ }\href@noop
  {} {\enquote {\bibinfo {title} {{A Yu-Shiba-Rusinov qubit}},}\ } (\bibinfo
  {year} {2021}),\ \Eprint {http://arxiv.org/abs/2106.01188} {arXiv:2106.01188
  [cond-mat.mes-hall]} \BibitemShut {NoStop}%
\bibitem [{\citenamefont {Fishman}\ \emph {et~al.}(2020)\citenamefont
  {Fishman}, \citenamefont {White},\ and\ \citenamefont
  {Stoudenmire}}]{itensor}%
  \BibitemOpen
  \bibfield  {author} {\bibinfo {author} {\bibfnamefont {Matthew}\ \bibnamefont
  {Fishman}}, \bibinfo {author} {\bibfnamefont {Steven~R.}\ \bibnamefont
  {White}}, \ and\ \bibinfo {author} {\bibfnamefont {E.~Miles}\ \bibnamefont
  {Stoudenmire}},\ }\bibfield  {title} {\enquote {\bibinfo {title} {The
  \mbox{ITensor} software library for tensor network calculations},}\
  }\href@noop {} {\  (\bibinfo {year} {2020})},\ \Eprint
  {http://arxiv.org/abs/2007.14822} {arXiv:2007.14822} \BibitemShut {NoStop}%
\bibitem [{\citenamefont {Pustilnik}\ and\ \citenamefont
  {Glazman}(2004)}]{pustilnik}%
  \BibitemOpen
  \bibfield  {author} {\bibinfo {author} {\bibfnamefont {Michael}\ \bibnamefont
  {Pustilnik}}\ and\ \bibinfo {author} {\bibfnamefont {Leonid}\ \bibnamefont
  {Glazman}},\ }\bibfield  {title} {\enquote {\bibinfo {title} {Kondo effect in
  quantum dots},}\ }\href {\doibase 10.1088/0953-8984/16/16/r01} {\bibfield
  {journal} {\bibinfo  {journal} {J. Phys.: Condens. Matter}\ }\textbf
  {\bibinfo {volume} {16}},\ \bibinfo {pages} {R513--R537} (\bibinfo {year}
  {2004})}\BibitemShut {NoStop}%
\bibitem [{\citenamefont {Choi}\ \emph {et~al.}(2004)\citenamefont {Choi},
  \citenamefont {Lee}, \citenamefont {Kang},\ and\ \citenamefont
  {Belzig}}]{choi2004}%
  \BibitemOpen
  \bibfield  {author} {\bibinfo {author} {\bibfnamefont {Mahn-Soo}\
  \bibnamefont {Choi}}, \bibinfo {author} {\bibfnamefont {Minchul}\
  \bibnamefont {Lee}}, \bibinfo {author} {\bibfnamefont {Kicheon}\ \bibnamefont
  {Kang}}, \ and\ \bibinfo {author} {\bibfnamefont {W.}~\bibnamefont
  {Belzig}},\ }\bibfield  {title} {\enquote {\bibinfo {title} {Kondo effect and
  {Josephson} current through a quantum dot between two superconductors},}\
  }\href {\doibase 10.1103/PhysRevB.70.020502} {\bibfield  {journal} {\bibinfo
  {journal} {Phys. Rev. B}\ }\textbf {\bibinfo {volume} {70}},\ \bibinfo
  {pages} {020502} (\bibinfo {year} {2004})}\BibitemShut {NoStop}%
\end{thebibliography}%

\pagebreak 
\appendix 

\begin{widetext}

\section{Transformation to mirror-symmetry-adapted (gerade/ungerade) basis}
\label{appA}

\subsection{Transformation of the BCS Hamiltonian} 

The mirror-symmetric problem of a quantum dot described by the single-impurity Anderson model that is coupled to two BCS superconductors with the same gap $\Delta$ and the same superconducting phase $\phi$ maps to a single-channel problem using a simple transformation to the mirror-symmetry basis (gerade and ungerade parity):
\begin{equation}
\begin{split}
g^\dag_{i,\sigma} &= \frac{1}{\sqrt{2}} \left( c^\dag_{i,\sigma,L} +
c^\dag_{i,\sigma,R} \right), \\
u^\dag_{i,\sigma} &= \frac{1}{\sqrt{2}} \left( c^\dag_{i,\sigma,L} -
c^\dag_{i,\sigma,R} \right).
\label{eq:even-odd_transform}
\end{split}
\end{equation}
One finds
\begin{equation}
    \Delta c^\dag_{i,\uparrow,L} c^\dag_{i,\downarrow,L}
+   \Delta c^\dag_{i,\uparrow,R} c^\dag_{i,\downarrow,R}
+ \text{H.c.}
=
    \Delta g^\dag_{i,\uparrow} g^\dag_{i,\downarrow}
+   \Delta u^\dag_{i,\uparrow} u^\dag_{i,\downarrow}
+ \text{H.c.}
\end{equation}
There are no terms mixing the g/u superconductor modes, thus the ungerade superconductor fully decouples from the problem, resulting in a single-channel problem, as is the case for normal-state contacts \cite{pustilnik} . At mathematical level this is trivial, but physically it is less so. The cancellation of the mixed gerade/ungerade terms is due to the mean-field decoupling of the effective electron-electron interaction terms and implies the assumption of complete coherence between the condensates in both SC contacts. 

If there exists a flux bias $\phi$, one proceeds in two steps. First one performs a gauge transformation
\begin{equation}
\begin{split}
    \tilde{c}^\dag_{i,\sigma,L} &= e^{i \phi/4} c^\dag_{i,\sigma,L}, \\
    \tilde{c}^\dag_{i,\sigma,R} &= e^{-i \phi/4} c^\dag_{i,\sigma,R}, \\
    \tilde{d}^\dag_\sigma &= e^{i \phi/4} d^\dag_\sigma.
\end{split}
\end{equation}
One then performs a transformation to g/u basis in terms of $\tilde{c}$ operators:
\begin{equation}
\begin{split}
g^\dag_{i,\sigma} &= \frac{1}{\sqrt{2}} \left( \tilde{c}^\dag_{i,\sigma,L} +
\tilde{c}^\dag_{i,\sigma,R} \right), \\
u^\dag_{i,\sigma} &= \frac{1}{\sqrt{2}} \left( \tilde{c}^\dag_{i,\sigma,L} -
\tilde{c}^\dag_{i,\sigma,R} \right).
\end{split}
\end{equation}
One finds 
\begin{equation}
 \begin{split}
 &   \Delta e^{i\phi/2} c^\dag_{i,\uparrow,L} c^\dag_{i,\downarrow,L}
+   \Delta e^{-i\phi/2} c^\dag_{i,\uparrow,R} c^\dag_{i,\downarrow,R}
+    t \sum_\sigma c^\dag_{i,\sigma,L} d_\sigma 
+    t \sum_\sigma c^\dag_{i,\sigma,R} d_\sigma
+ \text{H.c.}\\
&= 
    \Delta g^\dag_{i,\uparrow} g^\dag_{i,\downarrow}
+   \Delta u^\dag_{i,\uparrow} u^\dag_{i,\downarrow}
+ \sqrt{2} t \cos(\phi/4) \sum_\sigma g^\dag_{i,\sigma} d_\sigma
+ i \sqrt{2} t \sin(\phi/4) \sum_\sigma u^\dag_{i,\sigma} d_\sigma
+ \text{H.c.}
\end{split}
\end{equation}
The QD is coupled to the $g$ modes with the hybridisation multiplied by
a factor $2\cos^2(\phi/4)$ and to the $u$ modes with a factor $2\sin^2(\phi/4)$ \cite{choi2004}.
In the presence of phase-bias, the $u$ modes thus do not decouple. This leads to the second singlet YSR state \cite{Kirsanskas_2015}.

More generally, for an Anderson-model QD coupled to an arbitrary number of BCS superconductors with arbitrary hybridisation strengths and arbitrary gap parameters (amplitude and phase), the BCS superconductors can be integrated out resulting in a single hybridisation function in the Nambu space. In this sense the problem is always effectively a single-channel problem. Nevertheless, the non-trivial structure of the hybridisation matrix as a function of frequency reflects the complexity of the setup and can lead to, for example, the existence of multiple singlet Yu-Shiba-Rusinov states in the presence of flux biases.

\subsection{Transformation of the Richardson model}
Given that the Richardson model is equivalent to the BCS model in the limit of large $N$, a similar procedure is possible. 
Applying the even odd transformation \eqref{eq:even-odd_transform} to the number conserving pairing hamiltonian
\begin{equation}
    \sum_{i,j}^N \ciupL^\dagger \cidnL^\dagger \cjdnL \cjupL + \sum_{i,j}^N \ciupR^\dagger \cidnR^\dagger \cjdnR \cjupR,
\end{equation}
one finds three types of terms. There are intra-channel pairing terms:
\begin{equation}
    \uiup^\dagger \uidn^\dagger \ujdn \ujup + \gjup^\dagger \gjdn^\dagger \gidn \giup,
\end{equation}
the terms that pair levels between channels:
\begin{equation}
    \uiup^\dagger \uidn^\dagger \gjdn \gjup + \gjup^\dagger \gjdn^\dagger \uidn \uiup,
\end{equation}
and finally the terms that mix single particle states from both channels:
\begin{equation}
    \uiup^\dagger \gidn^\dagger \ujdn \gjup + \gjup^\dagger \ujdn^\dagger \gidn \uiup,
\end{equation}
\begin{equation}
    \uiup^\dagger \gidn^\dagger \gjdn \ujup + \gjup^\dagger \ujdn^\dagger \uidn \giup.
\end{equation}
The terms which couple $u$ and $g$ modes have random phase and give zero contribution in the limit of $N \to \infty$. The two channels therefore decouple in the Richardson model too. 
\section{MPO representation}
\label{appB}

We provide the matrix product operator representation of the Hamiltonian studied in this work. It is implemented with the impurity in the middle of the system, with $N$ sites on each side of it. An alternative implementation is possible with the impurity as the left-most site. We obtain exactly the same results with both approaches and do not observe very significant differences in computational efficiency of between implementations.
Left-most site (1 denotes the first level in the left channel):
\begin{equation}
W_1 = \begin{pmatrix}
I & [\epsilon_1 + \ECL (1-2\nL)] \hat{n}_1 + (g+2\ECL) \hat{n}_{1\uparrow}\hat{n}_{1\downarrow} &
v c^\dag_{1\uparrow} F_1 &
v c^\dag_{1\downarrow} F_1 &
-v c_{1\uparrow} F_1 &
-v c_{1\downarrow} F_1 &
g c_{1\downarrow} c_{1\uparrow} &
g c^\dag_{1\uparrow} c^\dag_{1\downarrow} &
2\ECL n_1
\end{pmatrix}.
\end{equation}
Here $F_i=(-1)^n$ is the local fermionic-parity operator, which gives phase of $-1$ if there is an odd number of
electrons on the site. Charge operators are 
\begin{equation}
\hat{n}_{i\sigma} = c^\dag_{i\sigma} c_{i\sigma}, \quad
\hat{n}_i = \sum_\sigma \hat{n}_{i\sigma},
\end{equation}

Generic site in the left half of the system ($i$ denotes a level in the left channel):
\begin{equation}
W_i = \begin{pmatrix}
1 & [\epsilon_i + \ECL (1-2\nL)] \hat{n}_i + (g+2\ECL) \hat{n}_{i\uparrow} \hat{n}_{i\downarrow} & 
v c^\dag_{i\uparrow} F_i & v c^\dag_{i\downarrow} F_i & -v c_{i\uparrow} F_i & -v c_{i\downarrow} F_i &
g c_{i\downarrow} c_{i\uparrow} & g c^\dag_{i\uparrow} c^\dag_{i\downarrow} & 2 \ECL \hat{n}_{i} \\
0 & I & 0 & 0 & 0 & 0 & 0 & 0 & 0 \\
0 & 0 & F_i & 0 & 0 & 0 & 0 & 0 & 0 \\
0 & 0 & 0 & F_i & 0 & 0 & 0 & 0 & 0 \\
0 & 0 & 0 & 0 & F_i & 0 & 0 & 0 & 0 \\
0 & 0 & 0 & 0 & 0 & F_i & 0 & 0 & 0 \\
0 & c^\dag_{i\uparrow} c^\dag_{i\downarrow} & 0 & 0 & 0 & 0 & I & 0 & 0 \\
0 & c_{i\downarrow} c_{i\uparrow} & 0 & 0 & 0 & 0 & 0 & I & 0 \\
0 & \hat{n}_i & 0 & 0 & 0 & 0 & 0 & 0 & I
\end{pmatrix}.
\end{equation}

Impurity site:
\begin{equation}
W_\mathrm{imp} = \begin{pmatrix}
1 & \epsilon_\mathrm{imp} \hat{n}_\mathrm{imp} + U \hat{n}_{\mathrm{imp}\uparrow}
\hat{n}_{\mathrm{imp}\downarrow} &
-d^\dag_{\uparrow} F & -d^\dag_{\downarrow} F & d_{\uparrow} F & d_{\downarrow} F &
0 & 0 & 0 \\
0 & I & 0 & 0 & 0 & 0 & 0 & 0 & 0 \\
0 & d_\uparrow & 0 & 0 & 0 & 0 & 0 & 0 & 0 \\
0 & d_\downarrow & 0 & 0 & 0 & 0 & 0 & 0 & 0 \\
0 & d^\dag_\uparrow & 0 & 0 & 0 & 0 & 0 & 0 & 0 \\
0 & d^\dag_\downarrow & 0 & 0 & 0 & 0 & 0 & 0 & 0 \\
0 & 0 & 0 & 0 & 0 & 0 & 0 & 0 & 0 \\
0 & 0 & 0 & 0 & 0 & 0 & 0 & 0 & 0 \\
0 & 0 & 0 & 0 & 0 & 0 & 0 & 0 & 0
\end{pmatrix}.
\end{equation}

Generic site in the right half of the system ($i$ denotes a level in the right channel):
\begin{equation}
W_i = \begin{pmatrix}
1 & [\epsilon_i + \ECR (1-2\nR)] \hat{n}_i + (g+2\ECR) \hat{n}_{i\uparrow} \hat{n}_{i\downarrow} & 
0 & 0 & 0 & 0 & 
g c_{i\downarrow} c_{i\uparrow} & g c^\dag_{i\uparrow} c^\dag_{i\downarrow} & 2 \ECR \hat{n}_{i} \\
0 & I & 0 & 0 & 0 & 0 & 0 & 0 & 0 \\
0 & v c^\dag_{i\uparrow} & F_i & 0 & 0 & 0 & 0 & 0 & 0 \\
0 & v c^\dag_{i\downarrow} & 0 & F_i & 0 & 0 & 0 & 0 & 0 \\
0 & v c_{i\uparrow} & 0 & 0 & F_i & 0 & 0 & 0 & 0 \\
0 & v c_{i\downarrow} & 0 & 0 & 0 & F_i & 0 & 0 & 0 \\
0 & c^\dag_{i\uparrow} c^\dag_{i\downarrow} & 0 & 0 & 0 & 0 & I & 0 & 0 \\
0 & c_{i\downarrow} c_{i\uparrow} & 0 & 0 & 0 & 0 & 0 & I & 0 \\
0 & \hat{n}_i & 0 & 0 & 0 & 0 & 0 & 0 & I
\end{pmatrix}.
\end{equation}

Right-most site ($N$ denotes the last level in the right channel):
\begin{equation}
W_N = \begin{pmatrix}
[\epsilon_N + \ECR(1-2\nR)] \hat{n}_N + (g+2\ECR) \hat{n}_{N\uparrow} \hat{n}_{N\downarrow} \\
I \\
v c^\dag_{N\uparrow} \\
v c^\dag_{N\downarrow} \\
v c_{N\uparrow} \\
v c_{N\downarrow} \\
c^\dag_{N\uparrow} c^\dag_{N\downarrow} \\
c_{N\downarrow} c_{N\uparrow} \\
\hat{n}_N
\end{pmatrix}.
\end{equation}

\end{widetext}

\section{Zero-bandwidth calculation of changing nature of subgap states}
\label{appC}

\begin{figure}
\centering
\includegraphics[width=\linewidth]{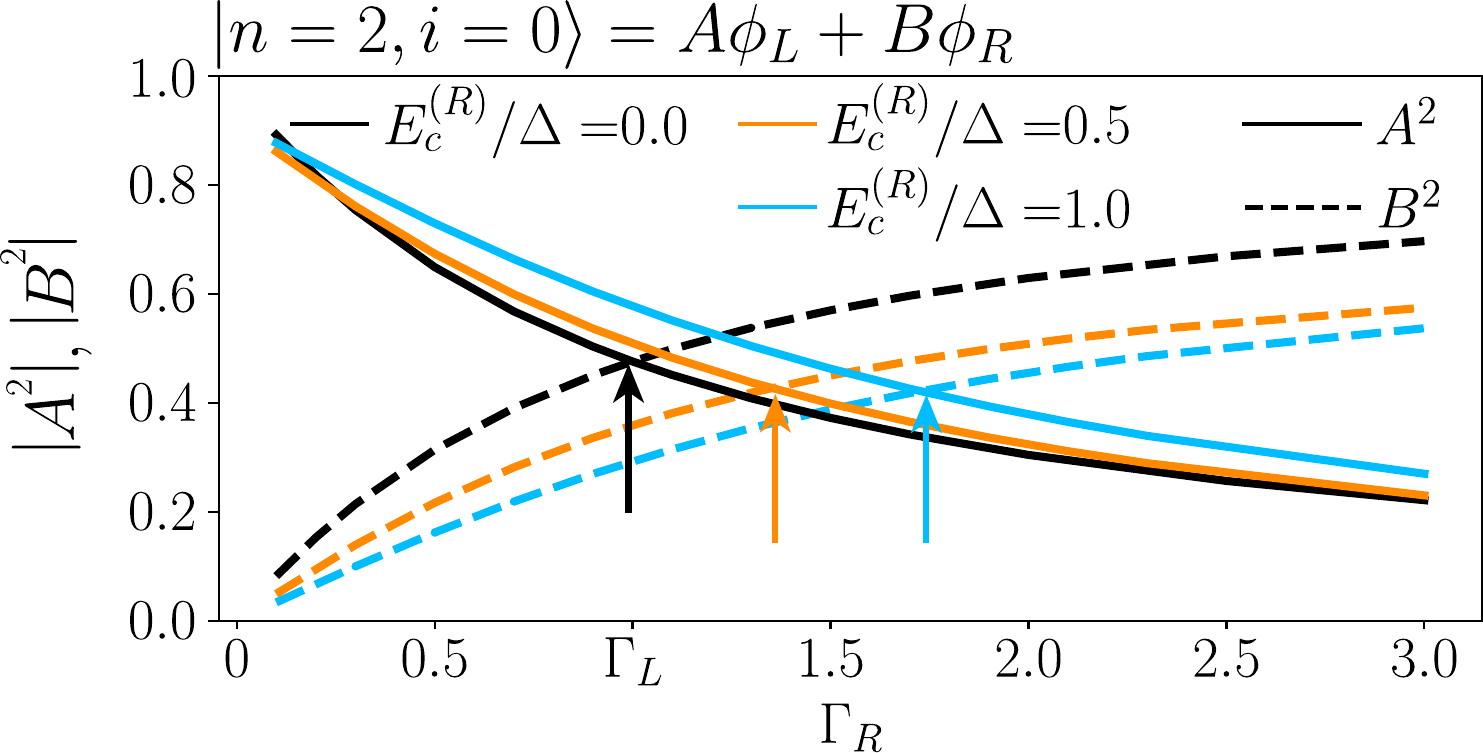}
\caption{
Probabilities of the $\phiL$ and $\phiR$ states for the lowest lying singlet state in the zero-bandwidth approximation ($\mathcal{N} = 1$). The colored arrows point to the $\vert A^2 \vert = \vert B^2 \vert$ points. $\ECL=0$, various $\ECR$. $N=1$, $U=10$, $\Gamma_L=1$.
}
\label{fig:ZBA}
\end{figure}

The contribution of the left and right channel singlets can be directly calculated in the zero bandwidth limit (the SC islands are represented by a single level $\mathcal{N}=1$), where the singlet ground state with two particles is $\ket{n=2, i=0} \approx A \phiL + B \phiR$. Fig.~\ref{fig:ZBA} shows the $\GR$ dependence of the probability amplitudes $A$ and $B$. 
Colored arrows point to the $\vert A \vert = \vert B \vert$ points, where the state is an equal superposition of the left and right channel singlets. When $\ECL=\ECR=0$, the equal superposition point occurs exactly at $\GL=\GR$, while it is pushed towards larger $\GR$ with increasing $\ECR$.  

\end{document}